\documentclass[floats,floatfix,amssymb,prd,twocolumn,superscriptaddress,nofootinbib,longbibliography]{revtex4-1}

\usepackage{graphicx,amssymb,amsmath,amsthm,amsfonts,epsfig,epsf}
\usepackage[linktocpage]{hyperref}
\usepackage[usenames]{color}
\usepackage{epstopdf}
\usepackage{bm}
\usepackage{dcolumn}
\usepackage{rotating}
\usepackage{hyperref}
\usepackage{color}
\usepackage{longtable}

\usepackage{enumerate}
\usepackage{tensor,multirow}
\usepackage{url}

\DeclareMathOperator{\diag}{diag} 

\makeatletter
\def\l@subsubsection#1#2{}
\makeatother

\begin{document}

\title{Extreme mass-ratio inspirals into black holes surrounded by scalar clouds}

\author{Richard Brito}
\affiliation{CENTRA, Departamento de F\'{\i}sica, Instituto Superior T\'ecnico -- IST, Universidade de Lisboa -- UL, Avenida Rovisco Pais 1, 1049-001 Lisboa, Portugal}
\author{Shreya Shah}
\affiliation{Institut f\"ur Astronomie und Astrophysik, Universit\"at T\"ubingen, Auf der Morgenstelle 10, T\"ubingen 72076, Germany}

\begin{abstract}
We study extreme mass-ratio binary systems in which a stellar mass compact object spirals into a supermassive black hole surrounded by a scalar cloud. Scalar clouds can form through superradiant instabilities of massive scalar fields around spinning black holes and can also serve as a proxy for dark matter halos. Our framework is fully relativistic and assumes that the impact of the cloud on the geometry can be treated perturbatively. As a proof of concept, here we consider a point particle in circular, equatorial motion around a non-spinning black hole surrounded either by a spherically symmetric or a dipolar non-axisymmetric scalar cloud, but the framework can in principle be generalized to generic black hole spins and scalar cloud geometries. We compute the leading-order power lost by the point particle due to scalar radiation and show that, in some regimes, it can dominate over gravitational-wave emission. We confirm the presence of striking signatures due to the presence of a scalar cloud that had been predicted using Newtonian approximations, such as resonances that can give rise to sinking and floating orbits, as well as ``sharp features'' in the power lost by the particle at given orbital radii. Finally, for a spherically symmetric scalar cloud, we also compute the leading-order corrections to the black-hole geometry and to the gravitational-wave energy flux, focusing only on axial metric perturbations for the latter. We find that, for non-compact clouds, the corrections to the (axial) gravitational-wave fluxes at high frequencies can be understood in terms of a gravitational-redshift effect, in agreement with previous works. 
\end{abstract}

\date{\today}

\maketitle
\tableofcontents
\section{Introduction}
The direct detection of gravitational waves (GWs) can give us exceptional insights about binary black hole (BH) systems~\cite{Barack:2018yly}. Current observations~\cite{LIGOScientific:2021djp} already provided crucial new information regarding the population of binary BH systems~\cite{KAGRA:2021duu} and allowed us to perform novel tests of gravity in strong and highly dynamical regimes~\cite{LIGOScientific:2021sio}. However, with the planned construction of next-generation ground-based~\cite{Maggiore:2019uih,Evans:2021gyd,Kalogera:2021bya} and space-based detectors~\cite{LISA:2017pwj,TianQin:2015yph,Ruan:2018tsw}, the information that GW observations will be able to give us will be pushed to new limits. For example, we expect that such detectors will allow us to perform highly precise tests of gravity, orders of magnitude better than what is currently possible~\cite{Babak:2017tow,Barack:2018yly,Berry:2019wgg,Perkins:2020tra,Kalogera:2021bya,LISA:2022kgy}. It has also been shown that, for some sources, we might be able to detect GW signatures from the environment in which binary BH systems live in~\cite{Yunes:2011ws,Barausse:2014tra,Cardoso:2019rou,Derdzinski:2020wlw,Zwick:2022dih,Cardoso:2022whc}. 

Intermediate and extreme mass-ratio inspirals (IMRIs and EMRIs, respectively) are among the most interesting sources for these purposes. Such sources are expected to be observed with the upcoming space-based detector LISA~\cite{Babak:2017tow} and possibly with next generation ground-based detectors in the case of IMRIs~\cite{Amaro-Seoane:2018gbb}. I/EMRIs are binary systems with highly asymmetric component masses, and are typically classified as IMRIs for mass ratios ranging between $\sim 10^{-2}-10^{-4}$ and EMRIs for mass ratios smaller than $\sim 10^{-4}$~\cite{Amaro-Seoane:2007osp,Babak:2017tow,Amaro-Seoane:2018gbb}. Such systems will typically complete a large number of GW cycles in band, making them ideal probes to not only perform highly precise tests of gravity but also to probe the environment surrounding astrophysical BHs~\cite{Barausse:2014tra,Babak:2017tow,Berry:2019wgg,Cardoso:2019rou,Cardoso:2022whc}.

An exciting prospect which has received some attention recently, is the possibility that GW signals from these systems could carry information about the dark matter environment surrounding massive BHs, see e.g.~\cite{Eda:2013gg,Macedo:2013qea,Hannuksela:2019vip,Hannuksela:2018izj,Baumann:2018vus,Cardoso:2019rou,Baumann:2019ztm,Annulli:2020lyc,Kavanagh:2020cfn,Coogan:2021uqv,Baumann:2021fkf,Baumann:2022pkl,Cole:2022fir,Kim:2022mdj,Figueiredo:2023gas,Tomaselli:2023ysb}, which could be partially or fully composed of new light bosonic fields~\cite{Hu:2000ke,Arvanitaki:2009fg,Robles:2012uy,Schive:2014dra,Hui:2016ltb,Ferreira:2020fam}. Light bosonic fields are especially interesting because they can significantly impact the dynamics of BHs. In particular, boson masses in the range $10^{-21}$ eV~--~$10^{-11}$ eV have Compton wavelengths of the order of the size of astrophysical BHs with masses in the range  $10 M_{\odot}$~--~$10^{10} M_{\odot}$, where the lower BH mass in this range corresponds to the heavier bosons and the upper end to the lighter bosons. This feature enhances wave-like phenomena making it possible for unique effects to occur when an ultralight boson interacts with an astrophysical BH. For example, such fields can extract rotational energy from a spinning BH through superradiant scattering and can render spinning BHs unstable against energy and angular momentum extraction~\cite{Detweiler:1980uk,Cardoso:2005vk,Dolan:2007mj,Arvanitaki:2010sy,Pani:2012bp,Brito:2013wya,Brito:2014wla,East:2017ovw,Baryakhtar:2017ngi,East:2018glu,Dolan:2018dqv,Brito:2020lup,Dias:2023ynv} (see Ref.~\cite{Brito:2015oca} for an extensive review on the subject). In this process, up to $\sim 10\%$ of the BH's energy can be transferred to the bosonic field~\cite{East:2018glu,Herdeiro:2021znw}. This mechanism provides a natural scenario in which a macroscopic bosonic environment, also known as ``boson clouds'' or ``gravitational atoms'', can form around astrophysical BHs. For complex boson fields, the backreaction of these clouds on the BH metric leads to the existence of stationary geometries known as ``Kerr BHs with bosonic hair''~\cite{Herdeiro:2014goa,Herdeiro:2016tmi}, which can form dynamically through the superradiant instability~\cite{East:2017ovw,Herdeiro:2017phl}. Moreover, ultralight fields can also form solitonic self-gravitating structures, or boson stars~\cite{Ruffini:1969qy,Kaup:1968zz,Seidel:1993zk,Liebling:2012fv,Brito:2015pxa}, which could describe the inner cores of dark matter halos~\cite{Hu:2000ke,Robles:2012uy,Schive:2014dra,Veltmaat:2018dfz}. When interacting with BHs, such structures will typically form long-lived states that can either be slowly accreted by the BH~\cite{Barranco:2011eyw,Barranco:2012qs,Barranco:2017aes,Cardoso:2022vpj,Cardoso:2022nzc} or, if the BH is spinning, they could also possibly form Kerr BHs with bosonic hair, see e.g.~\cite{Sanchis-Gual:2020mzb}.

These motivations have sparked considerable interest in the study of binary systems, and more specifically EMRIs, evolving in bosonic environments that can come either in the form of boson clouds or as boson stars~\cite{Macedo:2013qea,Ferreira:2017pth,Hannuksela:2018izj,Baumann:2018vus,Zhang:2018kib,Kavic:2019cgk,Zhang:2019eid,Berti:2019wnn,Baumann:2019ztm,Annulli:2020lyc,Takahashi:2021yhy,Collodel:2021jwi,Tong:2022bbl,Baumann:2022pkl,Cole:2022fir,Xie:2022uvp,Takahashi:2023flk,Tomaselli:2023ysb,Delgado:2023wnj,Cao:2023fyv}. However, while these works suggest that such environments could be clearly identified through GW detections, most of these studies employed approximations that are inaccurate when considering EMRIs, such as considering Newtonian approximations or in some cases ignoring important effects such as dynamical friction. The main exception to this rule is the work of Refs.~\cite{Cardoso:2021wlq,Cardoso:2022whc,Destounis:2022obl,Figueiredo:2023gas} where a fully relativistic and self-consistent formalism to study GW emission from EMRIs in spherically symmetric, non-vacuum BH spacetimes was developed. Considering a fully relativistic formalism for such sources is crucial given that weak-field and post-Newtonian approximations are known to be inadequate to describe I/EMRIs in the regimes where they are expected to be observable~\cite{vandeMeent:2020xgc}. For such systems, a strong-field perturbation theory approach, in which the mass ratio is used as an expansion parameter, is essential~\cite{vandeMeent:2020xgc}.

The main goal of this work is to start developing a strong-field small-mass-ratio perturbation theory formalism to study EMRIs in a boson cloud environment. Here we will focus solely on boson clouds formed by a massive scalar field; hence we will usually refer to this environment as a ``scalar cloud'' for concreteness. However, our work can in principle be extended to the case of massive vector fields in a straightforward manner by using the results of Refs.~\cite{Rosa:2011my,Pani:2012bp,Herdeiro:2016tmi,Baryakhtar:2017ngi,East:2017ovw,Dolan:2018dqv}. We should also note that, despite the fact that small mass-ratio approximations have been historically developed to model EMRIs, recent work suggests that the range of applicability of this approximation also includes the IMRI range and can even provide a good approximation for nearly equal mass binaries~\cite{vandeMeent:2020xgc,Warburton:2021kwk,Wardell:2021fyy,Ramos-Buades:2022lgf,Albertini:2022rfe}. Therefore we expect that the approach taken in this paper can also be useful for IMRIs and provide useful qualitative insights for comparable mass systems.

When using BH perturbation theory to study the EMRIs in a given environment, the first difficulty that arises is that one would need in principle to start by building a non-vacuum BH background solution that includes the impact of the environment. This was the approach taken in Refs.~\cite{Cardoso:2021wlq,Cardoso:2022whc,Destounis:2022obl,Figueiredo:2023gas}, where the impact of matter environments in EMRIs was studied by constructing analytical spherically symmetric BH spacetimes with an anisotropic fluid ``hair''. For boson clouds, however, exact BH solutions that include the backreaction of the cloud are only known numerically. For example, stationary BH solutions surrounded by a (complex) boson clouds have been constructed numerically~\cite{Herdeiro:2014goa,Herdeiro:2016tmi}, whereas geometries describing BHs surrounded by slowly decaying  spherically bosonic structures have been obtained through Numerical Relativity simulations~\cite{Barranco:2017aes,Cardoso:2022nzc}. In addition to the difficulty of having to deal with numerical spacetimes, another difficulty that arises is the fact that generic boson clouds, and specifically the ones formed through superradiant instabilities, are not spherically symmetric~\cite{Brito:2014wla,Herdeiro:2014goa,Herdeiro:2016tmi,East:2017ovw}. Therefore perturbing a BH spacetime such as the Kerr BHs with bosonic hair constructed in Refs.~\cite{Herdeiro:2014goa,Herdeiro:2016tmi} would require not only to deal with perturbations of a highly non-trivial geometry constructed numerically, but, in general, would also require dealing with 
 solving a set of partial differential equations describing those perturbations, see e.g.~\cite{Ganchev:2017uuo}. 

To avoid these problems we will consider that the boson field, here described by a complex massive scalar field, only affects the BH geometry perturbatively. That is, our approach will be to consider a two-parameter perturbation expansion, one parameter being the mass ratio $q:=m_p/M \ll 1$, with $M$ the mass of the massive BH and $m_p$ the mass of the small orbiting compact object, here described as a point particle, and a second parameter $\epsilon \ll 1$ describing the amplitude of the scalar field. Schemes that are similar in spirit to the one we use here were proposed and used, for example, to compute tidal Love numbers of BHs surrounded by scalar clouds~\cite{DeLuca:2021ite}, to compute quasinormal modes of BHs accreting a scalar field environment~\cite{Bamber:2021knr} and also to compute quasinormal modes of BHs in beyond general relativity theories~\cite{Li:2022pcy,Hussain:2022ins,Ghosh:2023etd,Cano:2023tmv}. Such a perturbative scheme allows us to use standard tools from BH perturbation theory, since the background spacetime is now given by an analytically known vacuum BH solution on top of which the effects of the scalar field and the point particle are added perturbatively. As a proof of concept, here we will take the vacuum BH background to be given by Schwarzschild and then consider either a spherically symmetric or a dipolar scalar cloud. Even though this will require some additional non-trivial extensions of this work, we expect that the method can be generalized to Kerr BHs with arbitrary BH spins using the methods developed in Refs.~\cite{Li:2022pcy,Hussain:2022ins,Ghosh:2023etd,Cano:2023tmv}. In addition, another major advantage of this formalism is that it allows us to add the effect of the environment on top of known vacuum General Relativity results. We expect this feature to be especially important in the long term, given that it greatly simplifies the task of building EMRI waveform models that include the effects of scalar clouds. Therefore we expect that the method could also become useful for other types of environments.
As we will show, our approach captures and generalizes to a relativistic framework all the features that had been computed using Newtonian approximations and quantum mechanics analogies, namely, resonances at some specific orbital frequencies that can lead to sinking and floating orbits~\cite{Cardoso:2011xi,Macedo:2013qea,Zhang:2018kib,Baumann:2018vus,Baumann:2019ztm,Cardoso:2022fbq} as well as sharp features in the energy lost by the point particle due to scalar radiation~\cite{Baumann:2021fkf,Baumann:2022pkl,Tomaselli:2023ysb}.

The rest of this paper is organized as follows. In Sec.~\ref{sec:framework} we summarize our framework, present the perturbative scheme that we use to study EMRIs in the presence of a scalar cloud and give a short review of the quasi-bound states of a massive scalar field in a BH spacetime, focusing on a Schwarzschild BH background.
Then, in Sec.~\ref{sec:OqOepsilon} we compute the leading-order perturbation on the scalar cloud, induced by the presence of a point particle in circular, equatorial motion. We then show that the presence of the cloud introduces an additional source of energy loss in the form of scalar radiation and present results for the power lost through this radiation, focusing on a spherically symmetric and a dipolar non-axisymmetric scalar cloud. 
For a more complete test of this framework we then also compute part of the leading-order corrections to the GW flux in Sec.~\ref{sec:Oq1epsilon2}. For this case we focus solely on a spherically symmetric scalar cloud. We first compute the leading-order corrections to the BH metric due to the backreaction of the cloud on the geometry, and then then use these results to compute the leading-order corrections to the axial metric perturbations and corresponding GW fluxes. Finally, in Sec.~\ref{sec:conclusions} we conclude by summarizing our main results and identifying some promising avenues for future work. Some details of our computations are also shown in the Appendices. Throughout this work, we use units in which $G=c=1$.

\section{Framework}\label{sec:framework}
\subsection{Action and equations of motion}
We consider a complex massive scalar field $\Phi$ minimally coupled to gravity,~\footnote{EMRIs in theories in which the scalar field couples non-minimally to gravity, have also been considered in Refs.~\cite{Maselli:2020zgv,Maselli:2021men,Barsanti:2022ana,Barsanti:2022vvl}. We do not consider this case here.} described by the action 
\begin{equation}
   S = \int d^{4} x \sqrt{-g} \left(\frac{R}{16\pi}-\partial_{\mu} \Phi^{*} \partial^{\mu} \Phi - \mu^{2} \Phi^{*}\Phi + \mathcal{L}_{\rm m}\right)\,,
\end{equation}
where an asterisk denotes the complex conjugate and $\mathcal{L}_{\rm m}$ represents the Lagrangian density for additional matter fields, which are assumed to be minimally coupled to gravity as well. Varying this action with respect to the metric and the scalar field we get the Einstein-Klein-Gordon field equations: 
\begin{eqnarray}\label{EFEs}
    G_{\mu\nu} &=& 8\pi (T_{\mu\nu}^{\Phi} + T_{\mu\nu})\,,\\
    \Box \Phi &=& \mu^2 \Phi \label{KGeq}\,,
\end{eqnarray}
where $G_{\mu\nu}:=R_{\mu\nu}-g_{\mu\nu}R/2$, $\Box:= \nabla_{\mu}\nabla^{\mu}$ is the d'Alembertian operator, $T_{\mu\nu}^{\Phi}$ is the stress-energy tensor of the scalar field:
\begin{equation}\label{KG_ST}
    T_{\mu\nu}^{\Phi}[\Phi,\Phi^*] = 2\partial_{(\mu} \Phi\partial_{\nu)}\Phi^{*} - g_{\mu\nu}\left(\partial_{\alpha} \Phi\partial^{\alpha}\Phi^{*}+\mu^2\Phi^{*}\Phi\right)\,,
\end{equation}
and $T_{\mu\nu}$ represents the stress-energy tensor of any additional matter.
The scalar field possesses a global $U(1)$ symmetry which implies the existence of a conserved current given by
\begin{equation}
    j^{\mu} = -i\left(\Phi^*\partial^\mu\Phi-\Phi\partial^\mu\Phi^*\right)\label{KG_current}\,.
\end{equation}
In the absence of dissipation in the system, such as scalar radiation at the horizon or at infinity, the conserved current implies the existence of a conserved Noether charge:
\begin{equation}
    Q = \int_{\Sigma}d^3x \sqrt{-g}\,j^0\label{KG_charge}\,,
\end{equation}
with $\Sigma$ a space-like hypersurface.

\subsection{Perturbation scheme}\label{sec:pert_scheme}

We consider a small compact object (often times referred to as the ``secondary'' object) with mass $m_p$ orbiting a BH of mass $M$ surrounded by a scalar cloud, such that the mass ratio $q := m_p/M$ is small, i.e., $q\ll 1$. At leading order in a small-$q$ expansion, the secondary object can be modelled as a point particle moving on geodesics of the background spacetime, $\bar g_{\mu\nu}$, generated by the BH-scalar cloud system (see e.g. Refs.~\cite{Barack:2018yvs,Pound:2021qin} for recent reviews). The point particle's stress-energy tensor is given by
 \begin{equation}\label{eq:STpoint}
    T^{\mu\nu}[\bar g] = m_p\int u_p^{\mu}u_p^{\nu}\frac{\delta^{(4)}\left(x^{\mu}-x_p^{\mu}(\tau)\right)}{\sqrt{-\bar g}} d\tau\,,
\end{equation}
where $\tau$ is the particle's proper time, $x_p^{\mu}$ its worldline and $u_p^{\mu}:= dx_p^{\mu}/d\tau$ its four-velocity. 

In order to take into account the impact of the scalar field, we consider that its amplitude is small, such that the modifications to the BH spacetime induced by the scalar field can be treated using perturbation theory. Therefore, besides the mass ratio $q$ we consider an additional small parameter $\epsilon\ll 1$ that parameterizes the scalar field amplitude, such that in the limit $\epsilon\to 0$ we recover GR's vacuum solutions. Namely, keeping only the terms up to order $\mathcal{O}(q^1, \epsilon^2)$, we consider an expansion of the form~\cite{Bamber:2021knr,Hussain:2022ins}:
\begin{eqnarray}
     \Phi &=& \epsilon \Phi^{(1)} + \epsilon^2 \Phi^{(2)} + q \Phi^{(q)} +\ldots\,,\label{eq:Phi_expansion}\\
     g_{\mu\nu} &=& g_{\mu\nu}^{(0)} + \epsilon g_{\mu\nu}^{(1)} + \epsilon^2 g_{\mu\nu}^{(2)} + q h_{\mu\nu} +\ldots\,.\label{eq:metric_expansion}
\end{eqnarray}
Inserting these expressions into Eqs.~\eqref{EFEs} and~\eqref{KGeq} we find, up to order $\mathcal{O}(q^0,\epsilon^0)$:
\begin{equation}\label{EFEs_vacuum}
    G_{\mu\nu}[g^{(0)}]=0\,.
\end{equation}
These are just the vaccum Einstein field equations for which the most generic BH solution is the Kerr metric. 
Continuing this procedure, at order $\mathcal{O}(q^0,\epsilon^1)$ we find:
\begin{eqnarray}
        \delta G_{\mu\nu}[g^{(1)}]&=&0\,,\label{linearEFEs}\\
        \Box^{(0)} \Phi^{(1)} &=& \mu^2 \Phi^{(1)}\label{KG_ST_10}\,,
\end{eqnarray}
where $\Box^{(0)}:=g^{(0)\,\mu\nu}\nabla^{(0)}_{\mu}\nabla^{(0)}_{\nu}$ is the d'Alembertian operator computed with respect to the metric $g_{\mu\nu}^{(0)}$ and $\delta G_{\mu\nu}$ represents the standard linearized Einstein operator defined as~\cite{Hussain:2022ins}
\begin{align}
    \label{linearEFEs_def}
    \delta G_{\mu\nu}[h]:= 
    \frac{1}{2} &\left[ 2 \nabla^{(0)\,\alpha} \nabla^{(0)}_{(\mu}h_{\nu)\alpha} - \Box^{(0)} h_{\mu\nu} - \nabla^{(0)}_{\mu} \nabla^{(0)}_{\nu} h^{\alpha}{}_{\alpha}
      \right.  
\nonumber \\
& 
+ \left. g^{(0)}_{\mu\nu}\left(\Box^{(0)} h^{\beta}{}_{\beta} - \nabla^{(0)\,\alpha} \nabla^{(0)\,\beta} h_{\alpha\beta}\right) \right]\,.
\end{align}
Notice that there is no source term for the field equations of the perturbation $g_{\mu\nu}^{(1)}$. As we are only interested in metric perturbations sourced by the scalar field or by the secondary object, we set $g_{\mu\nu}^{(1)}=0$ since this trivially solves Eq.~\eqref{linearEFEs}. Therefore the spacetime metric is only deformed at order $\mathcal{O}(\epsilon^2)$, as one could have easily guessed given that the scalar field stress-energy tensor~\eqref{KG_ST} is quadratic in the field's amplitude.

Setting $g_{\mu\nu}^{(1)}=0$, at order $\mathcal{O}(q^0,\epsilon^2)$ we have:
\begin{eqnarray}
        \delta G_{\mu\nu}[g^{(2)}]&=& 8\pi T_{\mu\nu}^{\Phi(2)}[\Phi^{(1)},\Phi^{(1)\,*}]\,,\label{linearEFEs_20}\\
        \Box^{(0)} \Phi^{(2)} &=& \mu^2 \Phi^{(2)}\label{KG_ST_20}\,,
\end{eqnarray}
where we defined
\begin{align}
    &T_{\mu\nu}^{\Phi(2)}[\Phi^{(1)},\Phi^{(1)\,*}] = 2\partial_{(\mu} \Phi^{(1)}\partial_{\nu)}\Phi^{(1)\,*} \nonumber\\
    &- g^{(0)}_{\mu\nu}\left(g^{(0)\,\alpha\beta}\partial_{\alpha} \Phi^{(1)}\partial_{\beta}\Phi^{(1)\,*}+\mu^2\Phi^{(1)}\Phi^{(1)\,*}\right)\,.
\end{align}
Notice that Eqs.~\eqref{KG_ST_10} and ~\eqref{KG_ST_20} are exactly the same, meaning that the correction $\Phi^{(2)}$ can be reabsorbed into the definition of $\Phi^{(1)}$. Therefore, without loss of generality, we can set $\Phi^{(2)}=0$. In summary, up to order $\mathcal{O}(q^0,\epsilon^2)$, once we have a BH background solution that solves Eq.~\eqref{EFEs_vacuum}, the set of equations one needs to solve is Eqs.~\eqref{KG_ST_10} and \eqref{linearEFEs_20}.

We can now consider the impact of the point particle by computing the corrections at order $\mathcal{O}(q^1,\epsilon^2)$. Schematically, we can write those as~\cite{Bamber:2021knr,Hussain:2022ins}
\begin{eqnarray}
        &&\delta G_{\mu\nu}[\bar{g},h]= 8\pi T^{p}_{\mu\nu}[\bar{g}]\nonumber\\
        &&+8\pi\epsilon \left(T_{\mu\nu}^{\Phi(2)}[\Phi^{(1)},\Phi^{(q)\,*}]+ T_{\mu\nu}^{\Phi(2)}[\Phi^{(q)},\Phi^{(1)\,*}]\right)\nonumber\\
        &&+8\pi\epsilon^2 S^{h}_{\mu\nu}[h,\Phi^{(1)},\Phi^{(1)\,*}]\,, \label{EFEs_21}\\
        &&\left(\Box^{\bar g}- \mu^2\right)\Phi^{(q)} =  \epsilon S^{\Phi}[h,\Phi^{(1)}]\label{KG_21}\,,
\end{eqnarray}
where $\delta G_{\mu\nu}[\bar{g},h]$ represents the linearized Einstein field equations for the perturbation $h_{\mu\nu}$ but computed with the background metric $\bar{g}_{\mu\nu}:=g_{\mu\nu}^{(0)}+\epsilon^2 g_{\mu\nu}^{(2)}$, for convenience we factorized the $q$ dependence from the point particle's stress-energy tensor as $T_{\mu\nu}=q T^p_{\mu\nu}$, $\Box^{\bar g}$ represents the d'Alembertian operator written with respect to the metric $\bar{g}_{\mu\nu}$, $S^{h}_{\mu\nu}$ is given by
\begin{align}
    S^{h}_{\mu\nu}[h,\Phi^{(1)},\Phi^{(1)\,*}] =& 
     g^{(0)}_{\mu\nu}h^{\alpha\beta}\partial_{\alpha} \Phi^{(1)}\partial_{\beta}\Phi^{(1)\,*} \nonumber\\
   & - h_{\mu\nu}g^{(0)\,\alpha\beta}\partial_{\alpha} \Phi^{(1)}\partial_{\beta}\Phi^{(1)\,*}
   \nonumber\\
   & - \mu^2 h_{\mu\nu} \Phi^{(1)}\Phi^{(1)\,*}
   \,,
\end{align}
whereas $S^{\Phi}$ is given by
\begin{equation}
    S^{\Phi}[h,\Phi^{(1)}] = g^{(0)\,\mu\nu}\Gamma^{(q)\alpha}_{\mu\nu}\partial_{\alpha}\Phi^{(1)} + h^{\mu\nu}\nabla^{(0)}_{\mu}\partial_\nu\Phi^{(1)}\,,
\end{equation}
with
\begin{equation}
\Gamma^{(q)\alpha}_{\mu\nu}=\frac{1}{2}g^{(0)\,\alpha\beta}\left(\nabla^{(0)}_{\mu}h_{\nu\beta}+\nabla^{(0)}_{\nu}h_{\mu\beta}-\nabla^{(0)}_{\beta}h_{\mu\nu}\right)\,.
\end{equation}
Importantly, we see from Eqs.~\eqref{EFEs_21} and~\eqref{KG_21} that even though the secondary object does not interact directly with the scalar field, the metric perturbations $h_{\mu\nu}$ induced by the object will source scalar perturbations $\Phi^{(q)}$, as long as $\Phi^{(1)}\neq 0$. We also note that, for consistency, $\delta G_{\mu\nu}[\bar{g},h]$, $T^p_{\mu\nu}[\bar{g}_{\mu\nu}]$ and $\Box^{\bar g}$ should only be thought as being valid up to order $\mathcal{O}(\epsilon^2)$. This procedure can be continued to include $\mathcal{O}(q^2)$ effects, but for the purposes of this work we will stop at order $\mathcal{O}(q)$. 

At this point we could stop and use directly Eqs.~\eqref{EFEs_21} and~\eqref{KG_21}. However the problem can be further simplified by noticing that the source term in Eq.~\eqref{KG_21} is of order $\mathcal{O}(\epsilon)$. Therefore for the scalar field perturbations we can seek solutions of the form
\begin{equation}
         \Phi^{(q)} = \epsilon \Phi^{(q,1)} +\ldots\,.\label{eq:Phiq_exp}
\end{equation}
Plugging in Eq.~\eqref{EFEs_21} we see that, up to order $\mathcal{O}(\epsilon^2)$, the metric perturbations $h_{\mu\nu}$ will be sourced only by terms of order $\mathcal{O}(\epsilon^0)$ and $\mathcal{O}(\epsilon^2)$. Therefore we expand $h_{\mu\nu}$ as
\begin{equation}\label{eq:expansion_h}
         h_{\mu\nu} = h^{(0)}_{\mu\nu}+\epsilon^2  h^{(2)}_{\mu\nu} +\ldots\,;
\end{equation}
Applying these expansions in Eqs.~\eqref{EFEs_21} and~\eqref{KG_21} we have:
\begin{itemize}
\item at order $\mathcal{O}(q^1,\epsilon^0)$:
\begin{equation}\label{EFEs_ep0q1}
       \delta G_{\mu\nu}[h^{(0)}]= 8\pi T^p_{\mu\nu}[g^{(0)}]\,;
\end{equation}
\item at order $\mathcal{O}(q^1,\epsilon^1)$:
\begin{equation}\label{KG_ep1q1}
        \left(\Box^{(0)}- \mu^2\right)\Phi^{(q,1)} =  S^{\Phi}[h^{(0)},\Phi^{(1)}]\,.
\end{equation}
\end{itemize}
At order $\mathcal{O}(q^1,\epsilon^2)$ one can get an equation for $h^{(2)}_{\mu\nu}$, but for the purposes of this work we will not need this equation, therefore we do not derive it explicitly here.

Besides the equations of motion for the scalar field and the metric we also need the equations of motion for the point particle, which can be derived from the conservation equation $\nabla_\nu T^{\mu\nu}=0$~\cite{Poisson:2011nh}. For the purposes of this work it will be enough to state that, up to order $\mathcal{O}(\epsilon^2)$, the equations of motion for the particle's worldline can be written as (see e.g.~\cite{Poisson:2011nh})
\begin{equation}\label{eq:geodesic}
u_p^{\nu}\nabla^{\bar g}_{\nu}u_p^{\mu} = 0+\mathcal{O}(q)\,,
\end{equation}
where $\nabla^{\bar g}$ is the covariant derivative computed with the metric $\bar{g}_{\mu\nu}$. The terms of order of $\mathcal{O}(q)$, commonly known as the ``self-force'', are generated by the metric perturbation $h_{\mu\nu}$ and express the fact that the particle's motion can be thought as being accelerated in the background spacetime $\bar{g}_{\mu\nu}$ due to the perturbations induced by the point particle~\cite{Poisson:2011nh}. However, for the purposes of solving Eqs.~\eqref{EFEs_21} and~\eqref{KG_21}, those terms are not needed since when solving Eqs.~\eqref{EFEs_21} and~\eqref{KG_21} one only needs to consider that the point particle moves along geodesics of the metric $\bar{g}_{\mu\nu}$ [or of $g_{\mu\nu}^{(0)}$, if working only up to order $\mathcal{O}(\epsilon)$].

Noticeably, if one stops at order $\mathcal{O}(q^1,\epsilon^1)$, the only corrections to the vacuum case will occur due to $\Phi^{(q,1)}$ which is sourced by the non-trivial background scalar profile $\Phi^{(1)}$ and by the metric perturbation $h_{\mu\nu}^{(0)}$ [cf. Eq.~\eqref{KG_ep1q1}], simplifying the problem considerably since $h_{\mu\nu}^{(0)}$ can be obtained using standard \emph{vacuum} BH perturbation theory [cf. Eq.~\eqref{EFEs_ep0q1}]. Importantly, the perturbative scheme we just summarized is generic and can in principle be applied for any background BH metric $g^{(0)}_{\mu\nu}$, including the case in which this metric is given by the Kerr geometry, which is the most general stationary BH solution of Eq.~\eqref{EFEs_vacuum}. The main difficulty in the Kerr case, however, is that perturbations are more easily studied using the Teukolsky formalism~\cite{Teukolsky:1973ha}, which provides separable equations for certain spin-weighted scalars that are related to the Weyl curvature scalars~\cite{Teukolsky:1973ha}. From those, it is possible to reconstruct the metric perturbations $h_{\mu\nu}^{(0)}$, but the procedure is highly non-trivial (we refer the reader to the review~\cite{Pound:2021qin} for a list of references regarding the metric reconstruction procedure). Given that the main goal of this paper is to serve as a first stepping stone towards tackling the full Kerr BH case, in this work we will instead start by considering the much simpler case in which the background metric $g^{(0)}_{\mu\nu}$ is given by a Schwarzschild BH.
\subsection{Quasi-bound states in black hole spacetimes}\label{sec:QBSs}
 Let us start by considering solutions to order $\mathcal{O}(q^0,\epsilon^1)$ that solve Eq.~\eqref{KG_ST_10} when the background metric $g^{(0)}_{\mu\nu}$ describes a BH spacetime, focusing on a Schwarzschild BH as mentioned above. This problem has been widely discussed in the literature (see Ref.~\cite{Brito:2015oca} for a review), so let us just briefly review the problem.
 
 In a BH spacetime, massive scalar fields admit quasi-bound state solutions that oscillate with a frequency $\omega\sim\mu$~\cite{Detweiler:1980uk,Cardoso:2005vk,Dolan:2007mj}. 
 For Schwarzschild BHs such quasi-bound states always decay in time due to absorption at the horizon, but they can be extremely long-lived when $M\mu \ll 1$~\cite{Detweiler:1980uk,Dolan:2007mj} and can form under quite generic initial conditions~\cite{Barranco:2011eyw,Barranco:2012qs,Witek:2012tr,Barranco:2017aes,Cardoso:2022vpj,Cardoso:2022nzc}. For Kerr BHs instead, some bound states can become superradiantly unstable when their oscillation frequency $\omega$ satisfies the superradiant condition, $\omega<m_i\Omega_H$~\cite{Detweiler:1980uk,Cardoso:2005vk,Dolan:2007mj,Brito:2015oca}, where $\Omega_H$ is the horizon's angular velocity~\footnote{In Boyer-Lindquist coordinates, $\Omega_H=J/(2M^2 r_+)$ where $J$ is the BH's angular momentum and $r_+$ is the outer event horizon of the Kerr metric.} and $m_i$ is the azimuthal index of a spheroidal harmonic function used to separate the Klein-Gordon equation in a Kerr BH background. 
 The evolution of this instability leads to the formation of scalar clouds~\cite{Brito:2014wla,East:2017ovw,Herdeiro:2017phl,East:2018glu}. For clouds that only grow through superradiance, the backreaction of the cloud on the metric is generically small~\cite{Brito:2014wla,East:2017ovw,Herdeiro:2017phl,Herdeiro:2021znw} and the resulting configuration is very well described by a bound state in a Kerr BH background with spin that saturates the condition $\Omega_H = \omega/m_i$~\cite{Brito:2014wla,East:2017ovw,Herdeiro:2017phl}. 
The BH spin is essential to form scalar clouds through superradiance, however some important remarks should be made: (i) the condition for the superradiant instability to occur, $\omega<m_i\Omega_H$, with $\omega\sim \mu$ and $M\Omega_H\leq 1/2$ (the equality corresponding to extremal Kerr BHs), implies $M\mu\lesssim m_i/2$. Therefore for $m_i=1$, which corresponds to the most unstable mode, one always had $M\mu \lesssim 1/2$; (ii) even after a $m_i=1$ cloud that saturates $\Omega_H \sim \omega$ has formed, higher modes with $m_i>1$ will still keep growing. However the instability timescale of those modes can be sufficiently long such that the $m_i=1$ cloud is effectively stable over very long timescales, if $M\mu$ is small enough, see e.g.~\cite{Degollado:2018ypf}; (iii) the profile of the bound states peak at a radius that scales with $M/(M\mu)^2$, which implies that for small $M\mu$, the cloud is localized far from the horizon where BH spin effects are small; (iv) given that $M\Omega_H \sim M\omega \sim M\mu$ for $m_i=1$ scalar clouds that have saturated the superradiant instability, the BH dimensionless spin $J/M^2$ is small when $\mu M\ll 1$. Taking all these points into consideration, makes us confident that considering the background BH metric to be a Schwarzschild BH, provides a reasonably good approximation of what one should expect for scalar clouds formed around Kerr BHs and grown out of the superradiant instability. 

Let us therefore consider $g^{(0)}_{\mu\nu}$ to be given by the Schwarzschild metric:
\begin{equation}\label{Sch_metric}
ds^{2} = -f(r)dt^{2} + f(r)^{-1}dr^{2} + r^{2}d\theta^{2}+r^{2}\sin^{2}\theta d\phi^{2}\,,
\end{equation}
with $f(r) = \left(1 - 2M/r\right)$. Since the metric is spherically symmetric, the scalar field can be decomposed as
\begin{equation}\label{background_scalar}
    \Phi^{(1)}(t,r,\theta,\phi) = R_{n_i\ell_i}(r) Y_{\ell_i m_i}(\theta,\phi)e^{-i\omega t}\,,
\end{equation}
where $Y_{\ell_i m_i}$ are scalar spherical harmonics, $n_i=0,1,2,\ldots$ is the analog to the radial quantum number in the hydrogen atom, describing the number of nodes in the radial wavefunction, and $\{\ell_i, m_i\}$ are the usual spherical harmonic quantum numbers specifying the total and the projection of the angular momentum along the $z-$axis of a given mode, respectively. Here we already anticipated that in a Schwarzschild background $R_{n_i\ell_i}(r)$ does not depend on the azimuthal number $m_i$ since, upon inserting~\eqref{background_scalar} in Eq.\eqref{KG_ST_10}, one can show that the radial function $R_{n_i\ell_{i}}(r)$ satisfies the differential equation
\begin{equation}\label{eq_radial_KG}
 \frac{d^{2}(r R_{n_i\ell_{i}})}{dr_{*}^{2}} + \left(\omega^{2} - V_{i}\right)(r R_{n_i\ell_{i}}) = 0 \,.
 \end{equation}
Here $r_*$ is the tortoise coordinate, defined through $dr_*/dr = 1/f(r)$ and the effective potential reads
\begin{equation}
      V_{i} = f(r)\left[\mu^{2} + \frac{\ell_i(\ell_i+ 1)}{r^{2}} + \frac{2M}{r^{3}}\right]\,.
 \end{equation}
Notice that $V_{i}$ does not depend on $m_i$ and therefore the radial function does not depend on it. This is only true in Schwarzschild; in a Kerr BH background this degeneracy is slightly broken~\cite{Dolan:2007mj}. 

For convenience, later on we will also make use of ingoing Eddington-Finkelstein coordinates $(v,r,\theta,\phi)$, with $v=t+r_*$, for which the Schwarzschild metric reads
\begin{equation}\label{Sch_metric_EF}
ds_{\rm EF}^{2} = -f(r)dv^{2} + 2dv dr + r^{2}d\theta^{2}+r^{2}\sin^{2}\theta d\phi^{2}\,.
\end{equation}
In these coordinates we can decompose the scalar field as
\begin{equation}\label{background_scalar_EF}
    \Phi^{(1)}(v,r,\theta,\phi) = \tilde{R}_{n_i\ell_i}(r) Y_{\ell_i m_i}(\theta,\phi)e^{-i\omega v}\,.
\end{equation}
Since $\Phi^{(1)}$ is a scalar function, it is locally invariant under a coordinate transformation. Therefore by equating Eqs.~\eqref{background_scalar} and~\eqref{background_scalar_EF} one finds that the radial functions are related by $\tilde{R}_{n_i\ell_i}(r)=e^{i\omega r_*}R_{n_i\ell_i}(r)$~\cite{Dolan:2007mj}. We will make use of this relation in order to compute the function $\tilde{R}_{n_i\ell_i}$ later on. It can be easily verified that $\tilde{R}_{n_i\ell_i}$ satisfies the following differential equation:
\begin{equation}\label{eq_radial_KG_EF}
 \frac{d^{2}(r \tilde{R}_{n_i\ell_i})}{dr_{*}^{2}} - 2i\omega\frac{d(r \tilde{R}_{n_i\ell_i})}{dr_*} - V_{i}\, r \tilde R_{n_i\ell_i} = 0 \,.
 \end{equation}

Imposing appropriate boundary conditions, solutions to Eq.~\eqref{eq_radial_KG} can be obtained numerically or semi-analytically, for example, by directly integrating the radial equation or by using a continued-fraction method~\cite{Cardoso:2005vk,Pani:2013pma,Dolan:2007mj}. Through this work we employ the continued-fraction method of Ref.~\cite{Dolan:2007mj} which provides a very efficient method to get accurate semi-analytical solutions.
For quasi-bound state solutions one imposes boundary conditions in which the field decays exponentially at spatial infinity, whereas close to the event horizon only ingoing waves are present:
\begin{equation}\label{QBsBCs}
    \lim_{r\to 2M}\, R(r) \sim  e^{-i\omega r_*}\,, \,\,\,
    \lim_{r\to \infty}\, R(r) \sim  \frac{e^{-q r_*}r^{M\mu^2/q}}{r} \,,
\end{equation}
where $\nu=M\mu^2/q$, $q = \sqrt{\mu^2 - \omega^2}$ and one requires $\Re(q)>0$ for quasi-bound state solutions. This pair of boundary conditions is satisfied for an infinite, discrete spectrum of complex eigenfrequencies~\cite{Dolan:2007mj} that can be labeled according to the three quantum numbers $\omega:= \omega_{n_i\ell_i m_i}$~\footnote{We note that in a Schwarzschild BH background, modes with the same $\{n_i,\ell_i\}$ but different $m_i$ are degenerate, but as alluded to above, this degeneracy is slightly broken in a Kerr BH background~\cite{Dolan:2007mj}.}. In the small $M\mu$ limit, the real and imaginary part of the eigenfrequencies reads~\cite{Detweiler:1980uk,Baumann:2019eav} 
\begin{eqnarray}
    \Re(\omega) &\approx& \mu-\frac{\mu}{2}\left(\frac{M\mu}{\ell_i+n_i+1}\right)^2\,, \label{omegaR}\\
    \Im(\omega) &\propto& -\left(M\mu\right)^{4\ell_i+5}\Re(\omega)\,.\label{omegaI}
\end{eqnarray}
In a Kerr spacetime, a similar expression can be found for $\Im(\omega)$ by doing the transformation $\Re(\omega)\to\Re(\omega)- m_i\Omega_H$ in Eq.~\eqref{omegaI}, such that when $\Re(\omega)<m_i\Omega_H$, $\Im(\omega)>0$ and the mode grows exponentially. Kerr BHs also admit true bound states with $\Im(\omega)=0$ when $\Re(\omega)= m_i\Omega_H$, which are a good approximation to the end state of the superradiant instability~\cite{Herdeiro:2017phl}.

One can see that, within our approximation of using a Schwarzschild BH background, the scalar bound states will typically slowly decay in time, since true bound states can only exist in a Kerr spacetime. However, Eq.~\eqref{omegaI} predicts that $|M\Im(\omega)| \ll 1$ when $M\mu\ll 1$ and therefore, even in Schwarzschild, quasi-bound states can be very long-lived, as already mentioned. In Sec.~\ref{sec:Oq1epsilon2} we show how this can be explicitly seen from the backreaction induced on the metric for a spherically symmetric quasi-bound state [cf. Eq.~\eqref{dMv2}]. Therefore, in this work we will neglect the slow decay of the cloud in our calculations and also assume that $\omega\approx \Re(\omega)$ when considering the perturbations induced by the point particle. While this might not always be a good approximation over the whole inspiral of an EMRI, especially for large $M\mu$, this serves a good proxy for what happens in a Kerr spacetime where true bound states that do not decay over time exist and can be formed through the superradiant instability. Even without this approximation, we note that, as long as the decay timescale $\tau_{\rm inst}=1/|\Im(\omega)|$ is much larger than the typical orbital period of the point particle $T_{\rm orb}=2\pi\sqrt{r_p^3/M}$, with $r_p$ the orbital radius, the decay of the cloud can be neglected when computing the (orbital averaged) scalar and GW energy fluxes emitted due to the orbital motion of the point particle. The decay can be included {\it a posteriori} using a flux-balance law when considering the slow inspiral of the secondary object. This condition requires $r_p/M\ll (2\pi)^{-2/3}\left(M\mu\right)^{-4-8\ell_i/3}$, where we took $\Re(\omega)\sim \mu$. Considering the mode with the smallest decay timescale, $\ell_i=n_i=0$, $r_p/M\ll 3\times 10^3 \left(M\mu/0.1\right)^{-4}$. Therefore, as long as we consider sufficiently small orbital radii, we can neglect the decay of the cloud for the purposes of computing fluxes at given orbital radii. If we instead require the cloud's decay timescale to be sufficiently slow such that one can neglect it during the whole inspiral we get a stricter bound on $M\mu$. Approximating the orbital decay as being due solely to GW emission and using the quadrupole formula in the EMRI limit, the typical orbital decay is given by Peter's formula $t_{\rm GW} \sim 5 r_{p,0}^4/(256 M^3 q)$~\cite{Peters:1964zz}, where $r_{p,0}$ represents the initial orbital radius. Therefore, if $r_{p,0}/M\ll 4(q/5)^{1/4}\left(M\mu\right)^{-3/2-\ell_i}$, the cloud's decay can be neglected throughout the whole inspiral. For $\ell_i=n_i=0$ this gives $r_{p,0}/M\ll 15 \left(q/10^{-3}\right)^{1/4}\left(M\mu/0.1\right)^{-3/2}$, whereas for $\ell_i=n_i=1$ we find $r_{p,0}/M\ll  150\left(q/10^{-3}\right)^{1/4}\left(M\mu/0.1\right)^{-5/2}$.

Finally, we should note that the metric perturbations $h_{\mu\nu}$ and $g_{\mu\nu}^{(2)}$ should also induce small corrections of order $\mathcal{O}(q)$ and $\mathcal{O}(\epsilon^2)$, respectively, to the quasi-bound state eigenfrequencies (see e.g.~\cite{Baumann:2018vus,Zhang:2018kib,Siemonsen:2022yyf,Hussain:2022ins}). These small frequency shifts can be computed perturbatively by expanding the eigenfrequencies as $\omega=\omega^{(0)}+q \omega^{(q)}+\epsilon^2\omega^{(2)}$ (see e.g. Ref.~\cite{Hussain:2022ins}), where $\omega^{(0)}$ are the eigenfrequencies in the vacuum BH background, whereas the frequency shifts $\omega^{(q)}$ and $\omega^{(2)}$ can be obtained employing a formalism similar to perturbation theory in quantum mechanics (see e.g. Sec. IV.A of Ref.~\cite{Hussain:2022ins}, and also App.~\ref{app:freq_shift} for more details). However, for the purposes of computing the leading-order power lost by a point particle moving inside a scalar cloud, these frequency shifts can be neglected. Therefore we will not consider these corrections here, leaving their computation for future work.
\section{Point particle in circular, equatorial motion: leading-order metric and scalar perturbations}\label{sec:OqOepsilon}
Using the framework presented in the previous section, we now consider the leading-order perturbations induced by a point particle in circular, equatorial motion around a Schwarzschild BH surrounded by a scalar cloud, here modeled as a quasi-bound state solution of the Klein-Gordon equation in a Schwarzschild background. That is, we will consider perturbations to the metric and the scalar field up to order $\mathcal{O}(q^1,\epsilon^1)$. As we saw above, those are described by Eqs.~\eqref{EFEs_ep0q1} and~\eqref{KG_ep1q1}. Therefore, we first need to find the metric perturbations using Eq.~\eqref{EFEs_ep0q1} and then use those solutions in the source of the scalar field Eq.~\eqref{KG_ep1q1}.
\subsection{Metric perturbations}\label{sec:GRperts}
The problem of solving Eq.~\eqref{EFEs_ep0q1} in a Schwarzschild BH background has been widely studied in the literature (see e.g. Refs.~\cite{Regge:1957td,Zerilli:1970se,Zerilli:1970wzz,Sago:2002fe,Martel:2005ir} for classical papers on the subject), therefore let us just briefly review the main equations here. Some additional details can also be found in Appendices~\ref{app:PP_SEtensor} and~\ref{app:grav_perts}.

In a spherically symmetric background, $h_{\mu\nu}$ can be expanded in a complete basis of tensor spherical harmonics. Those are labeled by spherical-harmonic indices $l$ and $m$, and can be classified as axial and polar perturbations, depending on their properties under parity transformations~\cite{Regge:1957td,Zerilli:1970se,Zerilli:1970wzz}. In this basis, the metric perturbations can be decomposed as: 
\begin{align}
\label{decom}
h^{(0)}_{\mu\nu}(t,r,\theta,\phi)&=\sum_{l,m}\Re\Big\{\int_{-\infty}^{+\infty}e^{-i\sigma t}\left[h^{{\rm axial},lm}_{\mu\nu}(\sigma,r,\theta,\phi)\right.\nonumber\\
&\left.+h^{{\rm polar},lm}_{\mu\nu}(\sigma,r,\theta,\phi)\right]d\sigma \Big\}\,,
\end{align}
where $h^{{\rm axial},lm}_{\mu\nu}$ represent axial perturbations and $h^{{\rm polar},lm}_{\mu\nu}$ represent polar perturbations, and notice that here we work in the frequency domain. The explicit form of the polar and axial perturbations in Regge-Wheeler gauge~\cite{Regge:1957td} can be found in App.~\ref{app:grav_perts}, see Eqs.~\eqref{polar} and~\eqref{axial}. Similarly the point particle's stress-energy tensor can be decomposed in terms of the tensor spherical harmonics basis (see App.~\ref{app:PP_SEtensor}) which allows to separate the equations of motion. Because of the spherical symmetry of the background, polar and axial perturbations completely decouple. 
\subsubsection{Master equations for $l\geq 2$}
For $l\geq 2$, polar and axial perturbations can be reduced to two scalar and gauge-invariant master functions, $\Psi^{lm}_{\rm pol}(t,r)$ and $\Psi^{lm}_{\rm ax}(t,r)$, which can be computed from the metric perturbations~\cite{Martel:2005ir}. In the frequency domain, those functions satisfy second-order ordinary differential equations given by
\begin{eqnarray}
   \left[\frac{d^{2} }{dr_{*}^{2}} + \sigma^{2} - V_{\rm pol}\right]\psi_{\rm pol}^{lm}(r) &=& S_{\rm pol}^{lm}(r)\,,\label{mastereven} \\
   \left[\frac{d^{2} }{dr_{*}^{2}} + \sigma^{2} - V_{\rm ax}\right]\psi_{\rm ax}^{lm}(r) &=& S_{\rm ax}^{lm}(r)\,,\label{masterodd}
\end{eqnarray} 
where $\psi_{\rm pol}^{lm}$ and $\psi_{\rm ax}^{lm}$ are the Fourier transforms of $\Psi _{\rm pol}$ and $\Psi _{\rm ax}$, respectively, defined here as:
\begin{eqnarray}
\Psi^{lm}_{\rm pol/ax}(t,r) &=& \int_{-\infty}^{+\infty}e^{-i \sigma t} \,\psi_{\rm pol/ax}^{lm}(\sigma,r)\,d\sigma\,,\label{inversefourier}\\
\psi_{\rm pol/ax}^{lm}(\sigma,r) &=& \frac{1}{2\pi}\int_{-\infty}^{+\infty}e^{i \sigma t} \,\Psi^{lm}_{\rm pol/ax}(t,r)\,dt\,.\label{fourier}
\end{eqnarray}
The potentials read
\begin{align}
    &V_{\rm pol} = \frac{f \left(18 M^3+18 \lambda  M^2 r+6 \lambda ^2 M
   r^2+2 \lambda ^2 (\lambda +1) r^3\right)}{r^3 (3 M+\lambda  r)^2} \,,\\
    &V_{\rm ax}= f\left(\frac{l(l+1)}{r^2}-\frac{6M}{r^2}\right)\,,
\end{align} 
where $\lambda = (l-1)(l+2)/2$, whereas the source terms $S_{\rm pol}^{lm}$ and $S_{\rm ax}^{lm}$ can be found in App.~\ref{app:grav_perts} [cf. Eqs.~\eqref{eq:sourcepol} and ~\eqref{eq:sourceax}]. 

These equations can be solved using a standard Green’s function approach. Namely, for equations of the type~\eqref{mastereven} and~\eqref{masterodd} we can construct two independent solutions of the homogeneous part of the equations, which satisfy the following boundary conditions (using the notation in Ref.~\cite{Sasaki:2003xr}):
\begin{eqnarray}
\psi^{\rm p/a}_{\rm in}
& \to &
\begin{cases}
e^{-i\sigma r_*}\,,&
 r\to 2M, \\
B^{\rm p/a}_{\rm inc} e^{-i \sigma r_*} +
B^{\rm p/a}_{\rm ref} e^{i \sigma r_*}\,,&
 r\to \infty, \\
\end{cases}
\label{psiin} 
\\
~\nonumber\\
\psi^{\rm p/a}_{\rm up}
&\to&
\begin{cases} 
C^{\rm p/a}_{\rm ref} e^{-i \sigma r_*}+ 
C^{\rm p/a}_{\rm up} e^{i \sigma r_*}\,,& 
 r\to 2M, \\
e^{i \sigma r_*} \,,& 
 r\to \infty, \\
\end{cases}
\label{psiup} 
\end{eqnarray} 
where here the superscript ``p'' and ``a'' refer to a solution to the (homogeneous) polar and axial master equation, respectively, and one should remember that there is an implicit dependence on $l$ and $m$. The Wronskian of these two solutions is constant and given by
\begin{equation}\label{wronskian_grav}
W(\psi^{\rm p/a}_{\rm in},\psi^{\rm p/a}_{\rm up}) = \frac{d\psi^{\rm p/a}_{\rm up}}{dr_*}\psi^{\rm p/a}_{\rm in}-\frac{d\psi^{\rm p/a}_{\rm in}}{dr_*}\psi^{\rm p/a}_{\rm up}=2i\sigma B^{\rm p/a}_{\rm inc}\,.
\end{equation}
With these ingredients one can then construct a solution to Eqs.~\eqref{mastereven} and~\eqref{masterodd} which behaves as a purely outgoing wave at infinity and purely ingoing wave at the horizon:
\begin{align}\label{full_sol_grav}
\psi_{\rm pol/ax}(r)=&\frac{\psi^{\rm p/a}_{\rm up}(r)}{W}
\int^r_{2M} 
 \frac{\psi^{\rm p/a}_{\rm in}(r')S_{\rm pol/ax}(r')}{f(r')} dr'\nonumber\\
&+ \frac{\psi^{\rm p/a}_{\rm in}(r)}{W}\int^{\infty}_{r}
 \frac{\psi^{\rm  p/a}_{\rm up}(r')S_{\rm pol/ax}(r')}{f(r')}dr' \,,
\end{align}
For circular orbits, the functions $S_{\rm pol/ax}(r)$ only contain terms proportional to Dirac delta functions $\delta(r - r_{p})$ and derivatives of it (see App.~\ref{app:grav_perts}). Therefore the integrals can be easily computed analytically (derivatives of the Dirac delta function can be dealt with by integrating by parts). Notice in particular that this allows to rewrite the integrals as
\begin{align}
\int^r_{2M} 
&\frac{\psi^{\rm p/a}_{\rm in}S_{\rm pol/ax}}{f} dr' = \Theta(r-r_p)\int^{\infty}_{2M} 
 \frac{\psi^{\rm p/a}_{\rm in}S_{\rm pol/ax}}{f} dr'\,,\nonumber\\
 \int^{\infty}_{r}
&\frac{\psi^{\rm  p/a}_{\rm up}S_{\rm pol/ax}}{f}dr' = \Theta(r_p-r)\int^{\infty}_{2M}
 \frac{\psi^{\rm  p/a}_{\rm up}S_{\rm pol/ax}}{f}dr'\,.
\end{align}

Once solutions for $\psi_{\rm pol}^{lm}(r)$ and $\psi_{\rm ax}^{lm}(r)$ are obtained, the metric perturbations can be reconstructed in a given gauge. Explicit equations to reconstruct the metric perturbations in the standard Regge-Wheeler gauge, which we use throughout this work, are given in App.~\ref{app:grav_perts}.
\subsubsection{Monopolar $l=0$ and dipolar $l=1$ perturbations}
To complete the computation of metric perturbations we also need to consider modes with $l=0$ and $l=1$. Those modes do not contribute to the gravitational radiation that travels towards future null infinity and the BH horizon, however they need to be included for a complete description of the metric perturbations. For a point particle moving in a Schwarzschild BH they were first computed in Ref.~\cite{Zerilli:1970wzz} in a particular gauge that we shall call the ``Zerilli gauge'' following Ref.~\cite{Detweiler:2003ci}. In what follows we will mostly use the Zerilli gauge in which the solutions take their simplest form. For other possible gauge choices see e.g. Ref.~\cite{Detweiler:2003ci}. Since we will need them for later use, let us briefly review the solutions in the case of a point particle in circular, equatorial motion in Schwarzschild.\footnote{Zerilli's original work~\cite{Zerilli:1970wzz} has some sign errors as noticed in~\cite{Sago:2002fe}, so to verify our computations we checked that our solutions reproduce the ones shown in Ref.~\cite{Detweiler:2003ci}. When comparing with~\cite{Detweiler:2003ci} we should also remember that all solutions we show are in the frequency domain.}
%
\paragraph{Monopolar perturbations.}
%
Monopolar $l=0$ metric perturbations are purely polar as can be easily inferred setting $l=m=0$ in Eqs.~\eqref{polar} and~\eqref{axial}. In this case, the Zerilli gauge can be obtained from Eq.~\eqref{polar} by setting the polar functions $H^{l=0}_1(r)=0$ and $K^{l=0}(r)=0$. Following Refs.~\cite{Zerilli:1970wzz,Sago:2002fe}, we find the following analytical solution for the functions $H^{l=0}_0(r)$ and $H^{l=0}_2(r)$~\cite{Detweiler:2003ci}:
\begin{eqnarray}
    H^{l=0}_0(r)&=& \frac{\sqrt{16\pi}\,(r_p-r)\, E}{r(r_p-2M)}\Theta(r-r_p)\,,\\
    H^{l=0}_2(r)&=& \frac{\sqrt{16\pi}\, r\, E}{(r-2M)^2}\Theta(r-r_p)\,,
\end{eqnarray}
where $E:= m_p (r_p-2M)/\sqrt{r_p(r_p-3M)}$ is the particle's conserved energy, $\Theta(r-r_p)$ is the Heaviside step function and we recall that $m_p$ is the mass of the point particle. It is easy to check that in the region $r>r_p$ the perturbed metric simply describes another Schwarzschild geometry with mass $M+E$~\cite{Zerilli:1970wzz,Detweiler:2003ci}.
\paragraph{Dipolar, polar perturbations.}
 Dipolar $l=1$ metric perturbations exist both in the polar and axial sector. In the polar sector, the Zerilli gauge corresponds to setting $K^{l=1}(r)=0$ in Eq.~\eqref{polar}. In the frequency domain, the solutions found in~\cite{Zerilli:1970wzz} can be written as:
\begin{eqnarray}
    H^{l=1}_0(r)&=& Y^*_{1m}(\pi/2,0)\frac{8\pi E (r_p-2M)}{3Mr(r-2M)}\nonumber\\
    &&\times \left(M-r^3\sigma^2\right)\Theta(r-r_p)\delta(\sigma-m\Omega_p)\,,\label{eq:dipH0}\\
    H^{l=1}_1(r)&=& i\,Y^*_{1m}(\pi/2,0)\frac{ 8\pi E r (r_p-2M)}{(r-2M)^2}\sigma\nonumber\\
    &&\times \Theta(r-r_p)\delta(\sigma-m\Omega_p)\,,\label{eq:dipH1}\\
    H^{l=1}_2(r)&=& Y^*_{1m}(\pi/2,0)\frac{8\pi E r(r_p-2M)}{(r-2M)^3}\nonumber\\
    &&\times \Theta(r-r_p)\delta(\sigma-m\Omega_p)\label{eq:dipH2}\,,
\end{eqnarray}
where $Y_{lm}(\theta,\phi)$ are scalar spherical harmonics and $\Omega_p= \pm \sqrt{M/r_p^3}$ is the particle's orbital frequency, with the plus (minus) sign corresponding to prograde (retrograde) orbits~\footnote{In a vacuum Schwarzschild BH, the distinction between prograde and retrograde orbits is purely conventional, given that, in a spherically symmetric spacetime, observables cannot depend on the direction of the orbit. However this distinction becomes relevant when including a rotating environment, as we do below.}. As discussed in Ref.~\cite{Zerilli:1970wzz}, for $r>r_p$, the resulting perturbed metric represents a Schwarzschild solution expressed in a noninertial coordinate system, and therefore one can find a gauge in which the perturbations vanish in the region outside $r=r_p$. However, as emphasized in Ref.~\cite{Detweiler:2003ci}, the perturbations in Eqs.~\eqref{eq:dipH0}--\eqref{eq:dipH2} are not pure gauge because of the presence of the particle at $r=r_p$. 

In fact, as shown in Refs.~\cite{Zerilli:1970wzz,Detweiler:2003ci} one can find a gauge in which the polar $l=1$ metric perturbations can be set to zero everywhere except at $r=r_p$. In this gauge, which we shall call the singular gauge following~\cite{Detweiler:2003ci}, the dipolar component of $h^{{\rm polar}}_{\mu\nu}$ takes the form
\begin{align}\label{polar_singular}
h_{\mu\nu}^{{\rm polar}, l=1} &= \nonumber\\
&
\begin{pmatrix}
0 & H^{s}_{1}(r) Y_{1m} & 0 &0\\
* & H^{s}_{2}(r) Y_{1m} & \eta^s_1(r) \partial_\theta Y_{1m}& \eta^s_1(r) \partial_\phi Y_{1m}\\
*&*&0&0\\
*&*&*&0
\end{pmatrix}\,,
\end{align}
where the superscripts `$s$' emphasize that these quantities are in the singular gauge, asterisks represent symmetric components and $Y_{1m}:=Y_{1m}(\theta,\phi)$. Under this gauge the radial functions are given by~\cite{Detweiler:2003ci}:
\begin{align}
    H^{s}_1(r)&= i\,Y^*_{1m}(\pi/2,0)\frac{4\pi r_p^2  E\sigma}{3M} \delta(r-r_p)\delta(\sigma-m\Omega_p)\,,\label{eq:H1s}\\
    H^{s}_2(r)&= Y^*_{1m}(\pi/2,0)\frac{8\pi r_p^2 E}{3M(r_p-2M)}
    \delta(r-r_p)\delta(\sigma-m\Omega_p)\,,\label{eq:H2s}\\
    \eta^{s}_1(r)&= Y^*_{1m}(\pi/2,0)\frac{4\pi r_p^2 E }{3M}\delta(r-r_p)\delta(\sigma-m\Omega_p)\label{eq:eta1s}\,.
\end{align}
One can verify that plugging these expressions in Eq.~\eqref{decom} reproduces Eqs.~(5.8)--(5.11) in~\cite{Detweiler:2003ci}. As we shall discuss, we will use both the Zerilli and singular gauge to check the consistency of some of our results when computing the (gauge-invariant) scalar fluxes.

\paragraph{Dipolar, axial perturbations.}
Finally, for completeness, let us also discuss $l=1$ perturbations in the axial sector. The Zerilli gauge for these perturbations can be found by setting $h^{l=1}_1(r)=0$ in Eq.~\eqref{axial} and the only free function is therefore $h^{l=1}_0(r)$. Following Refs.~\cite{Zerilli:1970wzz,Detweiler:2003ci} we find that it reads:
\begin{align}
    h^{l=1}_0(r)&= -\left[\frac{8\pi L}{3r}\Theta(r-r_p)+\frac{8\pi L r^2}{3r_p^3}\Theta(r_p-r)\right]\nonumber\\
    &\times \delta(\sigma-m\Omega_p)\partial_\theta Y^*_{1m}(\theta,0)|_{\theta=\pi/2}
    \,,\label{eq:dih0}
\end{align}
where $L= m_p\sqrt{M r_p/(1-3M/r_p)}$ is the particle's conserved angular momentum. As discussed in~\cite{Zerilli:1970wzz,Detweiler:2003ci} this perturbation describes the shift in the spacetime's angular momentum that occurs at $r=r_p$ due to the presence of the point particle.
\subsubsection{Gravitational-wave flux}
The metric perturbations computed using the procedure above can be used to analyse the gravitational radiation emitted towards (future null) infinity and the BH horizon~\cite{Martel:2005ir}. As already mentioned, only the modes with $l\geq 2$ contribute to this radiation. The procedure to compute the energy and angular momentum fluxes can be found in Ref.~\cite{Martel:2005ir}, therefore here we only provide the main equations. The energy flux at infinity $\dot E^{g,\infty}$ and at the BH horizon $\dot E^{g,H}$ can be written as
\begin{equation}
    \dot E^{g,H/\infty} = \dot E^{\rm ax}_{H/\infty} + \dot E^{\rm pol}_ {H/\infty}\,,
\end{equation}
where
\begin{eqnarray}
    \dot E^{{\rm pol/ax}}_{\infty} &=& \lim_{r\to \infty}\frac{1}{64 \pi}  \sum_{l,m}\frac{(l+2)!}{(l-2)!} \left|\dot\Psi_{\rm pol/ax}^{lm}\right|^{2}\,,\label{fluxgrav_inf}\\
     \dot E^{{\rm pol/ax}}_{H} &=& \lim_{r\to 2M}\frac{1}{64 \pi}  \sum_{l,m}\frac{(l+2)!}{(l-2)!} \left|\dot\Psi_{\rm pol/ax}^{lm}\right|^{2}\,.\label{fluxgrav_hor}
\end{eqnarray}
For circular orbits, the computation simplifies considerably since the source terms in Eqs.~\eqref{mastereven} and~\eqref{masterodd} can be factorized as $S_{\rm pol/ax} =\tilde{S}_{\rm pol/ax}\delta(\sigma-m\Omega_p)$, where $\Omega_p$ is the particle's orbital frequency. Therefore the solutions computed using Eq.~\eqref{full_sol_grav} can be similarly factorized and one finds that $|\dot\Psi_{\rm pol/ax}^{lm}(t,r)|^2=(m\Omega_p)^2 \psi_{\rm pol/ax}^{lm}(m\Omega_p,r)$ after using Eq.~\eqref{inversefourier}. In addition, for circular orbits, one finds that polar (axial) perturbations are non-zero only for modes for which the sum $l+m$ is even (odd). 

Finally, we note that angular momentum fluxes can be similarly computed~\cite{Martel:2005ir}. For circular orbits those can be easily obtained from the energy flux through the relation $\dot L^g= \dot E^g/\Omega_p$.
\subsection{Scalar perturbations}\label{sec:KGperts}
At order $\mathcal{O}(q^1,\epsilon^1)$ the only equation one needs to solve to describe perturbations to the scalar field configuration is Eq.~\eqref{KG_ep1q1}. This equation reduces to a Klein-Gordon equation with a source term that depends on $h_{\mu\nu}^{(0)}$ and $\Phi^{(1)}$.  To find solutions for $\Phi^{(q,1)}$, we follow Ref.~\cite{Annulli:2020lyc}~\footnote{Note that slightly different {\it ans\"atze} for the metric and scalar perturbations can be found in Refs.~\cite{Yoshida:1994xi,Kojima:1991np,Macedo:2013jja}. These ans\"atze are ultimately equivalent to the ones used in Ref.~\cite{Annulli:2020lyc} that we here follow.} and decompose the perturbations as
\begin{align}\label{eq:Phiq_decom}
\Phi^{(q,1)}=&\frac{1}{2r}\sum_{\ell_j,m_j} \int d\sigma \left[Z^{\ell_j m_j}_+(r)Y_{\ell_j m_j}(\theta,\phi)e^{-i\sigma t}\right.\nonumber\\
&\left.+\,(Z^{\ell_j m_j}_-(r))^*Y_{\ell_j m_j}^*(\theta,\phi)e^{i\sigma t}\right] e^{-i\omega t}\,.
\end{align}
Using this {\it ansatz} in Eq.~\eqref{KG_ep1q1}, together with~\eqref{background_scalar} and~\eqref{decom}, we find that the resulting equation can separated into two independent pieces, one that only contains factors of $e^{-i(\sigma+\omega)t}$ and another piece that only contains factors of $e^{i(\sigma-\omega)t}$. Equating each of these pieces to zero, allows to find the following equations for $Z^{\ell_j m_j}_{\pm}$: 
\begin{equation}
\label{KG_nonsepar}
\sum_{\ell_j,m_j} Y_{\ell_j m_j}\left[\frac{d^2}{dr^2_*}+(\omega\pm\sigma)^2-V_j\right]Z^{\ell_j m_j}_{\pm} = \sum_{l,m} S^{\pm}_{lm}\,,
\end{equation}
where 
\begin{equation}
V_j(r) = \left(1- \frac{2M}{r}\right)\left(\mu^2+\frac{\ell_j(\ell_j+1)}{r^2}+\frac{2M}{r^3}\right)\,,
\end{equation}
the source term $S^+_{lm}$ is schematically given by
\begin{align}\label{l1_source}
&S^+_{lm}(r,\theta,\phi) = P_{lm}(r) Y_{lm} Y_{\ell_i m_i} \nonumber \\
&+ \hat P_{lm}(r) \left(Y_{,\theta}^{lm}Y_{,\theta}^{\ell_i m_i}+ \frac{Y_{,\phi}^{lm}Y_{,\phi}^{\ell_i m_i}}{\sin^2\theta}\right) \nonumber \\
&+ A_{lm}(r) \frac{Y_{,\theta}^{lm}Y_{,\phi}^{\ell_i m_i}-Y_{,\phi}^{lm}Y_{,\theta}^{\ell_i m_i}}{\sin\theta} \,,
\end{align}
whereas $S^-_{lm}$ can be obtained from Eq.~\eqref{l1_source} by doing the transformation $\{\omega,Y_{\ell_i m_i},R_{n_i\ell_i}\} \to \{-\omega,Y^*_{\ell_i m_i},R^*_{n_i\ell_i}\}$. Here the radial functions $P_{lm},\hat P_{lm}$ only depend on polar functions whereas $A_{lm}$ only depends on axial functions. Their explicit form can be found in App.~\ref{app:scalar_source}.

Let us first focus on the equation for $Z^{\ell_j,m_j}_+$. In order to separate the angular part in Eq.~\eqref{KG_nonsepar}, we project it onto the basis of scalar spherical harmonics. Namely, we multiply Eq.~\eqref{KG_nonsepar} by $Y^*_{\ell'_j m'_j}$ and integrate over the solid angle. Using the orthogonality properties of the spherical harmonics, we find one radial equation for each pair of angular numbers $\{\ell'_j,m'_j\}$ with a source term that contains the following integrals:
\begin{eqnarray}
    \mathcal{P}^{\ell'_j,l,\ell_i}_{m'_j,m,m_i} &:= & \int d\Omega\,Y^*_{\ell'_j m'_j}Y_{lm}Y_{\ell_i m_i}\,,\label{IntP}\\
    \hat{\mathcal{P}}^{\ell'_j,l,\ell_i}_{m'_j,m,m_i} &:= &  \int d\Omega\, Y^*_{\ell'_j m'_j}\mathbf{Y}_a^{lm}\mathbf{Y}_b^{\ell_i m_i}\gamma^{ab}\,,\label{IntHatP}\\
    \mathcal{A}^{\ell'_j,l,\ell_i}_{m'_j,m,m_i} & := &  \int d\Omega\, Y^*_{\ell'_j m'_j}\mathbf{X}_a^{lm}\mathbf{Y}_b^{\ell_i m_i}\gamma^{ab}\label{IntA}\,,
\end{eqnarray}
where we defined $\gamma^{ab}=\diag(1,1/\sin^2\theta)$ and introduced the polar, $\mathbf{Y}_a^{l m}(\theta,\phi)$, and axial, $\mathbf{S}_a^{l m}(\theta,\phi)$, vector spherical harmonics, given by
\begin{eqnarray}
    \mathbf{Y}_a^{l m}(\theta,\phi) & = & \left(Y_{,\theta}^{lm}\, , \,Y_{,\phi}^{lm}\right)\\
    \mathbf{X}_a^{l m}(\theta,\phi) & = & \left(-\frac{Y_{,\phi}^{lm}}{\sin\theta}\,,\, \sin\theta Y_{,\theta}^{lm}\right)\,.
\end{eqnarray}
For the equation that $Z^{\ell_j,m_j}_-$ satisfies, the same procedure can be done and we find integrals in its source term that can be obtained from the ones above by replacing $Y_{\ell_i m_i}$ by its complex conjugate $Y^*_{\ell_i m_i}$. To simplify the notation below we relabel $\{\ell'_j, m'_j\} \to \{\ell_j, m_j\}$.  

As we discuss in App.~\ref{app:sphharmonics}, the integrals~\eqref{IntP}~--~\eqref{IntA} can be computed explicitly in terms of the Wigner 3-j symbols that satisfy known rules (see e.g. Chapter 34 in Ref.~\cite{DLMF}). In particular we find that these integrals vanish unless they satisfy the following selection rules:
\begin{itemize}
    \item $\pm m_i+m-m_j=0$;
    \item $|\ell_j-\ell_i|\leq l\leq \ell_j+\ell_i$;
    \item $\ell_j+\ell_i + l = 2p$ with $p \in \mathbb{N}$ for the integrals~\eqref{IntP} and~\eqref{IntHatP};
    \item $\ell_j+\ell_i + l = 2p + 1$ with $p \in \mathbb{N}$ for the integral~\eqref{IntA}.
\end{itemize}
In the first selection rule, the $+$ and $-$ signs correspond to the selection rule when considering the equations for $Z^{\ell_j,m_j}_+$ and $Z^{\ell_j,m_j}_-$, respectively. In particular, when $\ell_i=m_i=0$, we trivially find that $\ell_j=l$ and $m_j=m$ and only the term proportional to $P_{lm}(r)$ contributes to the source term of the radial equations. 
Therefore in the case where the Schwarzschild BH is surrounded by a spherical cloud, the scalar field perturbation $\Phi^{(q,1)}$ does not couple to axial perturbations. In fact, for a spherical cloud, this seems to remain true also at higher orders in $\epsilon$. We will show this explicitly in Sec.~\ref{sec:Oq1epsilon2} by computing axial perturbations up to order $\mathcal{O}(\epsilon^2)$ when $\ell_i=m_i=0$. On the other hand, for the quasi-bound state $\ell_i=m_i=1$ one finds that $m_j=m\pm 1$ and from the selection rules it follows that a scalar perturbation with angular number $\ell_j$ couples to gravitational polar perturbations with angular number $l=\ell_j\pm 1$ and to axial perturbations with angular number $l=\ell_j$. Finally, we note that from the resulting radial equations and the selection rules we can infer that $Z^{\ell_j,m_j}_-(\sigma;r)^*=(-1)^{m_j} Z^{\ell_j,-m_j}_+(-\sigma;r)$~\footnote{This follows from using $Y^*_{\ell_j, m_j}=(-1)^{m_j}Y_{\ell_j, -m_j}$ and the symmetries of metric functions outlined in App.~\ref{app:grav_perts}.}. Therefore for practical purposes we will only need to compute $Z^{\ell_j,m_j}_+$.

In summary, this procedure allows us to obtain an ordinary differential equation for each pair $\{\ell_j,m_j\}$ of the form
\begin{equation}\label{KG_genericl}
\left[\frac{d^2}{dr^2_*}+(\omega+\sigma)^2-V_j\right]Z^{\ell_j m_j}_{+} =\tilde{S}^{\ell_j,\ell_i}_{m_j,m_i}\,,
\end{equation}
where the source term is given by
\begin{itemize}
\item if $\ell_i=m_i=0$:
\begin{equation}
\tilde{S}^{\ell_j,0}_{m_j,0}(r)=\mathcal{P}^{\ell_j,\ell_j,0}_{m_j,m_j,0} P_{\ell_j m_j}(r)\,;
\end{equation}
\item if $\ell_i=m_i=1$:
\begin{align}
\tilde{S}^{\ell_j,1}_{m_j,1}(r) &= \left[\mathcal{P}^{\ell_j,l,1}_{m_j,m,1} P_{lm}(r)\left(\delta_{l,\ell_j-1} + \delta_{l,\ell_j+1}\right)\right.\nonumber\\
&+\hat{\mathcal{P}}^{\ell_j,l,1}_{m_j,m,1} \hat{P}_{l m}(r)\left(\delta_{l,\ell_j-1} + \delta_{l,\ell_j+1}\right)\nonumber\\
& \left. +  \mathcal{A}^{\ell_j,l,1}_{m_j,m,1} A_{l m}(r)\delta_{l,\ell_j}\right]\delta_{m,m_j-1}\,.
\end{align}
\end{itemize}
Based on the selection rules above, similar expressions can be derived for other values of $\{\ell_i,m_i\}$, but we do not write them explicitly since we only consider the cases $\ell_i=m_i=0$ and $\ell_i=m_i=1$ in this paper. We also note that, from the orthogonality properties of the spherical harmonics, we have that $\mathcal{P}^{\ell_j,\ell_j,0}_{m_j,m_j,0}=Y_{00}=1/\sqrt{4\pi}$.

As done in the case of metric perturbations, solutions to Eq.~\eqref{KG_genericl} can be found using a Green’s function approach. In this case, however, the asymptotic solutions at infinity will depend on whether $\omega_+^2-\mu^2>0$ or $\omega_+^2-\mu^2<0$, where we defined $\omega_+=\omega+\sigma$. If $\omega_+^2-\mu^2>0$, one can construct one solution that behaves as an outgoing wave at spatial infinity and a second that behaves as an ingoing wave at the horizon (in the following we omit the superscript $\{\ell_j,m_j\}$ for convenience):
\begin{eqnarray}
Z_{\rm in}
& \to &
\begin{cases}
e^{-i\omega_+ r_*}\,,&
 r\to 2M, \\
A e^{i k_{+} r_*} r^{-\nu_+}+
B e^{-i k_{+} r_*}r^{\nu_+}\,,&
 r\to \infty, \\
\end{cases}
\label{Zin} 
\\
~\nonumber\\
Z_{\rm up}
&\to&
\begin{cases} 
C e^{i \omega_+ r_*}+ 
D e^{-i \omega_+ r_*}\,,& 
 r\to 2M, \\
e^{i k_{+} r_*} r^{-\nu_+}\,,& 
 r\to \infty, \\
\end{cases}
\label{Zup} 
\end{eqnarray} 
where $k_{+}=\mathop{\mathrm{sgn}}(\omega_+)\sqrt{\omega_+^2-\mu^2}$ and $\nu_+=-i M\mu^2/k_{+}$. Here the sign function $\mathop{\mathrm{sgn}}(\omega_+)$ ensures that $Z_{\rm up}$ describes an outgoing wave at infinity. 
On the other hand if $\omega_+^2-\mu^2<0$, no waves can propagate to infinity. In that case we require $Z_{\rm up}$ to be regular at infinity, in which case we set $k_{+}=\sqrt{\omega_+^2-\mu^2}$ in Eqs.~\eqref{Zin} and~\eqref{Zup}, such that $Z_{\rm up} \sim e^{-\sqrt{\mu^2-\omega_+^2} r_*} r^{-\nu_+}$ at infinity.
The solution to the inhomogeneous equation with appropriate boundary conditions is then given by
\begin{align}\label{full_sol_KG}
Z_{+}(r)=&\frac{Z_{\rm up}(r)}{W}
\int^r_{2M} 
 \frac{Z_{\rm in}(r')\tilde{S}(r')}{f(r')} dr'\nonumber\\
&+ \frac{Z_{\rm in}(r)}{W}\int^{\infty}_{r}
 \frac{Z_{\rm up}(r')\tilde{S}(r')}{f(r')}dr' \,,
\end{align}
where to ease the notation we defined $\tilde{S}(r'):=\tilde{S}^{\ell_j,\ell_i}_{m_j,m_i}(r)$ and the Wronskian is now given by
\begin{equation}\label{wronskian_scal}
W(Z_{\rm in},Z_{\rm up})=2ik_{+}B\,.
\end{equation}
From Eq.~\eqref{full_sol_KG}, one infers that for circular orbits the radial functions $Z_{\pm}$ can be factorized as $Z^{\ell_j m_j}_{\pm}= \tilde Z^{\ell_j m_j}_{\pm}\delta\left(\sigma-\Omega^{m_j}_\pm\right)$, where $\Omega^{m_j}_\pm=\left(m_j\mp m_i\right)\Omega_p$ according to the selection rules mentioned previously. This follows from the fact that all metric functions inside the source term of Eq.~\eqref{KG_genericl} can also be similarly factorized (see Sec.~\ref{sec:GRperts}). Notice the symmetries $\Omega^{-m_j}_\pm=-\Omega^{m_j}_\mp$. In the particular case $m_i=0$, one also has $\Omega_+^{m_j,m_i=0}=\Omega_-^{m_j,m_i=0}$. 

In order to compute the integrals in Eq.~\eqref{full_sol_KG} we notice that, once we get the metric perturbations using the procedure shown in Sec.~\ref{sec:GRperts} and App.~\ref{app:grav_perts}, one can separate the source $\tilde{S}^{\ell_j,\ell_i}_{m_j,m_i}(r)$ in different factors that depend on either $\delta(r-r_p)$, $\delta'(r-r_p)$, $\Theta(r - r_p)$ or $\Theta(r_p - r)$. We therefore separate the integrals in Eq.~\eqref{full_sol_KG} into different pieces, where the terms involving the Dirac delta function and its derivative are integrated analytically, whereas the terms containing the Heaviside step functions are instead integrated numerically using \texttt{Mathematica}'s built-in function \texttt{NDSolveValue} (see also Sec.~\ref{sec:scalarfluxresults} below for more details concerning the computation of those integrals).

As a final note of caution, we remark that in the case in which $\ell_j=\ell_i$ and $m_j= m_i$ one should be careful when solving Eq.~\eqref{KG_genericl}, because~\eqref{full_sol_KG} is ill defined when $\ell_j=\ell_i$ and $m_j=m_i$~\footnote{Notice that this also implies that the solution only has support at $\sigma=0$, given that $Z^{\ell_i m_i}_{+}= \tilde Z^{\ell_i m_i}_{+}\delta\left(\sigma\right)$.}, i.e., if we naively set $\omega=\omega_{n_i \ell_i m_i}$ to be the eigenfrequency of the quasi-bound state as computed in the Schwarzschild BH background, then the Wronskian~\eqref{wronskian_scal} identically vanishes when $\ell_j=\ell_i$, $m_j=m_i$ and $\sigma=0$~\footnote{This just follows from the fact that, in this case, the unique solution to the homogeneous part of Eq.~\eqref{KG_genericl} is simply the eigenstate $r R_{n_i\ell_i}$, which satisfies the boundary conditions~\eqref{QBsBCs} and therefore $B=C=0$ in~\eqref{Zin} and \eqref{Zup}.}. As we anticipated in Sec.~\ref{sec:QBSs}, the usual approach to circumvent this problem is to expand the eigenfrequencies as $\omega=\omega^{(0)}+q\omega^{(q)}$~\cite{Hussain:2022ins}, where we remind that here $\omega^{(0)}=\omega_{n_i \ell_i m_i}$ corresponds to the eigenfrequency in the background Schwarzschild BH. In App.~\ref{app:freq_shift} we show that by doing this expansion, Eq.~\eqref{KG_genericl} with $\ell_j=\ell_i$ and $m_j=m_i$, can be used to compute $\omega^{(q)}$. Obtaining this frequency shift is an interesting problem on its own, but here we will not compute it, since it does not affect the leading-order power lost by the point particle due to scalar radiation.
%
\subsubsection{Scalar energy and angular momentum fluxes}
In addition to GW emission, the scalar field perturbations will also contribute to the total energy and angular momentum radiated towards infinity and towards the BH horizon. The energy and angular momentum fluxes can be computed using the  scalar field's stress-energy tensor. Assuming a Schwarzschild BH background, the (orbital-averaged) energy flux towards infinity and at the horizon are given by~\cite{Teukolsky:1973ha,Teukolsky:1974yv,Annulli:2020lyc}
\begin{eqnarray}
    \dot E^{\Phi,\infty} &=& -\lim_{r\to +\infty} r^2\int d\Omega\, T^{\Phi}_{\mu r}\xi_{(t)}^{\mu}\,,\label{eq:dotEinf_phi}\\ 
    \dot E^{\Phi,H} &=& \lim_{r\to 2M} 4M^2 \int d\Omega\, T^{\Phi}_{\mu\nu}\xi_{(t)}^{\mu}l^{\nu}\,,\label{eq:dotEhor_phi}
\end{eqnarray}
where $\xi_{(t)}^{\mu}:=\partial/\partial t$ is the Killing vector field associated with the BH metric's invariance under time translations and $l^{\mu}$ is a null vector, normal to the horizon. The flux of angular momentum along the $z$ direction is instead given by
\begin{eqnarray}
    \dot L^{\Phi,\infty} &=& \lim_{r\to +\infty} r^2\int d\Omega\, T^{\Phi}_{\mu r}\xi_{(\phi)}^{\mu}\,, \label{eq:dotLinf_phi}\\ 
    \dot L^{\Phi,H} &=& - \lim_{r\to 2M} 4M^2 \int d\Omega\, T^{\Phi}_{\mu\nu}\xi_{(\phi)}^{\mu}l^{\nu}\,,\label{eq:dotLhor_phi}
\end{eqnarray}
where $\xi_{(\phi)}^{\mu}:=\partial/\partial\phi$ is the Killing vector field associated with the axisymmetry of the BH metric. 

In order to compute the scalar fluxes let us make some remarks about the approximations we employ: (i) following Ref.~\cite{Dolan:2007mj}, one can show that Eqs.~\eqref{eq:dotEhor_phi} and~\eqref{eq:dotLhor_phi} predict that at order $\mathcal{O}(q^0,\epsilon^2)$ one gets a horizon flux term proportional to $\epsilon^2 \Im(\omega)$, related to the slow decay of the background quasi-bound states $\Phi^{(1)}$, which occurs even in the absence of the point particle (see Sec.~\ref{sec:QBSs} and also Sec.~\ref{sec:Oq1epsilon2} where we show this explicitly for the case of a $\ell_i=m_i=0$ quasi-bound state). As already mentioned, throughout this work we neglect this decay. However, its inclusion can be done in a straightforward manner by just adding an additional term in the horizon fluxes computed below;
(ii) in general, the stress-energy tensor~\eqref{KG_ST} also contains cross terms that involve $\partial\Phi^{(q,1)}\partial\Phi^{(1)*}$ (and the complex conjugate of such terms). 
At spatial infinity these terms do not contribute to the fluxes, given that $\Phi^{(1)}$ decays exponentially there. 
On the other hand, at the horizon, one needs to deal with these cross terms more carefully. If $\ell_j\neq \ell_i$ or $m_j\neq m_i$, cross terms of the type $\partial\Phi^{(q,1)}\partial\Phi^{(1)*}$ end up vanishing after integrating over the solid angle, due to the orthonormality properties of the spherical harmonics. 
For $\ell_j=\ell_i$ and $m_j=\pm m_i$ instead, one needs to consider how the cross terms coming from $Z_+$ and $Z^*_-$ in~\eqref{eq:Phiq_decom} affect $\partial\Phi^{(q,1)}\partial\Phi^{(1)*}$, separately. 
If $\ell_j=\ell_i$ and $m_j=m_i$ ($m_j=-m_i$), cross terms related to $Z^*_-$ ($Z_+$) end up not contributing to the fluxes since they vanish after integrating over the solid angle. 
On the other hand, if $\ell_j=\ell_i$ and $m_j=m_i$ ($m_j=-m_i$) the terms in $\Phi^{(q,1)}$ coming from $Z_+$ ($Z^*_-$) are related to static perturbations (i.e. they have $\sigma=0$) of the scalar cloud profile and do not contribute directly to the power lost by the point particle at leading order, therefore we will not consider those terms;
(iii) if we use Eqs.~\eqref{eq:dotEinf_phi}
~--~\eqref{eq:dotLhor_phi} with $T^{\Phi}_{\mu\nu}$ computed in a Schwarzschild background and with $\Phi=\epsilon\Phi^{(1)}+\epsilon q\Phi^{(q,1)}$, up to order $\mathcal{O}(\epsilon^2,q^2)$, we are actually also neglecting terms that, schematically, involve $h^{(0)}_{\mu\nu}\partial\Phi^{(1)}\partial\Phi^{(1)*}$ and $h^{(0)}_{\mu\nu}\partial\Phi^{(q,1)}\partial\Phi^{(1)*}$ (and complex conjugates). The terms of the type $h^{(0)}_{\mu\nu}\partial\Phi^{(1)}\partial\Phi^{(1)*}$ oscillate with frequencies $e^{\pm i \sigma t}$ and therefore average to zero under an orbit average of the fluxes. On the other hand, terms of the type $h^{(0)}_{\mu\nu}\partial\Phi^{(q,1)}\partial\Phi^{(1)*}$ contain non-oscillating pieces that cannot be averaged to zero, and that, as far as we could check, do not vanish when integrating over the sphere, given that they essentially involve integrals over three harmonics of the type discussed above. While one can argue that these terms can be neglected at infinity, given that $\Phi^{(1)}$ is exponentially suppressed there, at the horizon in principle they should contribute, since none of the fields vanish there. We leave a more concrete understanding of this issue for future work, and make the simplifying assumption of only computing the part of the fluxes that are quadratic in $\Phi^{(q,1)}$.

In summary, for the purposes of computing the fluxes related to the scalar perturbations $\Phi^{(q,1)}$, we will simply use Eq.~\eqref{KG_ST} computed in a Schwarzschild BH background, with $\Phi \to \Phi^{(q,1)}$. Therefore, at leading order, the fluxes related solely to scalar perturbations will be of order $\mathcal{O}(q^2,\epsilon^2)$ since they involve terms quadratic in $\Phi^{(q,1)}$ [see Eqs.~\eqref{eq:Phi_expansion} and~\eqref{eq:Phiq_exp}]. In order to simplify the notation, in the following we will absorb the factor $q^2\epsilon^2$ inside the definition of $\Phi^{(q,1)}$, however one should keep in mind that all expressions we show below are proportional to $q^2\epsilon^2$.
After performing the integration over the solid angle, one can also see that the fluxes can be separated in different modes, due to the orthonormality properties of the spherical harmonics, with the total flux simply obtained by summing over all modes, i.e., $\dot E^{\Phi,\infty/H} = \sum_{\ell_j,m_j}\dot E^{\Phi, \infty/H}_{\ell_j m_j}$. 

Inserting Eq.~\eqref{eq:Phiq_decom} in Eq.~\eqref{eq:dotEinf_phi}, and using $Z^{\ell_j m_j}_{\pm}= \tilde Z^{\ell_j m_j}_{\pm}\delta\left(\sigma-\Omega^{m_j}_\pm\right)$ we find that the energy flux\footnote{As explained in the Erratum~\cite{Brito:2023pyl}, the original published version of this manuscript contained a factor 2 error in the flux formulas. This error has been corrected here.} for a given mode $\{\ell_j,m_j\}$ at infinity is given by~\cite{Annulli:2020lyc}\footnote{For convenience, in the following we drop the superscript $m_j$ in $\Omega^{m_j}_\pm$, i.e. $\Omega_\pm:=\Omega^{m_j}_\pm$.}
\begin{align}\label{dEphi_inf}
    &\dot E^{\Phi, \infty}_{\ell_j m_j} = \left|\omega+\Omega_+\right|\Re\left[\sqrt{\left(\Omega_+ +\omega\right)^2-\mu^2}\right]\left|\tilde Z^{\ell_j m_j}_+(\Omega_+)\right|^{2} \nonumber\\
   & + \left|\omega-\Omega_-\right|\Re\left[\sqrt{\left(\Omega_- -\omega\right)^2-\mu^2}\right]\left|\tilde Z^{\ell_j m_j}_-(\Omega_-)\right|^{2}\,,
\end{align}
where here one should compute $\tilde Z^{\ell_j m_j}_\pm(\Omega_\pm)$ in the limit $r\to\infty$. Similarly, after using Eq.~\eqref{eq:dotEhor_phi} for the flux at the horizon we find
\begin{align}\label{dEphi_hor}
    \dot E^{\Phi, H}_{\ell_j m_j} & = \left(\omega+\Omega_+\right)^2\left|\tilde Z^{\ell_j m_j}_+(\Omega_+)\right|^{2} \nonumber\\
    & +\left(\omega-\Omega_-\right)^2\left|\tilde Z^{\ell_j m_j}_-(\Omega_-)\right|^{2}\,,
\end{align}
where now $\tilde Z^{\ell_j m_j}_\pm(\Omega_\pm)$ is computed in the limit $r\to 2M$.
On the other hand, the angular momentum fluxes are given by
\begin{align}\label{dLphi_inf}
    \dot L^{\Phi, \infty}_{\ell_j m_j} &= m_j s_+\Re\left[\sqrt{\left(\Omega_+ +\omega\right)^2-\mu^2}\right]\left|\tilde Z^{\ell_j m_j}_+(\Omega_+)\right|^{2} \nonumber\\
   & - m_j s_-\Re\left[\sqrt{\left(\Omega_- -\omega\right)^2-\mu^2}\right]\left|\tilde Z^{\ell_j m_j}_-(\Omega_-)\right|^{2}\,,\\
    \dot L^{\Phi, H}_{\ell_j m_j} &=  m_j\left(\omega+\Omega_+\right)\left|\tilde Z^{\ell_j m_j}_+(\Omega_+)\right|^{2} \nonumber\\
   & -m_j\left(\omega-\Omega_-\right)\left|\tilde Z^{\ell_j m_j}_-(\Omega_-)\right|^{2}\,,\label{dLphi_hor}
\end{align}
where, for convenience, we defined $s_{\pm}=\mathop{\mathrm{sgn}}(\omega\pm\Omega_\pm)$.

In order to understand how the scalar radiation affects the secondary object, one needs to take into account the fact that scalar perturbations should also affect the scalar cloud configuration. Here we follow the arguments of Ref.~\cite{Annulli:2020lyc} where a similar problem was studied but for the case of boson stars. To compute the rate at which the total mass of the scalar cloud $M_b$ changes due to the motion of the point particle, we use the fact that $M_b$ is related to the cloud's Noether charge by $M_b = \omega Q$, and similarly for the cloud's total angular momentum $J_b = m_i Q$ (see App.~\ref{app:mass_vs_charge} for a proof of these relations). Using these relations we can compute the rate of change of the cloud's mass and angular momentum using
\begin{equation}
\dot M_b = \omega \dot Q\,,\quad \dot J_b = m_i \dot Q\,.\label{dMbdJb}
\end{equation}
Notice that here we have neglected accretion onto the small secondary object~\cite{Baumann:2021fkf}, which would add a term in the right-hand side of the equations above related to the rate of change of the mass and spin of the secondary. For very small mass ratios we expect this accretion term to be subdominant~\cite{Baumann:2021fkf}, however an accurate evolution of the orbits should include it, as done in Ref.~\cite{Baumann:2021fkf}.
Using the fact that the current given by Eq.~\eqref{KG_current} is conserved, one can use the divergence theorem to compute the rate of change of the scalar charge $\dot Q$~\cite{Annulli:2020lyc}:
\begin{eqnarray}
    \dot Q^{\infty} &=& -\lim_{r\to +\infty} r^2\int d\Omega\, j_{r}\,, \\ 
    \dot Q^{H} &=&  \lim_{r\to 2M} 4M^2 \int d\Omega\, j_{\mu}l^{\mu}\,.
\end{eqnarray}
Plugging the {\it ansatz} for the scalar perturbations in those expressions, and making again the same approximations mentioned below Eq.~\eqref{eq:dotLhor_phi}, we find~\cite{Annulli:2020lyc}
\begin{align}\label{dQ_inf}
     \dot Q^{\infty}_{\ell_j m_j} &= - \,s_+\Re\left[\sqrt{\left(\Omega_+ +\omega\right)^2-\mu^2}\right]\left|\tilde Z^{\ell_j m_j}_+(\Omega_+)\right|^{2} \nonumber\\
   & -  s_-\Re\left[\sqrt{\left(\Omega_- -\omega\right)^2-\mu^2}\right]\left|\tilde Z^{\ell_j m_j}_-(\Omega_-)\right|^{2}\,,\\
    \dot Q^{H}_{\ell_j m_j} &= -\left(\omega+\Omega_+\right)\left|\tilde Z^{\ell_j m_j}_+(\Omega_+)\right|^{2} \nonumber\\
   & -\left(\omega-\Omega_-\right)\left|\tilde Z^{\ell_j m_j}_-(\Omega_-)\right|^{2}\,,\label{dQ_hor}
\end{align}
where again it is implicitly assumed in these expressions that $\tilde Z_{\pm}$ should be computed in the limit $r\to \infty$ ($r\to 2M$) when computing $\dot Q^{\infty}$ ($\dot Q^{H}$).

Given the fluxes computed above, we are now interested in understanding how those are related to the energy lost by the point particle. Under an adiabatic approximation, and by conservation of the system's total energy and angular momentum, we expect that the particle's energy $E$ and angular momentum $L$ will change according to~\cite{Annulli:2020lyc}:
\begin{eqnarray}
    \frac{dE}{dt} &=& -\dot E^{g,\infty}-\dot E^{g,H}- \dot E^{s,\infty}- \dot E^{s,H}\,,\label{totaldE}\\
    \frac{dL}{dt} &=& -\dot L^{g,\infty}-\dot L^{g,H}- \dot L^{s,\infty}- \dot L^{s,H}\label{totaldL}\,,
\end{eqnarray}
where $\dot E^{g,\infty/H}$ and $\dot L^{g,\infty/H}$ are the GW energy and angular momentum fluxes at infinity and at the BH horizon, respectively, and we defined
\begin{eqnarray}
    \dot E^{s,\infty/H} &=&\dot E^{\Phi,\infty/H}+\omega\,\dot Q^{\infty/H}\,,\\
    \dot L^{s,\infty/H} &=&\dot L^{\Phi,\infty/H}+m_i\,\dot Q^{\infty/H}\,.
\end{eqnarray}
The quantities $\dot E^{s}=\dot E^{s,\infty}+\dot E^{s,H}$ and $\dot L^{s}=\dot L^{s,\infty}+\dot L^{s,H}$ represent the total rate of change of the energy and angular momentum of the point particle due to the presence of the scalar field configuration. In particular, $\dot E^{s,\infty}$ is equivalent to the \emph{ionization power} computed in a Newtonian approximation in Refs.~\cite{Baumann:2021fkf,Baumann:2022pkl,Tomaselli:2023ysb} and can also be thought as the power lost due to \emph{dynamical friction}~\cite{Zhang:2019eid,Annulli:2020lyc,Tomaselli:2023ysb}, whereas we will see later that $\dot E^{s,H}$ encodes information about the resonant \emph{transitions} between different states of the cloud considered in Refs.~\cite{Baumann:2018vus,Baumann:2019ztm}. For lack of a better name, we will often times simply refer to $\dot E^{s}$ as the scalar power. Using the expressions for the fluxes derived above we find
\begin{align}
    \dot E^{s, \infty}_{\ell_j m_j} &= \Omega_+ s_+\Re\left[\sqrt{\left(\Omega_++\omega\right)^2-\mu^2}\right]\left|\tilde Z^{\ell_j m_j}_+(\Omega_+)\right|^{2} \nonumber\\
   & - \Omega_-s_-\Re\left[\sqrt{\left(\Omega_- -\omega\right)^2-\mu^2}\right]\left|\tilde Z^{\ell_j m_j}_-(\Omega_-)\right|^{2}\,, \label{dotEsinf}\\
    \dot E^{s, H}_{\ell_j m_j} &= \Omega_+\left(\omega+\Omega_+\right)\left|\tilde Z^{\ell_j m_j}_+(\Omega_+)\right|^{2} \nonumber\\
   & -\Omega_-\left(\omega-\Omega_-\right)\left|\tilde Z^{\ell_j m_j}_-(\Omega_-)\right|^{2}\,, \label{dotEshor}
\end{align}
and
\begin{align}
    \dot L^{s, \infty}_{\ell_j m_j} &=
     m_+ s_+\Re\left[\sqrt{\left(\Omega_+ +\omega\right)^2-\mu^2}\right]\left|\tilde Z^{\ell_j m_j}_+(\Omega_+)\right|^{2} \nonumber\\
    & - m_- s_-\Re\left[\sqrt{\left(\Omega_- -\omega\right)^2-\mu^2}\right]\left|\tilde Z^{\ell_j m_j}_-(\Omega_-)\right|^{2}\,, \label{dotLsinf}\\
    \dot L^{s, H}_{\ell_j m_j} &=  m_+\left(\omega+\Omega_+\right)\left|\tilde Z^{\ell_j m_j}_+(\Omega_+)\right|^{2} \nonumber\\
   & -m_-\left(\omega-\Omega_-\right)\left|\tilde Z^{\ell_j m_j}_-(\Omega_-)\right|^{2}\,, \label{dotLshor}
\end{align}
where $m_\pm=m_j\mp m_i$. In practice, we can use the symmetries $\Omega^{m_j}_-=-\Omega^{-m_j}_+$ and $\left|\tilde Z^{\ell_j, m_j}_-(\Omega^{m_j}_-)\right|^{2}=\left|\tilde Z^{\ell_j, -m_j}_+(\Omega^{-m_j}_+)\right|^{2}$ such that only $Z_+$ needs to be computed. In addition, we note that these symmetries also imply that $\dot E^{s}_{\ell_j m_j}=\dot E^{s}_{\ell_j -m_j}$ and $\dot L^{s}_{\ell_j m_j}=\dot L^{s}_{\ell_j -m_j}$. For circular orbits, we also find that the source terms of Eq.~\eqref{KG_genericl} vanish for modes in which the sum $\ell_j+m_j$ is odd. Therefore those modes do not contribute to the scalar power.

We also notice that when $m_i=0$, the scalar power $\dot E^{s}$ is invariant under a transformation $\Omega_p \to -\Omega_p$, corresponding to changing from prograde to retrograde orbits, whereas the rate of change of the angular momentum $\dot L^{s}$ changes sign under this transformation, as one would expect for a point particle moving in a spherically symmetric environment. On the other hand, if $m_i\neq 0$, this symmetry is broken. In this case retrograde and prograde orbits~\footnote{Here prograde (retrograde) means that the orbit rotates in the same (opposite) direction of the cloud's rotation.} need to be considered separately, in agreement with Refs.~\cite{Baumann:2018vus,Baumann:2021fkf,Tomaselli:2023ysb}. 

As a final note, let us make the remark that for simplicity we have focused on the case where the cloud is composed by a complex scalar field. For real scalar fields there is nothing equivalent to the Noether charge and therefore our setup would need to be modified. However, we conjecture that our formulas for the scalar energy loss should still apply to real scalar fields in an averaged sense. We expect that the main difference for a real field is the need to include an additional term in the total energy and angular momentum loss budget, Eqs.~\eqref{totaldE} and~\eqref{totaldL}, describing the slow decay of the cloud due to the emission of GWs by the cloud itself~\cite{Yoshino:2013ofa,Brito:2014wla,Brito:2017zvb}, as done for example in Refs.~\cite{Kavic:2019cgk,Xie:2022uvp,Takahashi:2023flk,Cao:2023fyv}.
%
\subsection{Numerical procedure and results}\label{sec:scalarfluxresults}
In order to compute the scalar power for a point particle in circular, equatorial motion, we have implemented the procedure described above in a \texttt{Mathematica} code, considering the particular cases where the background scalar field is either a spherically symmetric $\ell_i=m_i=0$ or a dipolar $\ell_i=m_i=1$ cloud. We also consider that those clouds are in the fundamental state $n_i=0$. To find the solutions $\psi^{\rm p/a}_{\rm in/up}$ and $Z_{\rm in/up}$ we numerically integrate the corresponding homogeneous differential equations, requiring that they satisfy the boundary conditions outlined in Eqs.~\eqref{psiin}, \eqref{psiup}, \eqref{Zin} and \eqref{Zup}. The numerical domain is restricted to the range $r \in [2M(1+\epsilon_h),r_{\infty}]$ where, to achieve good enough accuracy, $\epsilon_h \ll 1$ and $r_{\infty}$ typically extends up to many wavelengths, i.e. $r_{\infty}\sigma \gg 1$ for metric perturbations and $r_{\infty}|\omega_+| \gg 1$ for scalar perturbations. Numerically integrating over such a large domain becomes however impractical when $\omega_+^2-\mu^2<0$, as first noticed in Ref.~\cite{Macedo:2013jja}. When $\omega_+^2-\mu^2<0$, the scalar perturbations are exponentially suppressed [cf. discussion below Eqs.~\eqref{Zin} and \eqref{Zup}], and therefore in that situation it becomes unfeasible to numerically integrate the scalar perturbation equation up to scales much larger than $1/\sqrt{\mu^2-\omega_+^2}$. This sets an upper limit on the value of $r_{\infty}$ that can be used in this case~\cite{Macedo:2013jja}. Results shown here were typically obtained with $\epsilon_h=10^{-4}$ and $r_{\infty} = 3000/\sigma$ for metric perturbations. On the other hand, for scalar perturbations, we used $r_{\infty} = 3000/|\omega_+|$ if $\omega_+^2-\mu^2>0$ and $r_{\infty} = 50/k_+$ if $\omega_+^2-\mu^2<0$. Notice that very close to the transition point when $\omega_+^2-\mu^2 \sim 0$, it becomes extremely challenging to compute the scalar power accurately, since one needs to use very large $r_{\infty}$, therefore we jump over the orbital radii for which $\omega_+^2-\mu^2\sim 0$ and assume that the scalar power changes smoothly as we approach that point from either side.

Following previous work (see e.g.~\cite{Pani:2011xj,Macedo:2013jja,Figueiredo:2023gas}), in order to reduce numerical truncation errors, we use a series expansion for the boundary conditions at the horizon and at infinity, namely,
 \begin{eqnarray}
    \psi^{p/a}_{\rm{in}}(r\to 2M)&=& e^{-i \sigma r_*} \sum_{i=0}^{n_{\rm in}} a_i (r-2M)^i\,,\label{psiIN_series}\\
    \psi^{p/a}_{\rm{up}}(r\to \infty)&=& e^{+i \sigma r_*} \sum_{i=0}^{n_{\rm up}} \frac{b_i}{r^i}\,,\label{psiUP_series}\\
     Z_{\rm{in}}(r\to 2M)&=& e^{-i \omega_+ r_*} \sum_{i=0}^{\tilde n_{\rm in}} \tilde{a}_i (r-2M)^i\,,\\
    Z_{\rm{up}}(r\to \infty)&=& e^{+i k_+ r_*}r^{-\nu_+} \sum_{i=0}^{\tilde n_{\rm up}} \frac{\tilde{b}_i}{r^i}\,,\label{Zup_series}
\end{eqnarray}
where we typically include up to ten terms in the series expansions. The coefficients $(a_i,b_i,\tilde{a}_i,\tilde{b}_i)$ are obtained by inserting these series expansions in the homogeneous equations and solving them at each order in $(r-2M)$ or $1/r$ and fixing the zeroth-order coefficients $a_0=b_0=\tilde{a}_0=\tilde{b}_0=1$. We also use a series expansion of the type shown in Eqs.~\eqref{psiUP_series} and~\eqref{Zup_series} to describe the asymptotic behavior of $\psi^{p/a}_{\rm{in}}$ and $Z_{\rm{in}}$ at $r_{\infty}$ [cf. Eqs.~\eqref{psiin} and~\eqref{Zin}] which are used to extract the coefficients $B^{\rm p/a}_{\rm inc}$ and $B$. Those coefficients are then used to compute the Wronskians according to Eqs.~\eqref{wronskian_grav} and~\eqref{wronskian_scal}. 

The integrals appearing in Eq.~\eqref{full_sol_KG} are computed within the same numerical domain we mentioned above when discussing scalar perturbations (at least for the pieces in those integrals that need to be evaluated numerically). Here we should remark that the integrands can extend up to the horizon when the source term of Eq.~\eqref{KG_genericl} contains metric perturbations with $l\geq 2$~\footnote{When the source term contains metric perturbations with $l\geq 2$, the integrands in Eq.~\eqref{KG_genericl} involve terms that are multiplied by $\Theta(r_p-r)$. Therefore those terms have support up to the horizon.}. Given that the integrand also depends on the radial function $R_{n_i\ell_i}$ and its first derivative (see App.~\ref{app:scalar_source}), there is a very mild divergence at $r=2M$ due to the fact that $R_{n_i\ell_i}$ diverges extremely slowly when approaching the horizon [see Eq.~\eqref{QBsBCs}, and remember that $\Im(\omega)<0$]. Therefore, in principle, one would need to implement some regularization procedure to perform such integrals. Similar non-convergent integrals are commonly found in calculations involving quasinormal modes and different regularization procedures have been proposed in such cases (see e.g.~\cite{Leaver:1986gd,Leung:1997was,Zhang:2013ksa,Sberna:2021eui,Green:2022htq,Hussain:2022ins}). We did not attempt at implementing a regularization scheme in our case, since we always have $M\Im(\omega)\ll 1$ for the values of $M\mu$ we considered here. Hence, the divergence of $R_{n_i\ell_i}$ when approaching $r\to 2M$ is sufficiently mild such that, in practice, $R_{n_i\ell_i}$ can be considered approximately regular since it behaves as $\sim e^{-i\Re(\omega) r_*}$ up to reasonably small values of $\epsilon_h$. Therefore we checked that, for the values of $M\mu$ and the multipoles of the scalar power that we considered (see results below), our results are indeed stable when varying $\epsilon_h$ between $\sim [10^{-2},10^{-5}]$,\footnote{For $\ell_i>0$ the lower end of this range can be pushed even further down while still obtaining stable results, given that $\Im(\omega)$ decreases very quickly with $\ell_i$ [cf. Eq.~\eqref{omegaI}].}, indicating that good convergence is obtained within that range.

We also checked that the results are stable under changes of the numerical infinity $r_{\infty}$ and that our code reproduces the GW fluxes available in the \texttt{Black Hole Perturbation Toolkit}~\cite{BHPToolkit}. The results shown below were obtained using the Regge-Wheeler and Zerilli gauges for the metric perturbations, however, to check the robustness of our results we also checked that, for the cases where a comparison is possible, our results do not change when using the singular gauge for $l=1$ metric perturbations (see Sec.~\ref{sec:GRperts}), which further supports the robustness of our results. This is discussed in App.~\ref{app:fluxsingular}.

We should also mention that since $M_b/M \propto \epsilon^2$, it follows that the scalar power scales linearly with $M_b/M$. Therefore we normalize all the results for the scalar power by $M_b$, which we compute using Eq.~\eqref{Mb_EF} in App.~\ref{app:mass_vs_charge}. The requirement that the scalar cloud only affects the geometry perturbatively, i.e., $\epsilon |\Phi^{(1)}|\ll 1$, implies that our approximation should be formally only valid when $M_b/M\ll 1$. We do note however that even for $M\mu=0.2$, the largest value of $M\mu$ we considered, we find that for a cloud with total mass $M_b/M\sim \mathcal{O}(1)$, the maximum scalar amplitude is $\epsilon |\Phi^{(1)}| \sim \mathcal{O}(10^{-2})$ in the case of spherical clouds and $\epsilon |\Phi^{(1)}| \sim \mathcal{O}(10^{-3})$ for dipolar clouds~\footnote{These numbers were obtained by numerically evaluating Eq.~\eqref{Mb_EF} and agree quite well with estimates using Newtonian approximations, see e.g. Eqs. (11) and (24) in Ref.~\cite{DeLuca:2021ite}.}. For smaller values of $M\mu$, one obtains even smaller values for $\epsilon |\Phi^{(1)}|$. Therefore we expect that our perturbative results should be a very good approximation even for $M_b/M\sim \mathcal{O}(1)$.
\subsubsection{Results for $\ell_i=m_i=0$}
%
\begin{figure}[thb!]
    \includegraphics[width=0.48\textwidth]{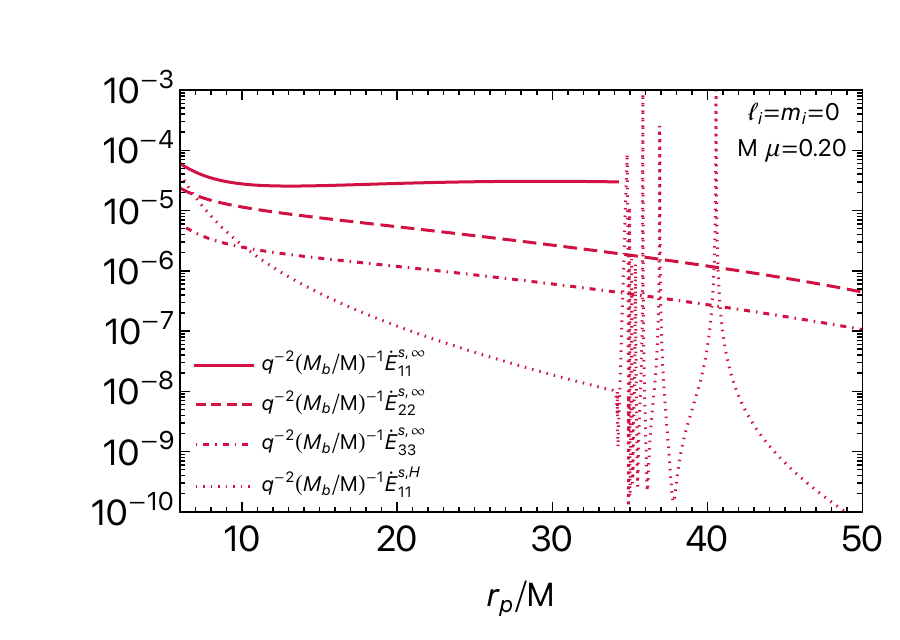}
    \caption{Contribution of the most important multipoles $\{\ell_j,m_j\}$ of the scalar power at infinity $\dot E^{s,\infty}_{\ell_j m_j}$ and at the horizon $\dot E^{s,H}_{\ell_j m_j}$, as a function of the orbital radius $r_p/M$. We consider a spherically symmetric $\ell_i=m_i=0$ background scalar cloud in the fundamental state $n_i=0$, with $M\mu=0.2$. We take orbital radii ranging from $r_p=50M$ up to the ISCO radius $r_p^{\rm ISCO}=6M$.}
    \label{fig:dEsljmj_l0back}
\end{figure}
\begin{figure*}[thb!]
    \centering
    \includegraphics[width=0.48\textwidth]{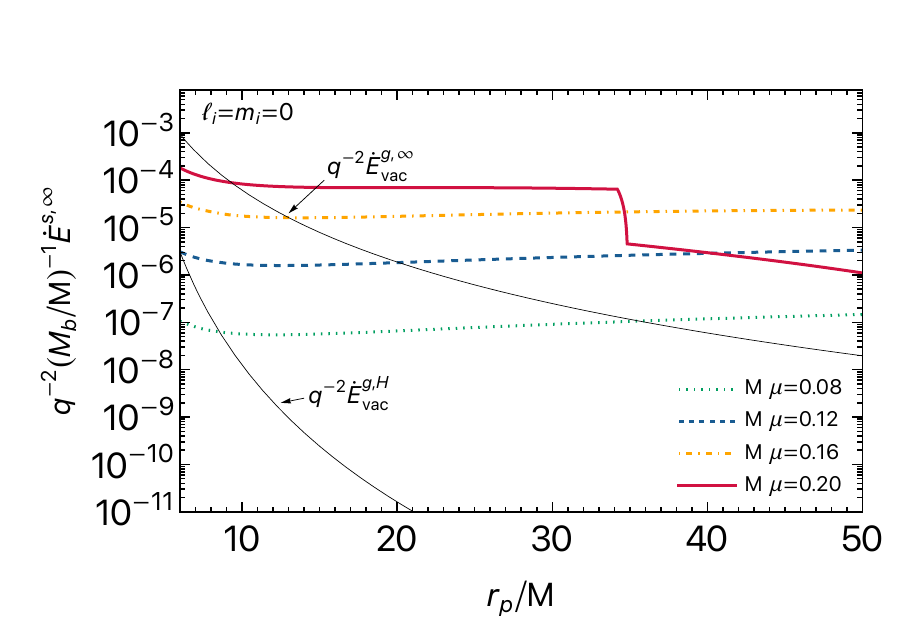}
     \includegraphics[width=0.48\textwidth]{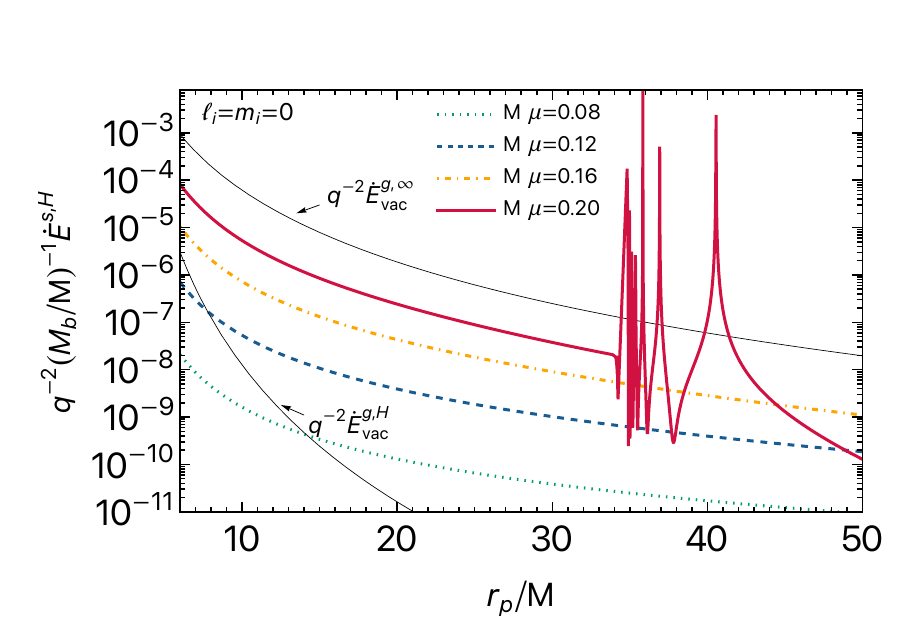}
    \caption{Total scalar power at infinity $\dot E^{s,\infty}$ (left panel) and at the horizon $\dot E^{s,H}$ (right panel) as a function of the orbital radius $r_p/M$ for different values of $M\mu$, when considering a spherically symmetric $\ell_i=m_i=0$ background scalar cloud in the fundamental state $n_i=0$. The solid black lines show the GW flux in the absence of a scalar cloud at infinity and at the horizon, as indicated in the figures.}
    \label{fig:dEssum_l0back}
\end{figure*}

Let us first discuss the case in which the background scalar cloud is spherically symmetric. Our main results are summarized in Figs.~\ref{fig:dEsljmj_l0back} and~\ref{fig:dEssum_l0back}.

Fig.~\ref{fig:dEsljmj_l0back} shows the multipoles $\{\ell_j,m_j\}$ that contribute the most to the scalar power $\dot E^{s}$, for a cloud with $M\mu=0.2$ and considering both the power at infinity and the horizon [cf. Eqs.~\eqref{dotEsinf} and ~\eqref{dotEshor}]. Several features should be highlighted: (i) for small orbital radii, the main contribution to $\dot E^{s}$ comes from the dipolar $\ell_j=m_j=1$ mode, with the power emitted towards infinity contributing the most to the energy loss budget; (ii) for $r_p/M \gtrsim 35$, the $\ell_j=m_j=1$ multipole does not contribute to the scalar power at infinity. This occurs because the terms inside the square roots of Eq.~\eqref{dotEsinf} become negative, i.e. for $r_p/M \gtrsim 35$, $\ell_j=m_j=1$ modes cannot propagate towards infinity because $(\Omega_{\pm}\pm\omega)^2-\mu^2<0$. In particular, for $\ell_j=m_j=1$, we find that $(\Omega_{-}-\omega)^2-\mu^2<0$, all the way up to the innermost stable circular orbit (ISCO) radius, $r_p^{\rm ISCO}=6M$~\footnote{The adiabatic inspiral regime, which we are assuming in our calculations, is only valid up to a transition regime in which the orbit gradually changes from an inspiral to a plunging regime. For quasi-circular orbits this transition regime starts approximately at a radius $r_{\rm trans}/M\approx r_{\rm ISCO}/M+3 q^{2/5} $~\cite{Ori:2000zn}, which can be taken to be the radius at which the inspiral ends.}, whereas $(\Omega_{+}+\omega)^2-\mu^2<0$ only when $r_p/M \gtrsim 35$. For larger values of $m_j$, the same happens but at larger values of $r_p/M$ than what is shown in Figs.~\ref{fig:dEsljmj_l0back} and~\ref{fig:dEssum_l0back}. When summing up all the modes (left panel of Fig.~\ref{fig:dEssum_l0back}), this leads to characteristic sharp features in the power lost towards infinity $\dot E^{s,\infty}$, as first noticed in Refs.~\cite{Baumann:2021fkf,Baumann:2022pkl,Tomaselli:2023ysb}; (iii) the contribution to $\dot E^{s}$ due to absorption of scalar waves at the BH horizon is almost always subdominant, except at specific orbital radii corresponding to orbital frequencies such that eigenmodes of Eq.~\eqref{KG_genericl} are resonantly excited. Specifically, for a cloud with generic quantum numbers $\{n_i,\ell_i,m_i\}$ these resonances should occur at an infinite set of orbital frequencies (that we label with $n_j$) given by:
\begin{equation}\label{resonances}
    \Omega^{n_j}_p=\frac{\Re(\omega_{n_j \ell_j m_j})\mp \Re(\omega_{n_i\ell_i m_i})}{m_j \mp m_i}\,,
\end{equation}
where the minus and plus signs are the resonances due to the term that depends on $Z_+$ and $Z_-$ in the expression for the fluxes [cf. Eq.~\eqref{dotEshor}], respectively, and we recall that $\omega_{n \ell m}$ corresponds to an eigenfrequency of the Klein-Gordon equation with quantum numbers $\{n,\ell,m\}$ (see Sec.~\ref{sec:QBSs}). For the results shown here only the resonances related to $Z_+$ are important, since the ones associated to $Z_-$ occur at frequencies above the ISCO frequency. Using an analogy with the hydrogen atom, these resonances were first discussed in~\cite{Baumann:2018vus,Baumann:2019ztm} in the context of clouds in the Newtonian regime. Analogous resonances were also found in Ref.~\cite{Macedo:2013jja} in the context of EMRIs around boson stars and in~\cite{Cardoso:2011xi,Yunes:2011aa} in the context of scalar-tensor theories of gravity. The widths of the resonances typically scales with $|M\Im(\omega_{n_j\ell_j m_j})|$~\cite{Cardoso:2011xi,Yunes:2011aa,Macedo:2013jja,Cardoso:2022fbq} and are therefore expected to be extremely narrow in the small $M\mu \ll 1$ limit [cf. Eq.~\eqref{omegaI}]. According to Eq.~\eqref{resonances}, the first three $n_j=\{0,1,2\}$ resonances for a cloud in the state $n_i=\ell_i=m_i=0$ should occur at orbital radii $r_p/M\approx\{40.56,36.93,35.84\}$ for the $\ell_j=m_j=1$ scalar power~\footnote{Those numbers were obtained by computing the eigenfrequencies numerically, using a continued-fraction method~\cite{Dolan:2007mj}. However a rough estimate can also be obtained using the analytical approximation~\eqref{omegaR} which is only accurate when $M\mu \ll 1$.}. This is in excellent agreement with what we find in Fig.~\ref{fig:dEsljmj_l0back}. Notice also that in the high $n_j$ limit, one has $\Re(\omega_{n_j\ell_j m_j})\to \mu$ [cf. Eq.~\eqref{omegaR}], so resonances with increasingly larger $n_j$ tend to accumulate close to a given radius (in this case $r_p/M\sim 35$).

The total scalar power when summing over different $\{\ell_j,m_j\}$ modes is shown in Fig.~\ref{fig:dEssum_l0back}, where we show the results for several values of $M\mu$. We consider both the scalar power at infinity (left panel) and at the horizon (right panel) and, for reference, we also compare the results to the (vacuum) GW flux at infinity and at the horizon (solid black lines)~\footnote{All data shown for the GW fluxes were taken from the \texttt{Black Hole Perturbation Toolkit}~\cite{BHPToolkit}.}. When computing the total scalar power we only considered the multipoles shown in Fig.~\ref{fig:dEsljmj_l0back}, including also the corresponding $m_j<0$ modes, which can be obtained using $\dot E^{s}_{\ell_j -m_j}=\dot E^{s}_{\ell_j m_j}$ [see discussion below Eq.~\eqref{dotLshor}]. We checked this to be a good approximation since higher modes tend to be further suppressed. In particular for the scalar power at the horizon we find that, for the range of orbital radii we consider, higher multipoles are several orders of magnitude smaller than the dominant $\ell_j=1$ mode. Therefore we did not include them here, since they are harder to compute accurately. In general, the scalar power increases with $M\mu$, which is to be expected given that the cloud becomes more compact as $M\mu$ increases. One can see that, at large orbital radii and for sufficiently large values of $M_b/M$, the total power lost by the secondary due to the presence of the scalar field can dominate over GW emission.
\subsubsection{Results for $\ell_i=m_i=1$}
\begin{figure*}[thb!]
    \includegraphics[width=0.48\textwidth]{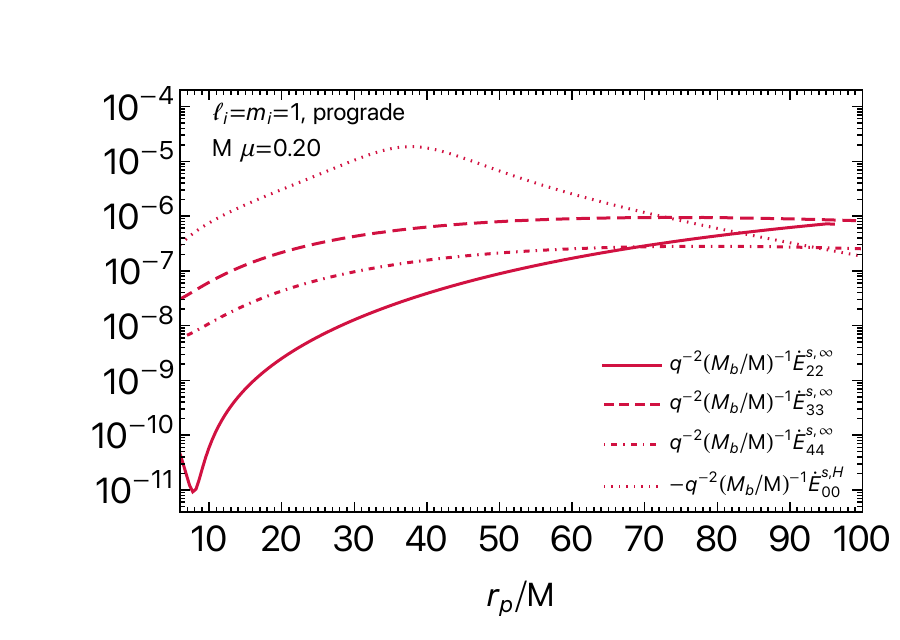}
    \includegraphics[width=0.48\textwidth]{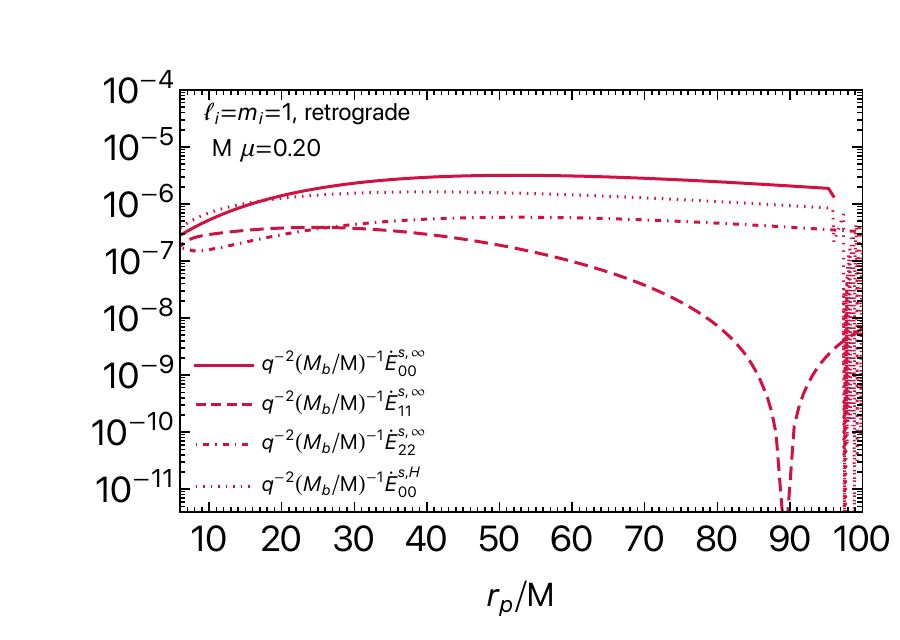}
    \caption{Same as Fig.~\ref{fig:dEsljmj_l0back}, but for a dipolar $\ell_i=m_i=1$ background scalar cloud. Both prograde (left panel) and retrograde (right panel) orbits are shown. The range of orbital radii is here extended up to $r_p/M=100$. In the right panel, the ``noise'' seen when $r_p/M \gtrsim 97$ in the multipole $\dot E^{s,H}_{00}$ is due to the excitation of many close-by resonances, which are hard to resolve numerically.}
    \label{fig:dEsljmj_l1back}
\end{figure*}

\begin{figure}[thb!]
    \includegraphics[width=0.48\textwidth]{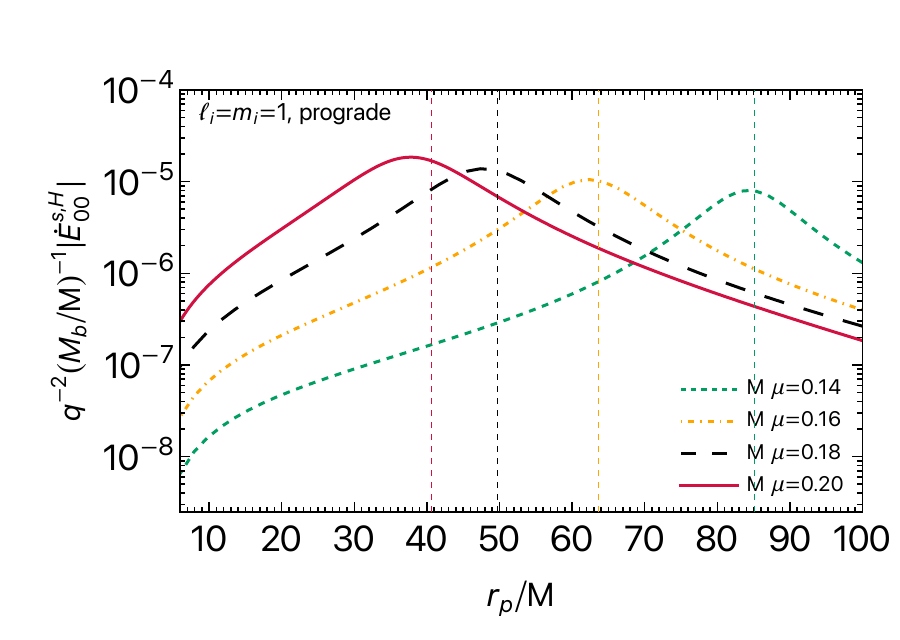}
    \caption{Contribution of the mode $\ell_j=m_j=0$ to the scalar power at the horizon, for a dipolar $\ell_i=m_i=1$ background scalar cloud, considering prograde orbits and different values of $M\mu$. The vertical dashed lines correspond to the expected orbital radius where a resonance with the fundamental mode $\ell_j=m_j=0$ should occur, according to Eq.~\eqref{resonances}.}
    \label{fig:dEhor00_res}
\end{figure}

\begin{figure*}[thb!]
    \centering
    \includegraphics[width=0.48\textwidth]{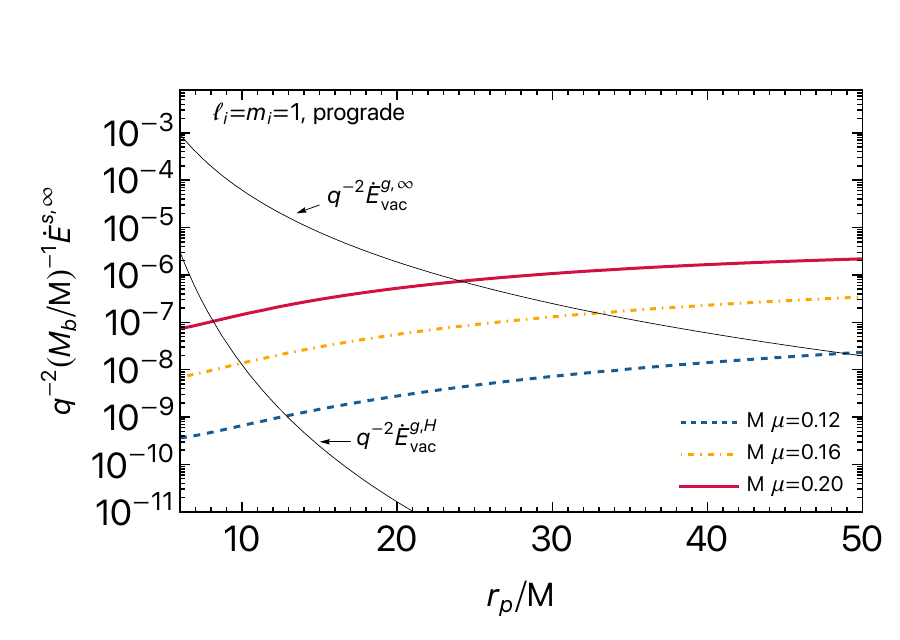}
     \includegraphics[width=0.48\textwidth]{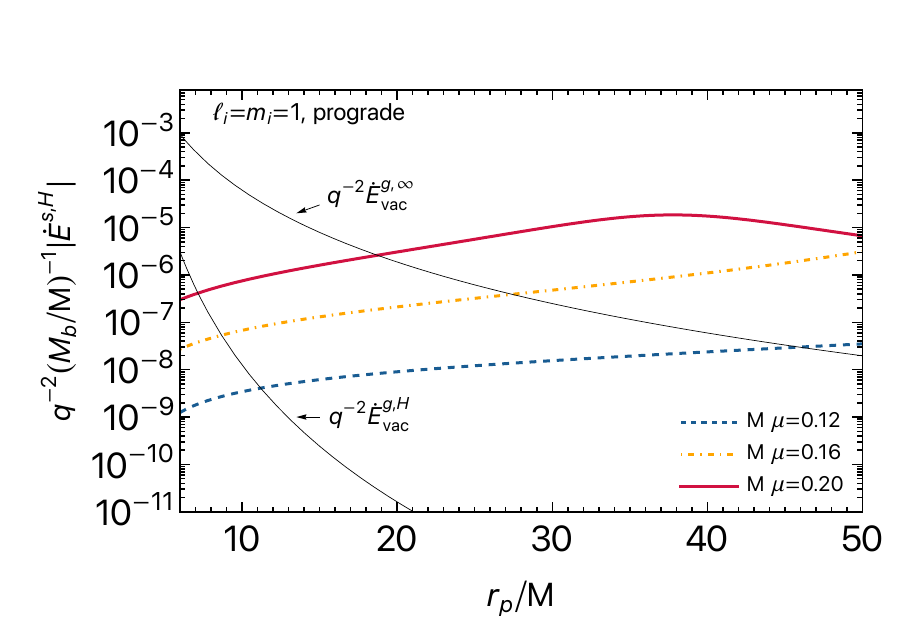}
     \includegraphics[width=0.48\textwidth]{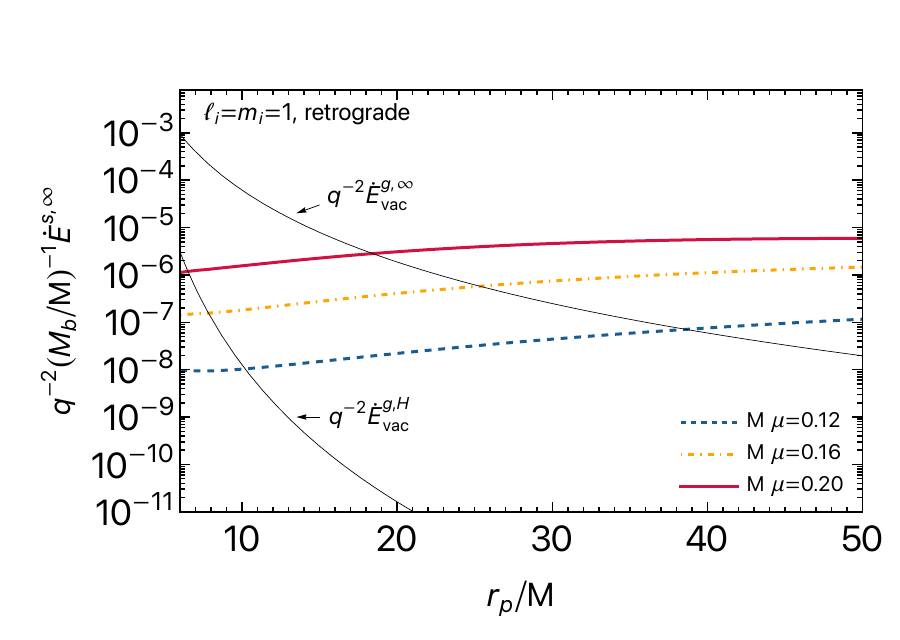}
     \includegraphics[width=0.48\textwidth]{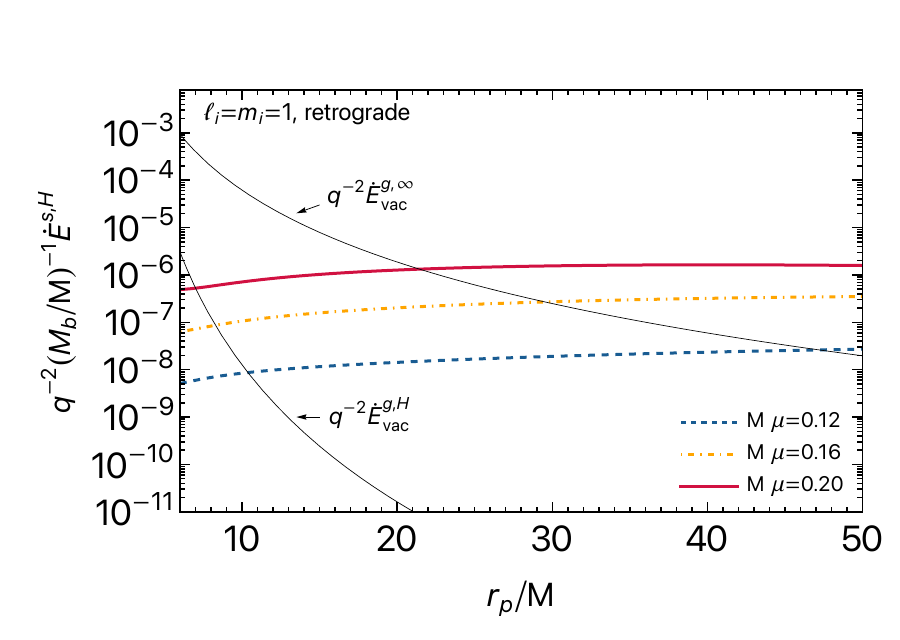}
    \caption{Same as Fig.~\ref{fig:dEssum_l0back}, but for a dipolar $\ell_i=m_i=1$ background scalar cloud. The top panels show results for prograde orbits, whereas in the bottom panels we show results for retrograde orbits. In order to obtain the total scalar power we summed over the modes shown in Fig.~\ref{fig:dEsljmj_l1back} (including also the $m_j<0$ modes) for prograde orbits, while for retrograde orbits we also added the $\ell_j=|m_j|=3$ and $\ell_j=|m_j|=4$ modes.}
    \label{fig:dEssum_l1back}
\end{figure*}

Let us now turn to the case of a dipolar background scalar cloud with $\ell_i=m_i=1$ and $n_i=0$. This case is particularly interesting since in a Kerr BH background, such clouds can form through the  superradiant instability (see Sec.~\ref{sec:QBSs}). The results are summarized in Figs.~\ref{fig:dEsljmj_l1back},~\ref{fig:dEhor00_res} and~\ref{fig:dEssum_l1back}, where we show results analogous to the ones we discussed above. There are however some important differences with respect to the case of a spherical cloud that are worth highlighting.

First of all, as already anticipated, the scalar power differs between prograde and retrograde orbits. This is expected given that a $\ell_i=m_i=1$ cloud has a non-zero angular momentum, which breaks the symmetry between prograde and retrograde orbits. The multipoles that contribute the most to the scalar power also depend on the direction of the orbit, as can be seen in Fig.~\ref{fig:dEsljmj_l1back} for the case where $M\mu=0.2$. In this example, for the case of prograde orbits, multipoles $\ell_j=0$ and $\ell_j=1$ do not contribute to the scalar power at infinity, $\dot E^{s,\infty}$, because: (i) the $\ell_j=m_j=0$ does not contribute given that $\left(\Omega_{\pm} \pm\omega\right)^2-\mu^2<0$ for all radii larger than the ISCO frequency; (ii) a similar situation arises for the $\ell_j=m_j=1$ ($\ell_j=1, m_j=-1$) multipole. For this multipole, the first (second) term on the right-hand side of Eq.~\eqref{dotEsinf} vanishes, i.e. $\Omega_+^{m_j=1}=0$ ($\Omega_-^{m_j=-1}=0$), whereas the second (first) term also vanishes because $\left(\Omega_- -\omega\right)^2-\mu^2<0$ [$\left(\Omega_+ +\omega\right)^2-\mu^2<0$] for all radii larger than the ISCO radius~\footnote{For sufficiently small $M\mu$ this last condition is no longer true for $r_p$ smaller than a given threshold. Therefore for sufficiently small $M\mu$ we do expect to have contributions from $\ell_j=0, 1$ close to the ISCO.}. These arguments do not apply in the case of retrograde orbits, at least for most of the orbital radii we show in Fig.~\ref{fig:dEsljmj_l1back} and therefore the multipoles $\ell_j=0,1$ do contribute to the scalar power for retrogade orbits~\footnote{In the example shown in Fig.~\ref{fig:dEsljmj_l1back}, the mode $\ell_j=m_j=0$ stops contributing to the power at infinity only when $r_p\gtrsim 97$.}.

The multipole that dominates the overall budget for the scalar power at infinity depends in general on the orbital radius, as well as the direction of the orbit. In the example shown in Fig.~\ref{fig:dEsljmj_l1back}, for prograde orbits the $\ell_j=m_j=3$ multipole is the most important one, whereas for retrograde orbits the $\ell_j=m_j=0$ multipole dominates. While we do find similar trends for other values of $M\mu$, it is unclear whether this trend remains for smaller values of $M\mu$ than the ones we considered here. We should also make the remark that, in the case of retrograde orbits, for $M\mu=0.2$ we see a minimum at $r_p/M\sim 89$ for the $\dot E^{s,\infty}_{11}$ multipole, which does not seem to be related to any obvious properties of the system. We have checked that this feature remains when increasing the numerical precision of our code, therefore we highly believe it to be a physical feature. 

Finally, perhaps the most important distinctive feature we find, is that the scalar power at the horizon, which is largely dominated by the $\ell_j=m_j=0$ multipole for both prograde and retrograde orbits, can dominate over the scalar power at infinity, especially for prograde orbits, as can be clearly seen in Fig~\ref{fig:dEsljmj_l1back}. Interestingly, for prograde orbits and the values of $M\mu$ that we considered, $\dot E_{00}^{s,H}$ is always negative [cf. Eq.~\eqref{dotEshor}], that is, the particle gains energy due to the presence of the term $\dot E_{00}^{s,H}$ in this case. These results therefore indicate that, for $M_b/M$ above a certain threshold, the total energy loss budget can vanish at certain orbital radii, $\dot E=0$ [cf. Eq.~\eqref{totaldE}], indicating the possible presence of floating orbits for prograde orbits~\footnote{Notice that the scalar radiation flux at the horizon, as given by Eq.~\eqref{dEphi_hor}, is always positive. Therefore the energy that sustains the floating orbit comes from the energy lost by the cloud, Eqs.~\eqref{dMbdJb} and~\eqref{dQ_hor}.}.

In the case of prograde orbits, the peak of $\dot E_{00}^{s,H}$ at $r_p/M \sim 40$ seen in Fig.~\ref{fig:dEsljmj_l1back} is in agreement with the expected location of a resonance with the mode $\{n_j=0,\ell_j=0,m_j=0\}$ which is the only mode that can be excited within the range of orbital radii shown here, according to Eq.~\eqref{resonances}. To give further support to the claim that the peak we see in $\dot E_{00}^{s,H}$ is related to a resonance, in Fig.~\ref{fig:dEhor00_res} we show $\dot E_{00}^{s,H}$ for different values of $M\mu$ and also show the expected orbital radius where a resonance should occur, according to Eq.~\eqref{resonances} (vertical dashed lines). As can be seen, $\dot E_{00}^{s,H}$ always peaks close to a resonance. However, we do note that the width is much larger than the resonances we found in Fig.~\ref{fig:dEsljmj_l0back}. A possible explanation for this behavior is that, for the values of $M\mu$ we considered in Fig.~\ref{fig:dEhor00_res}, one has $|M\Im(\omega_{000})|$ ranging from $\sim 10^{-4}$ (for $M\mu=0.14$) up to $\sim 10^{-3}$ (for $M\mu=0.2$) -- compared to $|M\Im(\omega_{011})|\sim 10^{-8}$ for the resonance with the largest width in Fig.~\ref{fig:dEsljmj_l0back}). Therefore for the cases considered here, $|M\Im(\omega_{000})|$ might be too large to see a clear distinction between the ``on-'' and ``off-resonance'' behavior, which might explain what we observe. For smaller $M\mu$ than what we considered in Fig.~\ref{fig:dEhor00_res}, one should expect narrower resonances, occurring at larger orbital radii. However, with our current code, we found it challenging to compute the scalar power accurately for much smaller values of $M\mu$ and at larger radii, and therefore we do not explore those cases here.

On the other hand, for retrograde orbits, Fig.~\ref{fig:dEsljmj_l1back} also shows that resonances arise in $\dot E_{00}^{s,H}$ for $r_p/M\gtrsim 97$~\footnote{Notice that for retrograde orbits we can still use Eq.~\eqref{resonances} to compute the location of the resonances, but in that case one should remember that $\Omega^{\rm retro}_p =-\sqrt{M/r_p^3}$, therefore the allowed set of resonances is different for prograde and retrograde orbits, as noticed in Ref.~\cite{Baumann:2018vus}.}. Those correspond to resonances with modes that have a high overtone number $n_j$. Notice that, just like in the case we saw for spherical clouds, for large $n_j$ one should get many close-by narrow resonances, therefore one needs to use a very high resolution when computing the scalar power close to these resonances, therefore we did not attempt to fully resolve them. Hence the $\dot E_{00}^{s,H}$ curve in Fig.~\ref{fig:dEsljmj_l1back} for $r_p/M\gtrsim 97$ can only be trusted at a qualitative level. 

We should emphasize that the possibility to excite a $\ell_j=m_j=0$ multipole was missed in previous works~\cite{Baumann:2018vus,Baumann:2019ztm}. According to the selection rules discussed in Sec.~\ref{sec:KGperts}, $\ell_j=m_j=0$ scalar perturbations are excited by $l=1,m=\pm 1$ metric perturbations which were not taken into account in Refs.~\cite{Baumann:2018vus,Baumann:2019ztm}~\footnote{This was corrected in more recent work~\cite{Tomaselli:2023ysb}, which uses the same Newtonian formalism of Refs.~\cite{Baumann:2018vus,Baumann:2019ztm}. However in that work they did not compute the resonant transitions related to the $\ell_j=m_j=0$ mode, as far as we are aware.}. As we emphasized in Sec.~\ref{sec:GRperts}, dipolar metric perturbations cannot be entirely gauged away in the presence of the 
point particle. Our results show that, due to the excitation of $\ell_j=m_j=0$ scalar perturbations, the scalar power at the horizon can dominate the overall energy loss/gain budget in a given range of orbital radii, especially for prograde orbits. This is a trend we seem to find for other values of $M\mu$, as can be seen in Fig.~\ref{fig:dEssum_l1back} where we show the total scalar power at infinity (left panel) and at the horizon (right panel) for different values of $M\mu$. 

Overall,one can see from these results that $\dot E^{s}$ is in general more important at larger orbital radii. This feature is in agreement with Refs.~\cite{Zhang:2019eid,Baumann:2021fkf,Tomaselli:2023ysb} and can be understood from the profile of dipolar scalar clouds which peak at a radius $1/(M\mu^2)$ (see e.g. Ref.~\cite{Brito:2014wla}). One can also see that $\dot E^{s,\infty}$ is in general slightly larger for retrograde orbits when compared to the prograde case, also in agreement with Refs.~\cite{Baumann:2021fkf,Tomaselli:2023ysb}.

Finally, we note that in order to fully understand the impact of the scalar power on the orbit and on the emitted gravitational waveform requires a self-consistent evolution of the orbit that also includes the evolution of the cloud's mass and angular momentum, according to the discussion presented in the previous subsection. It is likely that this requires starting the evolution at orbital radii much larger than what we considered here, given that one needs to understand how the cloud's mass and angular momentum evolved since the early formation of the binary. For example, it is not entirely clear what would be appropriate values for the cloud's initial mass when an EMRI enters the LISA band. Studies of the orbital evolution were explored in the Newtonian regime in e.g. Refs.~\cite{Baumann:2019ztm,Baumann:2021fkf,Tomaselli:2023ysb,Takahashi:2023flk} which showed that in some cases the effect of the cloud can significantly alter the orbital evolution, however, more accurate results would require joining our results with the Newtonian approximations employed in Refs.~\cite{Baumann:2018vus,Baumann:2019ztm,Baumann:2021fkf,Tomaselli:2023ysb,Takahashi:2023flk}. We will not attempt to do this in this work, since we believe this issue requires its own dedicated study. We therefore leave a detailed analysis of this problem for future work.
%
\section{Leading-order corrections to the black hole metric and the gravitational-wave flux}\label{sec:Oq1epsilon2}
Having laid down the foundation needed to compute the leading $\mathcal{O}(\epsilon)$ perturbations, the next step is to consider $\mathcal{O}(\epsilon^2)$ perturbations. At that order, one needs to compute corrections  of order $\mathcal{O}(q^0,\epsilon^2)$, which can be interpreted as modifications to the background metric due to the self-gravity of the cloud [cf. Eq.~\eqref{linearEFEs_20}], whereas at order $\mathcal{O}(q^1,\epsilon^2)$ there will be corrections to the GW flux [cf. Eq.~\eqref{EFEs_21}]. In order to fully compute all corrections up to order $\mathcal{O}(q^1,\epsilon^2)$ one needs to compute both of these corrections. As a proof of concept, here we will consider only part of the problem, leaving a more detailed study for future work. We will only consider the simpler case of perturbations around a spherically symmetric cloud, for which $\mathcal{O}(q^0,\epsilon^2)$ corrections can be computed following the procedure in Refs.~\cite{Babichev:2012sg,Bamber:2021knr,DeLuca:2021ite,Kimura:2021dsa}. Then, we compute corrections to the axial metric perturbations which, for a spherically symmetric cloud, completely decouple from the $\mathcal{O}(q^1,\epsilon^1)$ scalar perturbations that we computed in the previous section, greatly simplifying the problem.
\subsection{Corrections to the black hole metric due to a spherical scalar cloud}
Let us first compute the corrections to the BH metric at order $\mathcal{O}(q^0,\epsilon^2)$ due to a spherically symmetric scalar cloud. This problem was considered in Refs.~\cite{Bamber:2021knr,DeLuca:2021ite}, where it was shown that in the standard Schwarzschild-like coordinate system singularities appear at the horizon when considering the backreaction of the scalar field on the metric, due to the slow accretion of the scalar field by the BH. This problem can be circumvented~\cite{Bamber:2021knr,DeLuca:2021ite} by using ingoing Eddington-Finkelstein coordinates $(v,r,\theta,\phi)$. In this coordinate system, generic spherically symmetric perturbations of the Schwarzschild metric can be written as~\cite{Bamber:2021knr,DeLuca:2021ite}
\begin{equation}\label{EF_perts}
    ds^{2} = -F(v,r) e^{2\epsilon^2\delta\lambda(v,r)}dv^{2} + 2e^{\epsilon^2\delta\lambda(v,r)}dvdr + r^{2}d\Omega^2\,,
\end{equation}
where
\begin{equation}
 F(v,r) = f(r) - \frac{2\epsilon^2 \delta M (v,r)}{r}\,,\quad f(r)=1-\frac{2M}{r}\,.
\end{equation}
By inserting this metric in Einstein's field equations and expanding up to order $\mathcal{O}(\epsilon^2)$, one finds that the metric perturbations $\delta M$ and $ \delta \lambda$ satisfy the following differential equations:
\begin{eqnarray}
     \partial_{r} \delta M &=& - 4 \pi r^{2}  T^{v}_{v} \equiv 4 \pi r^{2} \rho_{\rm EF}\,,\label{dMr}\\
     \partial_{v} \delta M &=& 4 \pi r^{2} T^{r}_{v} \equiv \delta A(v,r)\,,\label{dMv} \\
    \partial_{r} \delta \lambda &=& 4 \pi r T_{rr} = 2 r  e^{2 \Im(\omega) v}|\tilde{R}|^2\,,\label{dlr}
\end{eqnarray}
where we defined $\rho_{\rm EF}$ to be the energy density measured by coordinate observers in ingoing Eddington-Finkelstein:
\begin{equation}\label{rhoEF}
    \rho_{\rm EF} = -T^{v}_{v} = \frac{e^{2 \Im(\omega) v}}{4\pi}\left(f |\partial_{r}\tilde R|^2 + \mu^{2} |\tilde R|^2\right)\,,
\end{equation}
whereas the function $\delta A$ is given by
\begin{equation}
\delta A =  e^{2 \Im(\omega) v}r^2 \left[2|\omega|^2|\tilde{R}|^2-2 f \Im\left(\omega \tilde{R}^* \partial_r \tilde{R}\right) \right]\,.
\end{equation}
Differentiating Eq.~\eqref{dMr} with respect to $v$ and Eq.~\eqref{dMv} with respect to $r$ and equating the two resulting equations we find the following relation between $\delta  A$ and $\rho_{\rm EF}$:
\begin{equation}
   \partial_r\delta  A  = 4\pi r^2 \partial_v\rho_{\rm EF}\,.
\end{equation}
Integrating and using Eq.~\eqref{rhoEF} we therefore have
\begin{equation}\label{dA}
\delta  A = 8\pi\Im(\omega)\int_{2M}^r \rho_{\rm EF}(v,r') r'^2 dr'\,.
\end{equation}
On the other hand, upon integrating Eq.~\eqref{dMr} we get that
\begin{equation}
\delta  M(v,r) =4\pi\int_{2M}^r \rho_{\rm EF}(v,r') r'^2 dr'\,.
\end{equation}
We therefore find that Eq.~\eqref{dMv} can be rewritten as 
\begin{equation}
\partial_{v} \delta M = 2\Im(\omega)\delta  M(v,r)\,.\label{dMv2}
\end{equation}
The function $\epsilon^2\delta M(v,r)$ gives us the mass contained in the scalar cloud on a radius $r$ and at given (advanced) time $v$. In particular, the total mass in the cloud can be defined as $M_b := \lim_{r\to \infty}\epsilon^2\delta M$. Therefore Eq.~\eqref{dMv2} tells us that that the cloud decays with a rate $2\Im(\omega)$. As already mentioned, for quasi-bound states in a Schwarzschild background one typically has $M\Im(\omega)\ll 1$. Therefore, as we did in the previous section, for the purpose of computing the GW fluxes we will take the approximation $\partial_{v} \delta M \approx 0$ and $\partial_{v} \delta\lambda \approx 0$, such that $\delta M$ and $\delta \lambda$ are constant in time and $\rho_{\rm EF}\approx \rho_{\rm EF}(0,r)$.

In summary, in order to find the perturbations to the metric at order $\mathcal{O}(q^0,\epsilon^2)$ for a $\ell_i=0$ scalar bound state, we first obtain a solution for the radial function $\tilde R(r)$ and then solve the following differential equations:
\begin{eqnarray}
    \partial_r\delta M &\approx& r^{2}\left[f |\partial_{r}\tilde R|^2 + \mu^{2} |\tilde R(r)|^2\right]\,,\label{dMr_final}\\
    \partial_r \delta \lambda &\approx& 2 r |\tilde R(r)|^2\,.\label{dlambdar_final}
\end{eqnarray}
In order to solve these equations we impose that $\delta M(2M)=0$ and that the metric asymptotically approaches the Minkowski metric in the form $ds^{2} = -dv^{2} + 2 dv dr + r^{2} d\Omega^2$, at spatial infinity. We therefore require $\lim_{r\to\infty} \delta\lambda=0$~\footnote{The function $\delta\lambda$ is defined up to an arbitrary function of $v$~\cite{Kimura:2021dsa}, which can always be chosen such that $\lim_{r\to\infty} \delta\lambda=0$.}.
\begin{figure*}[thb!]
    \centering
    \includegraphics[width=0.48\textwidth]{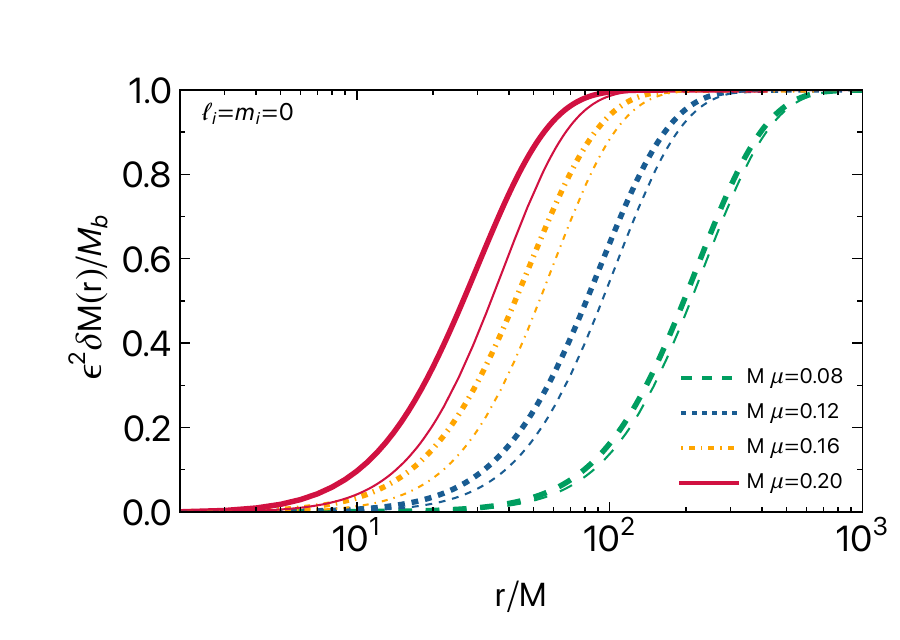}
    \includegraphics[width=0.48\textwidth]{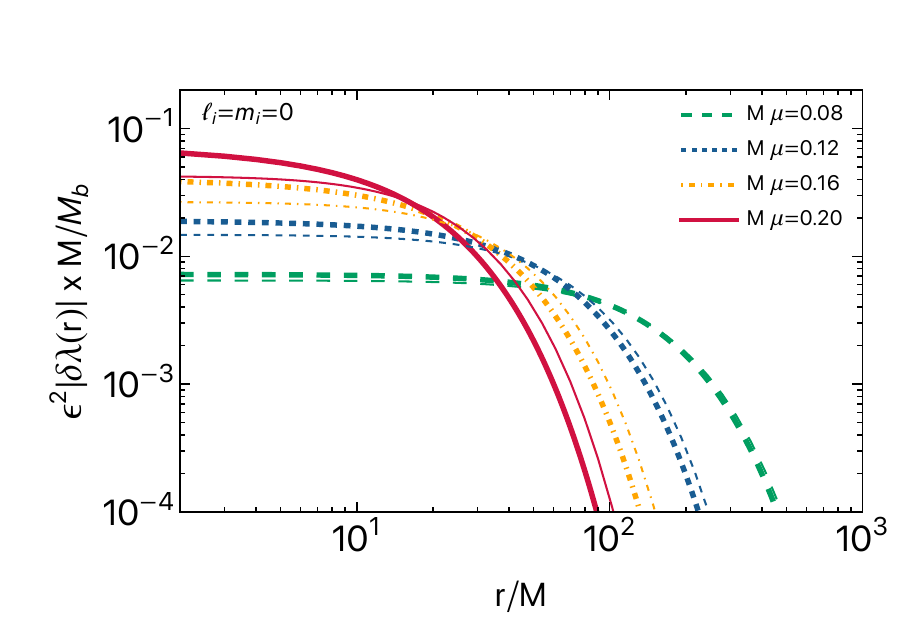}
    \caption{Solutions for the metric perturbations $\delta M$ (left panel) and $\delta\lambda$ (right panel), as a function of the radial coordinate for different values of $M\mu$ and assuming a spherically symmetric scalar cloud in the fundamental state, i.e. $\ell_i=m_i=n_i=0$. Notice that $\delta\lambda<0$, so in order to use a log scale in the y-axis of the right panel we plot $|\delta\lambda|$. Thick lines are obtained by computing the scalar field eigenstates numerically, whereas thin lines show the results when using the hydrogenic approximation $R_{00}\propto E^{-\mu^2 M r}$ for the eigenstates.}
    \label{fig:deltaMlambda}
\end{figure*}

Examples of solutions for different values of $M\mu$, obtained by numerically solving Eqs.~\eqref{dMr_final} and~\eqref{dlambdar_final}, are shown in Fig.~\ref{fig:deltaMlambda}. When $r \to \infty$, the spacetime approaches a Schwarzschild metric with total mass $M+M_b$, where the main effect of the constant $M\mu$ is to make the cloud more compact as $M\mu$ increases. In particular, from the curves in the left panel of~\eqref{dMr_final}, one can check that $\epsilon^2\delta M(r_c)/M_b \sim 0.999$ when $r_c\sim 5/(M\mu^2)$, which we can use as a definition of the typical size of the cloud. On the other hand, close to the the BH horizon, $\delta\lambda$ approaches a negative constant value $\delta\lambda(r\to 2M)\sim \delta\lambda_H<0$. Using a non-relativistic approximation for the scalar eigenstates, $R_{00}\propto E^{-\mu^2 M r}$ (see e.g.~\cite{DeLuca:2021ite}) we find $\epsilon^2\delta\lambda_H\approx -\mu^2 M M_b$, which is in good agreement with our numerical results in the limit $M\mu \ll 1$ (see thin lines in Fig.~\ref{fig:deltaMlambda}). Therefore, in the region close to the horizon, the difference with a vacuum Schwarzschild geometry can be essentially interpreted in terms of a constant redshift factor $e^{2\epsilon^2\delta\lambda_H}\approx 1-2\mu^2 M M_b$, where we notice that the factor $\mu^2 M M_b$ can be thought as a measure of the compactness of the cloud given that the cloud's size is set by the scale $1/(M\mu^2)$. Notice that this is in agreement to what was found in Refs.~\cite{Cardoso:2021wlq,Cardoso:2022whc,Figueiredo:2023gas} where different types of dark matter profiles around BHs were studied.

\subsubsection{Time-like circular geodesics}
Let us now assume a point particle moving along circular geodesics of the spacetime given by Eq.~\eqref{EF_perts}. Following the procedure outlined in App.~\ref{app:app_geo}, one can compute the leading-order corrections to the particle's energy $E$, angular momentum $L$ and orbital angular frequency $\Omega_p$, which can be written as~\footnote{Given that we are considering a spherical cloud, for this section we can take $\Omega_p>0$ without loss of generality.}:
\begin{eqnarray}
\frac{E}{m_p} &=& \frac{r_p-2M}{\sqrt{r_p(r_p-3M)}}+\epsilon^2\delta E(r_p)\,,\label{E_epsilon2}\\
\frac{L}{m_p} &=& \sqrt{\frac{M r_p^2}{r_p-3M}}+ \epsilon^2\delta L(r_p)\,,\label{L_epsilon2}\\
\Omega_p &=&  \sqrt{\frac{M}{r_p^3}}+\epsilon^2\delta \Omega(r_p)\,,\label{Omega_epsilon2}
\end{eqnarray}
where $\delta E(r_p)$, $\delta L(r_p)$ and $\delta \Omega(r_p)$ are explicitly given in terms of the perturbation functions $\delta M$ and $\delta\lambda$ in App.~\ref{app:app_geo} [cf. Eqs.~\eqref{deltaE}--~\eqref{deltaOmega}]. In vacuum, i.e. in the absence of the cloud, only the order $\mathcal{O}(\epsilon^0)$ term contributes and one recovers standard results. For concreteness, below we use the subscript or superscript  ``vac'' to refer to order $\mathcal{O}(\epsilon^0)$ quantities. 

\begin{figure}[thb!]
    \includegraphics[width=0.45\textwidth]{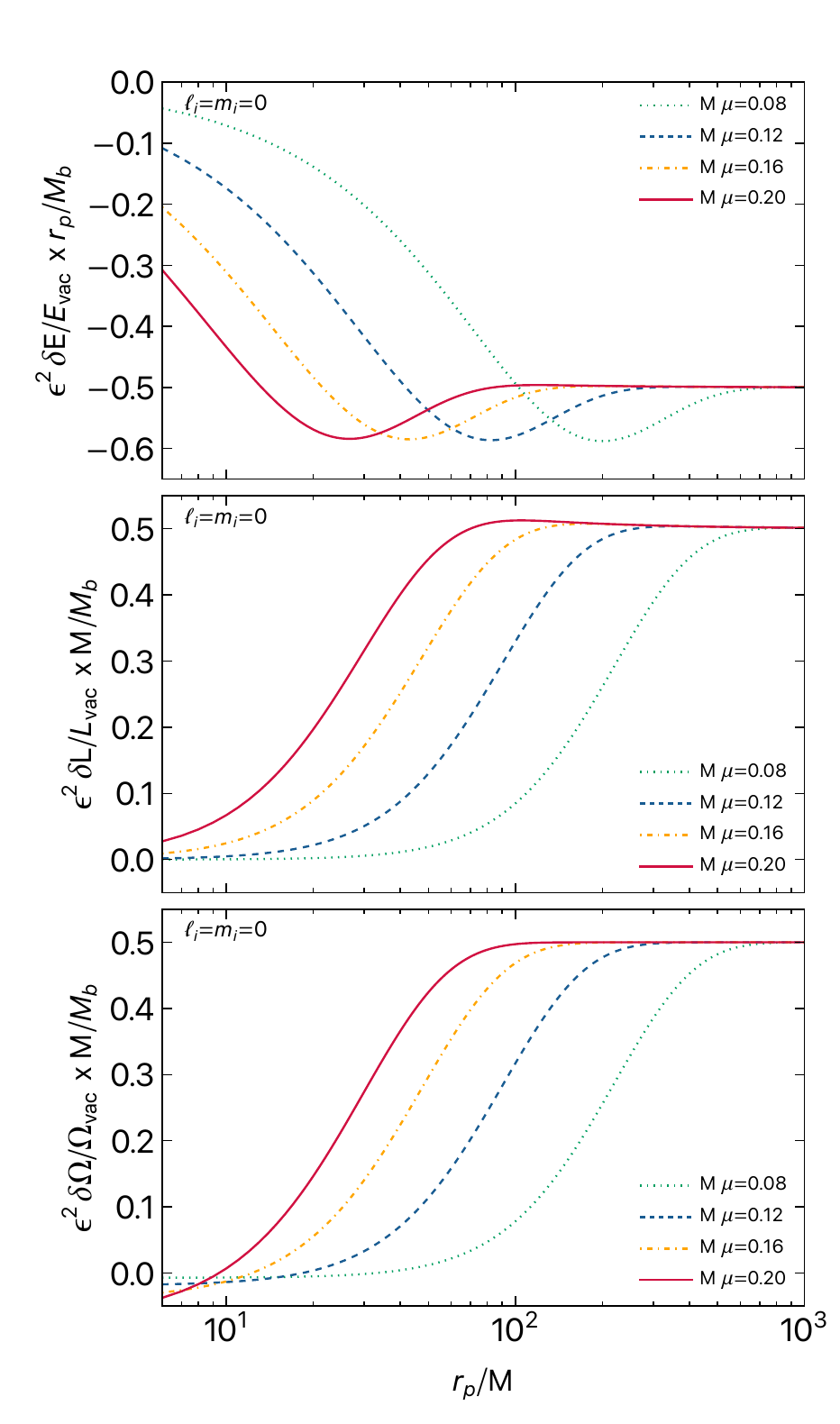}
    \caption{Leading-order corrections to the energy (top panel), angular momentum (middle panel) and orbital angular frequency (bottom panel) of a point particle moving along circular geodesics of the spacetime~\eqref{EF_perts}, when considering a spherically symmetric scalar cloud. We show the corrections as a function of the particle's orbital radius $r_p$ and consider different values of the constant $M\mu$. The normalization shown in the plots was chosen in such a way as to make the asymptotic behavior when $r_p \gg M$ clear (see main text).}
    \label{fig:deltaOMegaE}
\end{figure}

In Fig.~\ref{fig:deltaOMegaE} we compare the leading-order corrections to the particle's energy (top panel), angular momentum (middle panel) and orbital angular frequency (bottom panel) against the quantities computed in vacuum, for different values of $M\mu$. As one can see from the plots, we have that $\lim_{r_p\to\infty} |\epsilon^2 r_p\delta E/E_{\rm vac}|= \epsilon^2 M \delta L/L_{\rm vac}= \epsilon^2 M \delta \Omega/\Omega_{p,{\rm vac}}=M_b/2$. This is consistent with the fact that for $r_p \gg M$ the metric~\eqref{EF_perts} reduces to a Schwarzschild spacetime but with total mass given by $M+M_b$~\footnote{At large distances, the orbital frequency in such a spacetime should be given by $\Omega_p \approx \sqrt{(M+M_b)/r_p^3}\approx \sqrt{M/r_p^3}[1+M_b/(2M)]$ for $M_b\ll M$, consistent with what we find. Similar expressions can be obtained for $E$ and $L$ replacing $M \to M+M_b$ in the leading-order expressions of Eqs.~\eqref{E_epsilon2} and~\eqref{L_epsilon2}.}. 

\begin{figure}[thb!]
    \centering
    \includegraphics[width=0.48\textwidth]{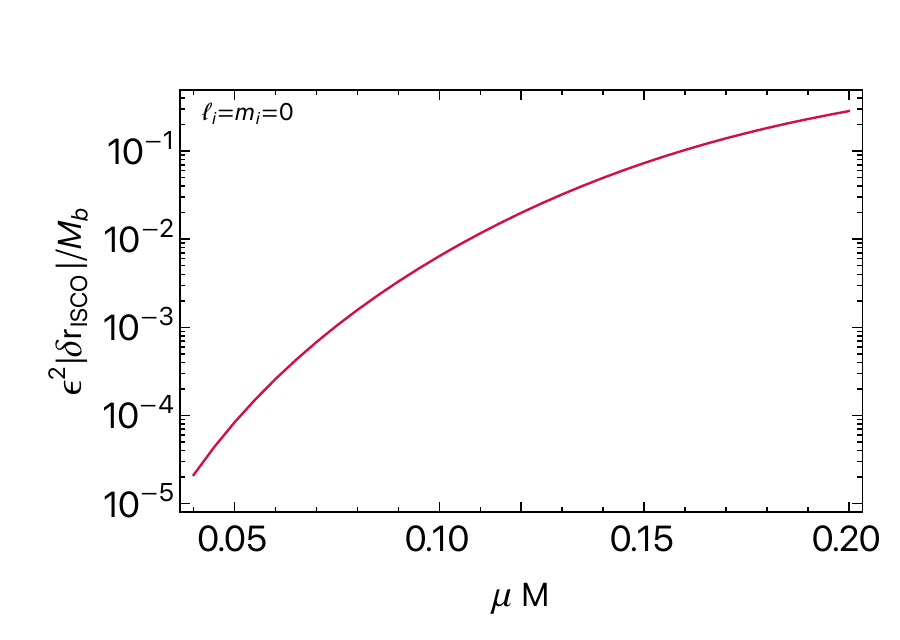}
    \caption{Leading-order corrections to the ISCO radius as a function of $M\mu$. Note that $\delta r_{\rm ISCO}<0$, therefore the absolute value of $\delta r_{\rm ISCO}$ is shown in order to use a log scale in the y-axis.}
    \label{fig:deltarISCO}
\end{figure}

Stable circular orbits exist for $r>r_{\rm ISCO}$, where $r_{\rm ISCO}$ defines the ISCO radius. As shown in App.~\ref{app:app_geo}, $r_{\rm ISCO}$ can be written as 
\begin{equation}\label{rISCO}
r_{\rm ISCO} = 6M +\epsilon^2\delta r_{\rm ISCO}\,,
\end{equation}
where the explicit form of $\delta r_{\rm ISCO}$ can be found in Eq.~\eqref{eq:deltarISCO}. The dependence of $\delta r_{\rm ISCO}$ with $M\mu$ is shown in Fig.~\ref{fig:deltarISCO}. Typically, we find $\delta r_{\rm ISCO}<0$, that is, the ISCO is located at a slightly smaller radius in the presence of the cloud, when compared to the vacuum case. We also find that $|\delta r_{\rm ISCO}|/M_b$ monotonically increases with $M\mu$, which can be understood from the fact that for larger $M\mu$ the cloud becomes more compact, leading to larger corrections to the spacetime's geometry in the BH's vicinity.
Finally, let us also notice that if one computes the orbital angular frequency at $r_{\rm ISCO}$ using Eq.~\eqref{Omega_epsilon2} and one uses the approximation $\delta M(r_{\rm ISCO})\sim 0$ and $\epsilon^2\delta \lambda(r_{\rm ISCO})\sim -\mu^2 M M_b$ (see discussion above and Fig.~\ref{fig:deltaMlambda}), we get $M\Omega_{\rm ISCO}\approx 1/(6\sqrt{6})\left(1 - M\mu^2 M_b\right)$, which is in good agreement with what we find numerically when $M\mu\ll 1$. The same arguments also show that the energy $E$ close to the ISCO is redshifted by the same factor $\left(1 - M\mu^2 M_b\right)$. This redshift factor is exactly the same as the one computed in Refs.~\cite{Cardoso:2021wlq,Cardoso:2022whc,Figueiredo:2023gas} if one identifies their $a_0$ with our $1/(M\mu^2)$. Finally, within this approximation Eq.~\eqref{deltaL} predicts $\delta L\sim 0$ close to the ISCO, again in good agreement with Fig.~\ref{fig:deltaOMegaE} when $M\mu\ll 1$.

\subsection{Corrections to axial metric perturbations due to a spherical scalar cloud}
Having computed the corrections to the background metric and to the circular orbits due to a spherical scalar cloud, we now consider the leading-order corrections to the GW flux induced by the motion of the point particle. In Sec.~\ref{sec:GRperts} we discussed how to compute gravitational perturbations at order $\mathcal{O}(\epsilon^0,q^1)$. We now wish to consider such perturbations up to $\mathcal{O}(\epsilon^2,q^1)$ focusing on axial gravitational perturbations which are much simpler to deal with. In fact, since the scalar cloud and the spacetime~\eqref{EF_perts} are spherically symmetric, one can verify that scalar perturbations only couple to polar gravitational perturbations (see e.g. Refs.~\cite{Bamber:2021knr,Cardoso:2022whc}). That is, the terms that depend on the perturbation $\Phi^{(q)}$ in Eq.~\eqref{EFEs_21} do not affect axial metric perturbations. 

As discussed in Sec.~\ref{sec:pert_scheme}, we consider an expansion for the metric of the form~\eqref{eq:metric_expansion}, where we recall that $g_{\mu\nu}^{(1)}=0$ by virtue of the field equations, $g_{\mu\nu}^{(2)}$ are the corrections to the Schwarzschild metric that we just computed in the previous subsection and $h_{\mu\nu}$ is what we wish to (partially) compute. If we consider perturbations only up to order $\mathcal{O}(\epsilon^2,q^1)$, we could in principle further expand $h_{\mu\nu}$ as done in Eq.~\eqref{eq:expansion_h}. However, in this case, it turns out to be simpler to work with the expansion~\eqref{eq:metric_expansion} as is and consider the perturbation equations in the form~\eqref{EFEs_21}. By doing so, our problem reduces to considering axial perturbations of the background metric~\eqref{EF_perts} (but neglecting the slow time variation of $\delta M$ and $\delta\lambda$ as discussed above), which are induced by a point particle moving along circular geodesics in this background. The details of the computation are given in App.~\ref{app:axial_epsilon2}. Here we only provide the main results.

We follow a procedure similar to the one discussed in Sec.~\ref{sec:GRperts}, namely we decompose the metric perturbation $h_{\mu\nu}$ using a similar decomposition to Eq.~\eqref{decom} but now, given that we are working in ingoing Eddington-Finkelstein coordinates, we replace $t$ by $v$ in Eq.~\eqref{decom}. We also only consider the axial part of the perturbations, as already mentioned. Doing so, one finds that $l\geq 2$ axial perturbations can be described in terms of a single master wave equation for the function $\bar{\psi}_{\rm ax}^{lm}(r)$ that can be written as\footnote{Here and in the following, quantities with a bar on top are used to refer to quantities that are valid up to order $\mathcal{O}(\epsilon^2)$.}
\begin{equation}
   \left[\frac{d^{2} }{d\bar{r}_{*}^{2}} + \sigma^{2} - \bar{V}_{\rm ax}\right]\bar{\psi}_{\rm ax}^{lm}(r) = \bar{S}_{\rm ax}^{lm}(r)\,,\label{masterodd_epsilon2}
\end{equation} 
where the ``generalized'' tortoise coordinate $\bar{r}_{*}$ related to the metric~\eqref{EF_perts} is now defined through $d\bar{r}_*/dr = 1/F_*(r)$ with the function $F_*(r)$ given by
\begin{equation}
    F_*(r) := \left(f(r) - \frac{2 \epsilon^{2}\delta M(r)}{r}\right) e^{\epsilon^{2}\delta\lambda(r)}\,.
 \end{equation}
The potential $\bar{V}_{\rm ax}$ is given by 
\begin{equation}
     \bar{V}_{\rm ax} = F_*\left(\frac{(l-1)(l+2)}{r^2}e^{\epsilon^2 \delta\lambda}-\frac{F_*'}{r}+\frac{2 F_*}{r^2}\right)\,,
\end{equation}
whereas the source term $\bar{S}_{\rm ax}^{lm}(r)$ can be found in App.~\ref{app:axial_epsilon2} [see Eq.~\eqref{axial_source_epsilon2}]. We should note that when using Eq.~\eqref{masterodd_epsilon2} one should remind ourselves that all quantities should be expanded up to $\mathcal{O}(\epsilon^2)$. Notice that when $r \to \infty$, where $\delta\lambda\sim 0$ and $\epsilon^2\delta M\sim M_b$, Eq.~\eqref{masterodd_epsilon2} essentially describes axial perturbations of a Schwarzschild spacetime with mass $M+M_b$. On the other hand, as discussed above, in the vicinity of the BH one has $\delta M\sim 0$ and $\epsilon^2\delta \lambda\sim \epsilon^2\delta \lambda_H$, where again $\epsilon^2\delta \lambda_H \approx -\mu^2 M M_b$ in the small-$M\mu$ limit. Within this approximation we find $\bar{V}_{\rm ax}\approx (1+2\epsilon^2\delta \lambda_H) V^{\rm vac}_{\rm ax}$, $dr/d\bar{r}_* \approx (1+\epsilon^2\delta \lambda_H)dr/dr^{\rm vac}_*$, whereas the source term satisfies $\bar{S}_{\rm ax}^{lm}(r)\approx (1+3\epsilon^2\delta \lambda_H)S_{\rm ax}^{{\rm vac},lm}(r)$, in agreement with Ref.~\cite{Cardoso:2022whc}. Therefore, at small orbital radii, the axial master equation can be approximated by~\cite{Cardoso:2022whc}
\begin{equation}
   \left[\frac{d^{2}}{d(r^{\rm vac}_*)^{2}} + \frac{\sigma^{2}}{\gamma^2} - V^{\rm vac}_{\rm ax}\right]\bar{\psi}_{\rm ax}^{lm}(r) = \gamma S_{\rm ax}^{{\rm vac},lm}(r)\,,\label{masterodd_approx}
\end{equation} 
where $\gamma=(1+\epsilon^2\delta \lambda_H)$. As already noticed in Refs.~\cite{Cardoso:2021wlq,Cardoso:2022whc,Figueiredo:2023gas}, this implies that the axial master equation in the vicinity of the BH is well approximated by the same master equation describing axial perturbations of a vacuum Schwarzschild BH but with the following rescalling:
\begin{equation}\label{redshift}
\sigma_{\rm vac}\to \sigma/\gamma\,,\quad \Omega_{p,{\rm vac}}\to \Omega_p/\gamma\,,\quad m_{p,{\rm vac}} \to m_p\gamma\,.
\end{equation}
That is, at small orbital radii, corrections to the axial GW flux can be fully understood in terms of a redshift effect.

\subsubsection{Gravitational-wave flux}
As we did in Sec.~\ref{sec:GRperts}, solutions to Eq.~\eqref{masterodd_epsilon2} can be constructed using a Green's function technique. The procedure is essentially identical to what we did there, so let us just highlight the main differences. As done in Sec.~\ref{sec:GRperts} one first computes two independent solutions of the homogeneous part of Eq.~\eqref{masterodd_epsilon2}, $\bar{\psi}^{\rm ax}_{\rm up}$ and $\bar{\psi}^{\rm ax}_{\rm in}(r)$, satisfying boundary conditions identical to~\eqref{psiin} and~\eqref{psiup} but with $r_*$ replaced by $\bar{r}_*$. From those two independent solutions one can construct a constant Wronskian given by an expression similar to Eq.~\eqref{wronskian_grav} but with $r_*$ again replaced by $\bar{r}_*$. We can then use these ingredients to construct a solution to Eq.~\eqref{masterodd_epsilon2} given by
\begin{align}\label{axial_sol_epsilon2}
\bar{\psi}_{\rm ax}(r)=&\frac{\bar{\psi}^{\rm ax}_{\rm up}(r)}{W}
\int^r_{2M} 
 \frac{\bar{\psi}^{\rm ax}_{\rm in}(r')\bar{S}_{\rm ax}(r')}{F_*(r')} dr'\nonumber\\
&+ \frac{\bar{\psi}^{\rm ax}_{\rm in}(r)}{W}\int^{\infty}_{r}
 \frac{\bar{\psi}^{\rm  ax}_{\rm up}(r')\bar{S}_{\rm ax}(r')}{F_*(r')}dr'\,.
\end{align}
Similarly to Sec.~\ref{sec:GRperts}, the integrals in the expression above can be computed analytically since the function $\bar{S}_{\rm ax}(r)$ only contains terms proportional to $\delta(r - r_{p})$ and derivatives of it. 

Having found a solution $\bar{\psi}_{\rm ax}(r)$ to Eq.~\eqref{masterodd_epsilon2} we can compute the (axial) GW flux at infinity and at the horizon. Since the metric~\eqref{EF_perts} is asymptotically flat, the procedure in Ref.~\cite{Martel:2005ir} still applies and the flux at infinity can be computed using an equation identical to~ Eq.~\eqref{fluxgrav_inf}. Similarly, given that the metric~\eqref{EF_perts} at the horizon is geometrically equivalent to Schwarzschild, as long as we neglect the slow time variation of $\delta M$, the procedure of Ref.~\cite{Martel:2005ir} to compute the axial flux at the horizon also applies to our case and we can arrive at an equation identical to Eq.~\eqref{fluxgrav_hor} to compute the flux at the event horizon. Again noticing that $\bar{S}_{\rm ax}(r)\propto \delta(\sigma-m\Omega_p)$, where now $\Omega_p$ contains corrections at order $\mathcal{O}(\epsilon^2)$ [see Eq.~\eqref{Omega_epsilon2}], we then get that the axial GW fluxes for each multipole can be computed using
\begin{equation}
    \dot E^{g,H/\infty}_{lm} = \lim_{r \to 2M/\infty}\,\frac{1}{64 \pi}\frac{(l+2)!}{(l-2)!} (m\Omega_p)^2\left|\bar{\psi}_{\rm ax}^{lm}\right|^{2}\,,
\end{equation}
where we recall that, for axial perturbations, only the multipoles for which the sum $l+m$ is an odd number contribute. 

Since one has $\Omega_p\approx \Omega_p^{\rm vac}+\epsilon^2\delta\Omega$ and $\bar{\psi}_{\rm ax}^{lm}\approx\psi_{{\rm ax}}^{{\rm vac},lm}+\epsilon^2 \psi_{{\rm ax}}^{(2),lm}$, the leading-order correction to the GW flux should also scale with $\epsilon^2$ or, equivalently, with the cloud's total mass $M_b$, that is: $\dot E^{g}\approx \dot E^{g,{\rm vac}}+(M_b/M)\,\dot E^{g,(2)}+\mathcal{O}(\epsilon^4)$ where $\dot E^{g,{\rm vac}}$ is the zeroth-order flux computed in Sec.~\ref{sec:GRperts} and $(M_b/M)\,\dot E^{g,(2)}$ is the correction at order $\epsilon^2$. In particular, the absolute relative difference between the GW flux in the presence of the cloud and in vacuum should scale as $\delta\dot E:=|1-\dot{E}^{g}/\dot{E}^{g,\rm{vac}}|\approx (M_b/M)|\dot E^{g,(2)}/\dot{E}^{g,\rm{vac}}|$.  

\subsection{Numerical procedure and results}\label{sec:resultsaxial}
We implemented the procedure outlined above in a \texttt{Mathematica} code that follows closely the discussion of Sec.~\ref{sec:scalarfluxresults}, the main difference being that now we also need to compute the metric perturbations $\delta M$ and $\delta \lambda$ by numerically integrating Eqs.~\eqref{dMr_final} and~\eqref{dlambdar_final}. As done in Sec.~\ref{sec:scalarfluxresults}, we use a series expansion for the boundary conditions at the horizon and at infinity, with the addition that now $\delta M$, $\delta \lambda$ and  $\tilde{R}$ also enter the equations we need to integrate. Therefore at the horizon we use
 \begin{align}
    \psi^{\rm ax}_{\rm in}(r\to 2M)&= e^{-i \sigma \bar{r}_*} \sum_{i=0}^{n_{\rm in}} (a^{(0)}_i+\epsilon^2 a^{(2)}_i) (r-2M)^i\,,\label{barpsiIN_series}\\
   \delta M(r\to 2M)& = \sum_{i=1}^{n_{\rm in}} c_i (r-2M)^i\,,\\
   \delta \lambda(r\to 2M)& = \sum_{i=0}^{n_{\rm in}} d_i (r-2M)^i\,,\\
   \tilde{R}(r\to 2M)& = \sum_{i=0}^{n_{\rm in}} e_i (r-2M)^i\,,
\end{align}
where we explicitly expanded the coefficients of the series expansion for $\bar{\psi}^{\rm ax}_{\rm in}$ in powers of $\epsilon$ given that Eq.~\eqref{masterodd_epsilon2} is only valid up to that order and because the computation of the coefficients turns out to be simpler by doing so. We adopt an overall normalization $a^{(0)}_0=1$ and $a^{(2)}_0=0$, whereas $d_0$ is fixed by requiring $\delta \lambda(r\to \infty)=0$. On the other hand, the arbitrary coefficient $e_0$ can be absorbed in $\epsilon$ which in turn we relate to the cloud's total mass $M_b$ using Eq.~\eqref{Mb_EF}. The rest of the coefficients can be obtained by inserting these series expansions in Eqs.~\eqref{eq_radial_KG_EF},~\eqref{dMr_final} and~\eqref{dlambdar_final}, in addition to (the homogeneous part of) Eq.~\eqref{masterodd_epsilon2}, and expanding the equations in $\epsilon$ and at $r=2M$. On the other hand, at infinity we fix $\epsilon^2\delta M(r\to\infty)=M_b$ and $\delta\lambda(r\to\infty)=0$ and expand $\psi^{\rm ax}_{\rm{up}}$ as:
\begin{equation}
     \psi^{\rm ax}_{\rm{up}}(r\to \infty)= e^{+i \sigma \bar{r}_*} \sum_{i=0}^{n_{\rm up}} \frac{(b^{(0)}_i+\epsilon^2 b^{(2)}_i)}{r^i}\,,\label{barpsiUP_series}
\end{equation}
where we use the normalization $b^{(0)}_0=1$ and $b^{(2)}_0=0$ and obtain the other coefficients by expanding and solving (the homogeneous part of) Eq.~\eqref{masterodd_epsilon2} order by order in powers of $\epsilon$ and $1/r$. As done in Sec.~\ref{sec:scalarfluxresults}, we keep up to ten terms in these series expansions and integrate the equations in the numerical domain described in Sec.~\ref{sec:scalarfluxresults}, checking {\it a posteriori} that our results are stable against changes in these choices. We also checked that if we set $M_b=0$, i.e., in the absence of the cloud, the absolute relative differences between the fluxes obtained with our code and with the \texttt{ReggeWheeler} package of the \texttt{Black Hole Perturbation Toolkit}~\cite{BHPToolkit} are typically of the order $\sim 10^{-9}~-~10^{-10}$ for the range of orbital radii we consider, indicating very good agreement.
\subsubsection{Results}
%
\begin{figure*}[thb!]
    \centering
    \includegraphics[width=0.45\textwidth]{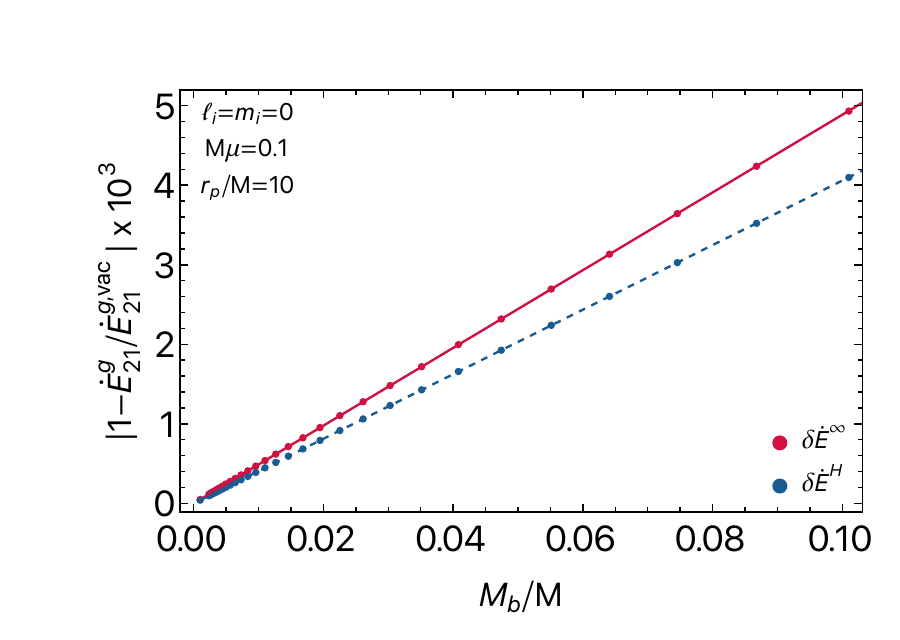}
    \includegraphics[width=0.45\textwidth]{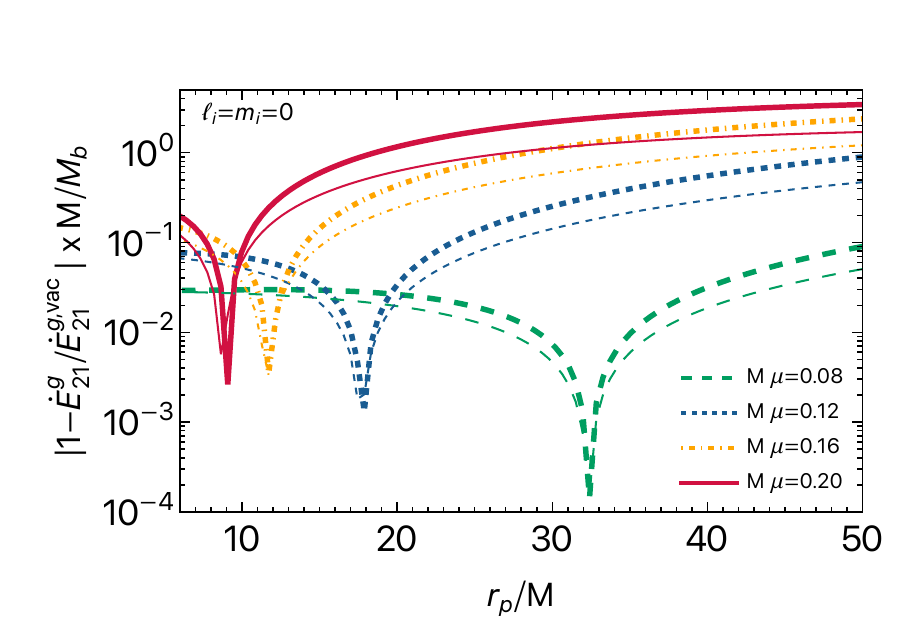}
    \caption{Left panel: absolute relative difference between the $l=2,m=1$ multipole of the GW energy flux, emitted in the presence of a $\ell_i=m_i=n_i=0$ scalar cloud, against the same quantity computed in vacuum, as a function of the cloud's total mass. We show both the relative difference for the horizon flux, $\delta\dot E^H$, and the flux at infinity $\delta\dot E^\infty$. Solid lines correspond to a linear fit to the data points shown in the figure. Right panel: same as the left panel but now the relative difference is normalized by $M_b$ and shown as a function of the point particle's location $r_p/M$, for different values of $M\mu$. Thicker lines are for the flux at infinity, whereas thin lines correspond to the horizon fluxes. We note that at large orbital radii we typically have $1-\dot{E}^{g}/\dot{E}^{g,\rm{vac}}<0$, whereas close to the ISCO we find that $1-\dot{E}^{g}/\dot{E}^{g,\rm{vac}}>0$. The peaks seen in the figure correspond to the transition between the two regimes.}
    \label{fig:fluxaxial}
\end{figure*}
Our main results are shown in Fig.~\ref{fig:fluxaxial} where we show the relative differences between the GW energy flux in the presence of a spherical scalar cloud against the GW energy flux in vacuum. We consider both the flux at infinity and at the horizon and focus on the $l=2,m=1$ multipole, the dominant axial GW multipole for a particle in circular orbit. 

The left panel of Fig.~\ref{fig:fluxaxial} shows how the relative difference scales with the cloud's total mass $M_b$, fixing $M\mu=0.1$ and the particle's orbital radius $r_p=10M$. As anticipated, we find that the relative difference $\delta\dot E:=|1-\dot{E}_{21}^{g}/\dot{E}_{21}^{g,\rm{vac}}|$ scales linearly with $M_b$, as long as $M_b$ is sufficiently small, which is a good consistency check of our procedure. Given this scaling, in the right panel of Fig.~\ref{fig:fluxaxial} we show the relative difference normalized\footnote{More concretely, we computed the non-vacuum fluxes by fixing $M_b/M=10^{-2}$ and then divided the computed relative difference by $10^{-2}$. Since $\delta\dot E$ scales linearly with $M_b$ the resulting values are valid for any value of $M_b/M$, as long as $M_b/M$ is not too large.} by $M_b$ as a function of $r_p/M$, and for different values of $M\mu$. For the values shown, we find that the maximum relative difference increases with $M\mu$, which is to be expected, given that the compactness of the cloud increases with $M\mu$. 

\begin{figure}[thb!]
    \centering
    \includegraphics[width=0.48\textwidth]{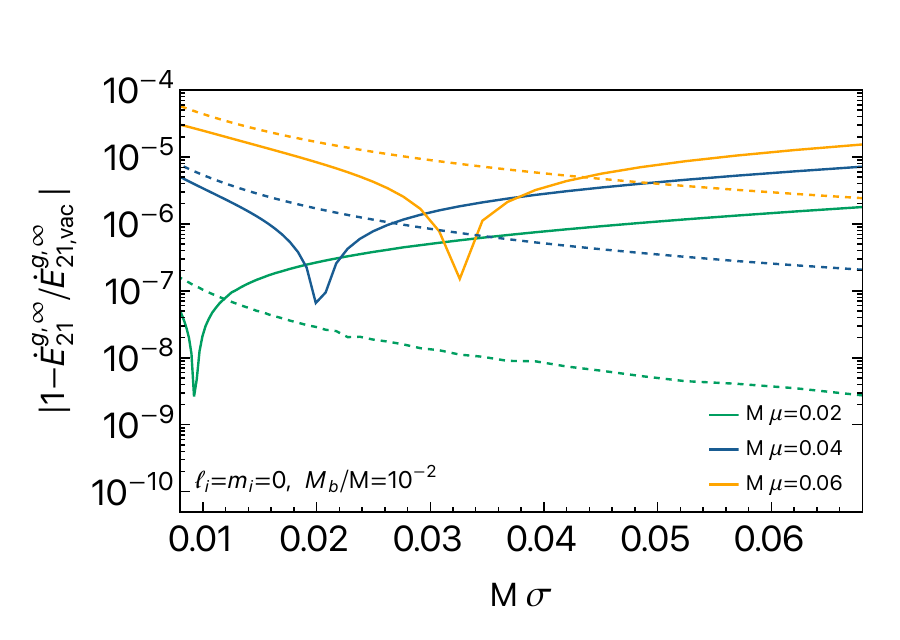}
    \caption{Same as Fig.~\ref{fig:fluxaxial} but now we show the relative difference of the flux at infinity as a function of the GW frequency (solid line). We assume $M_b = 10^{-2}M$ for the cloud's total mass. Solid (dashed) lines correspond to relative differences with respect to vacuum unredshifted (redshifted) results (see discussion in the main text).}
    \label{fig:fluxaxial_vs_GWfreq}
\end{figure}

As discussed above, and according to Ref.~\cite{Cardoso:2022whc} where GW fluxes for non-vacuum BH geometries for other types of environments were computed, we expect that at high orbital frequencies, or equivalently, at small orbital radii, the results can be understood in terms of a simple redshift effect. In order to check this claim, in Fig.~\ref{fig:fluxaxial_vs_GWfreq} we show the relative difference between the GW flux at infinity computed in the presence of the cloud against the vacuum case but now comparing the two cases at fixed GW frequency (solid lines). This is to be contrasted with the relative difference between the non-vacuum fluxes and the ``redshifted'' fluxes (dashed lines), which were computed assuming the vacuum perturbation equations after performing the rescaling given in Eq.~\eqref{redshift}. As expected, at high frequencies, the redshifted vacuum fluxes are a good description of our results for axial perturbations. This description is more accurate for small $M\mu$, that is, when the cloud is less compact, which is in agreement with the findings of Ref.~\cite{Cardoso:2022whc}.

\section{Conclusions and Outlook}\label{sec:conclusions}
In this work we presented a relativistic perturbation theory framework to study EMRIs in BHs surrounded by scalar clouds. Previous studies on this topic typically employed non-relativistic approximations~\cite{Baumann:2018vus,Baumann:2019ztm,Baumann:2021fkf,Baumann:2022pkl,Tomaselli:2023ysb}, which are known to be inaccurate to describe EMRIs in the regimes where such sources are expected to be detected by LISA (see e.g. Ref.~\cite{vandeMeent:2020xgc}). The main goal of this work was to give the first steps towards studying such systems using a fully relativistic setup. This can be seen as a natural follow-up of recent studies aimed at computing in a fully relativistic framework the GW emission by EMRIs in non-vacuum spacetimes~\cite{Cardoso:2021wlq,Cardoso:2022whc,Figueiredo:2023gas}, and is a crucial step towards the goal of including environmental effects in accurate EMRI waveforms. Two important contributions of our work when compared to previous studies are: (i) we showed that our framework allows to also consider non-spherically symmetric environments, such as non-spherically symmetric scalar clouds that can be formed through superradiant instabilities; (ii) we showed that the effect of the environment can be added in a modular way on top of ``vacuum'' results. We believe this to be an important feature, given that it will allow an easy integration of environmental effects onto the modular machinery being developed to compute fast EMRI waveforms~\cite{Chua:2020stf,Katz:2021yft}. 

We did not perform a full comparison with previous non-relativistic studies~\cite{Baumann:2018vus,Baumann:2019ztm,Baumann:2021fkf,Baumann:2022pkl,Tomaselli:2023ysb}, since we believe this requires a full dedicated study. However, overall, our results seem to be at least in qualitative agreement with those works. 
We confirm that in the presence of a scalar cloud, the energy lost by the orbit at infinity due to scalar perturbations can exceed the energy loss due to GW emission. We also found that the scalar power contains ``sharp features'' at given orbital radii, as described in Refs.~\cite{Baumann:2021fkf,Baumann:2022pkl,Tomaselli:2023ysb}, and also showed evidence that the modes of the cloud can be resonantly excited. These resonances, first studied in Refs.~\cite{Baumann:2018vus,Baumann:2019ztm} for the case of Newtonian scalar clouds, can lead to sinking or even floating orbits~\cite{Cardoso:2011xi,Zhang:2018kib,Baumann:2019ztm}. There are however some important differences, such as the relevance of the $\ell_j=m_j=0$ scalar power at the horizon in the presence of a dipolar non-axisymmetric scalar cloud background, which was missed in previous studies (see Sec.~\ref{sec:scalarfluxresults}).

This work is only meant as a first important step towards the long-term goal of building accurate EMRI waveforms that take into account the presence of boson clouds around supermassive BHs. There is plenty of room for significant improvements. First of all, as already mentioned in the main text, we did not compute the corrections to the eigenfrequencies of the background scalar cloud that should arise due to the secondary object as well as the self-gravity of the cloud, but this can in principle be done within the same perturbative framework we presented here (see App.~\ref{app:freq_shift}). When computing the scalar fluxes we neglected possible contributions that could arise due to the interaction between the metric and the scalar perturbations (see discussion in Sec.~\ref{sec:KGperts}). A better understanding of how to deal with such terms requires further work. Furthermore, we only computed corrections to the GW flux at order $\mathcal{O}(\epsilon^2)$ for axial perturbations and for spherically symmetric clouds. Besides extending this computation to non-spherical clouds, in order to fully describe the GW flux at this order we also need to compute the corrections to the polar GW fluxes, which we have not done in this work. Using the results of Ref.~\cite{Cardoso:2022whc} as a guide, we anticipate that the corrections to the polar fluxes will be comparable or slightly larger than what we found here for axial perturbations. Although the contribution of the scalar flux shown in Sec.~\ref{sec:scalarfluxresults} typically dominates over the corrections to the GW flux considered in Sec.~\ref{sec:resultsaxial}, we found that the corrections to the axial GW flux can be non-negligible. Therefore one might expect that accurate waveforms will need to include both effects. However a full understanding of this issue will require studying the evolution of the orbit and in turn studying how these different effects impact gravitational waveforms.
Another obvious and very important extension of this work is to generalize it to the case where the background BH geometry is given by the Kerr geometry. This is a much harder task to tackle, however we expect this to be possible using the Teukolsky formalism and the methods recently built in Refs.~\cite{Li:2022pcy,Hussain:2022ins,Ghosh:2023etd,Cano:2023tmv}. A possible intermediate step, before tackling the Kerr BH case, is to consider a slowly-rotating approximation~\cite{Pani:2012bp,Pani:2013wsa}, which might provide good insights into the impact that the BH spin has on the results. Other improvements include going beyond circular, equatorial orbits and even adding self-force corrections. Finally, another natural extension of this work is to consider clouds formed by massive vector fields~\cite{Dolan:2018dqv,Baumann:2019ztm}. We plan to come back to these problems in future work.
%
\begin{acknowledgments}
We are indebted to Vitor Cardoso and Francisco Duque for very valuable discussions and suggestions. We also thank Thomas Spieksma for discussions that allowed us to find a factor two error in our original derivation of the flux formulas. This work makes use of the Black Hole Perturbation Toolkit. R.B. acknowledges financial support provided by FCT – Fundação para a Ciência e a Tecnologia, I.P., under the Scientific Employment Stimulus -- Individual Call -- 2020.00470.CEECIND and under Project No. 2022.01324.PTDC.
\end{acknowledgments}

\appendix

\section{Decomposition of the point particle's stress-energy tensor}\label{app:PP_SEtensor}
The point particle's stress-energy tensor [Eq.~\eqref{eq:STpoint}] can be decomposed in terms of a tensor harmonics basis as~\cite{Zerilli:1970wzz,Sago:2002fe}:
%
\begin{align}
T_{\mu\nu}=&\sum_{l=0}^{\infty}\sum_{m =-l}^{l}
\Re\Bigg\{\bigg[{\cal A}^{(0)}_{l m }a^{(0)}_{l m,\mu\nu}(\theta,\phi)+{\cal A}^{(1)}_{l m}a^{(1)}_{l m,\mu\nu}(\theta,\phi)\nonumber\\
&+{\cal A}_{l m }a_{l m,\mu\nu}(\theta,\phi)+{\cal B}^{(0)}_{l m }b^{(0)}_{l m,\mu\nu}(r,\theta,\phi)\nonumber\\
&+{\cal B}_{l m }b_{l m,\mu\nu}(r,\theta,\phi)+{\cal Q}^{(0)}_{l m }c^{(0)}_{l m,\mu\nu}(r,\theta,\phi)
\nonumber\\
&+{\cal Q}_{l m }c_{l m,\mu\nu}(r,\theta,\phi)+{\cal D}_{l m }d_{l m,\mu\nu}(r,\theta,\phi)
\nonumber\\
&+{\cal G}_{l m}g_{l m,\mu\nu}(r,\theta,\phi)+{\cal F}_{l m }f_{l m,\mu\nu}(r,\theta,\phi)\bigg]\Bigg\}\,,\label{harmonicTmunu}
\end{align}
%
where $(c^{(0)}_{l m,\mu\nu};c_{l m,\mu\nu};d_{l m,\mu\nu})$ are axial tensor harmonics and $(a^{(0)}_{l m,\mu\nu};a^{(1)}_{l m,\mu\nu};a_{l m,\mu\nu};b^{(0)}_{l m,\mu\nu};b_{l m,\mu\nu};g_{l m,\mu\nu};f_{l m,\mu\nu})$ are polar tensor harmonics. Their explicit form can be found in Refs.~\cite{Zerilli:1970wzz,Sago:2002fe}. The tensor harmonics $(c^{(0)}_{l m,\mu\nu};c_{l m,\mu\nu},b^{(0)}_{l m,\mu\nu};b_{l m,\mu\nu})$ vanish for $l=0$, whereas the harmonics $(d_{l m,\mu\nu};f_{l m,\mu\nu})$ vanish for $l=0,1$. The expansion coefficients $({\cal A}^{(0)}_{l m },{\cal A}^{(1)}_{l m },\ldots,{\cal F}_{l m })$ are only functions of the time and the radial coordinate and can be computed making use of the orthonormality of the tensor harmonics under the inner product $(A,B)=\int\eta^{\mu\rho}\eta^{\nu\sigma}A^{*}_{\mu\nu}B_{\rho\sigma}d\Omega$,
where $A,B$ are two generic tensors and we defined the matrix $\eta^{\mu\nu}=\diag(1,1,r^2,r^2\sin^2\theta)$. One can then find the expansion coefficients by projecting the stress-energy tensor onto each tensor harmonic, e.g., ${\cal A}^{(0)}_{l m }=(a^{(0)}_{l m,\mu\nu},T_{\mu\nu})$ and similarly for other coefficients. The explicit form of these coefficients when considering a point particle moving in a Schwarzschild metric written in Schwarzschild coordinates, can be found in~\cite{Sago:2002fe} (see their Table I) and we explicitly checked that we recover their results using the procedure just outlined. Finally, since we work in the frequency domain, we Fourier transform the coefficients $\{{\cal A}^{(0)}_{l m }(x^{0},r),\ldots\}$ as in Eq.~\eqref{fourier}\footnote{When working in ingoing Eddington-Finkelstein coordinates we replace $t$ by $v$ in Eqs.~\eqref{inversefourier} and~\eqref{fourier}.}.
\subsection{The stress-energy tensor for circular orbits}
Considering a spherical coordinate system, $x^{\mu}=(x^0,r,\theta,\phi)$, where $x^{0}=t$ or $x^{0}=v$ depending on whether one uses Schwarzschild or ingoing Eddington-Finkelstein coordinates, Eq.~\eqref{eq:STpoint} can be computed by transforming the integral over $\tau$ into an integral over $x^{0}$:
 \begin{align}\label{eq:STpoint_v2}
    T^{\mu\nu} = & \frac{\gamma m_{p}}{\sqrt{-g}}\frac{dx_p^{\mu}}{dx^{0}}\frac{dx_p^{\nu}}{dx^{0}} \nonumber\\
    & \times \delta\left(r - R(x^{0})\right)\delta\left(\theta - \Theta(x^{0})\right)\delta\left(\phi - \Phi(x^{0})\right)\,,
\end{align}
where we defined the Lorentz factor $\gamma:=dx_p^{0}/d\tau$ and used the notation $x_p^{\mu}(\tau)=\left\{x_p^{0}(\tau),R(\tau),\Theta(\tau),\Phi(\tau)\right\}$ for the particle's worldline. For circular, equatorial orbits one has $R(x^0)=r_p$, $\theta(x^0)=\pi/2$ and $\phi(x^0)=\Omega_p x^0+\phi_0$, with $r_p$ the particle's orbital radius, $\Omega_p$ its angular orbital frequency and $\phi_0$ an arbitrary initial phase that we can set to zero without loss of generality (see App.~\ref{app:app_geo}). In order to compute the metric perturbations sourced by this stress-energy tensor, we need the explicit form of the coefficients $\{{\cal A}^{(0)}_{l m }(x^0,r),\ldots\}$ which depend on the coordinate system used for the computation, as we now discuss.
\subsubsection{Schwarzschild coordinates}
For the sake of generality, let us consider a generic spherically symmetric metric in Schwarzschild coordinates:
\begin{equation}\label{Sch_metric_general}
ds^{2} = -A(r)dt^{2} + \frac{dr^{2}}{B(r)} + r^{2}d\theta^{2}+r^{2}\sin^{2}\theta d\phi^{2}\,.
\end{equation}
The Schwarzschild BH metric can be recovered by simply setting $A(r)=B(r)=1-2M/r$. Circular orbits in this geometry are characterized by a Lorentz factor $\gamma=E/(m_p A(r))$, where the particle's energy $E$, angular momentum $L$ and orbital frequency $\Omega_p$ are given by~\cite{Cardoso:2008bp}
 \begin{eqnarray}
    \frac{E}{m_p} &=& \sqrt{\frac{2 A(r_p)^2}{2
   A(r_p)-r_p A'(r_p)}}\,,\\
   \frac{L}{m_p} &=& \pm\sqrt{\frac{r_p^3 A'(r_p)}{2 A(r_p)-r_p
   A'(r_p)}}\,,\\
   \Omega_p &=& \pm\sqrt{\frac{A'(r_p)}{2r_p}}\,.
\end{eqnarray}
Applying the method outlined above for such orbits, one finds that ${\cal A}_{l m }={\cal A}^{(1)}_{l m}={\cal B}_{l m}={\cal Q}_{l m}=0$, whereas the Fourier transforms of the non-vanishing coefficients read
\begin{eqnarray}\label{comp_Tmunu_Sch_1}
    \tilde{\cal A}^{(0)}_{l m} &=& \frac{E A(r)}{r^2}\delta_r\delta_{\sigma}Y_{lm}^*(\pi/2,0)\,,\\
    \tilde{\cal B}^{(0)}_{l m } &=& \frac{\sqrt{2}m\Omega_p E}{\sqrt{l(l+1)}r} \delta_r\delta_{\sigma}Y_{lm}^*(\pi/2,0)\,,\\
    \tilde{\cal Q}^{(0)}_{l m } &=& \frac{\sqrt{2}\Omega_p E}{\sqrt{l(l+1)}r} \delta_r\delta_{\sigma}\partial_\theta Y_{lm}^*(\pi/2,0)\,,\\
    \tilde{\cal D}_{l m } &=& -\frac{i E\Omega_p^2}{\sqrt{2\Lambda} A(r)} \delta_r\delta_{\sigma}X_{lm}^*(\pi/2,0)\,,\\
    \tilde{\cal G}_{l m } &=& \frac{E \Omega_p^2}{\sqrt{2} A(r)}\delta_r\delta_{\sigma}Y_{lm}^*(\pi/2,0)\,,\\
    \tilde{\cal F}_{l m } &=& -\frac{E \Omega_p^2}{\sqrt{2\Lambda} A(r)}\delta_r\delta_{\sigma}W_{lm}^*(\pi/2,0)\label{comp_Tmunu_Sch_2}\,,
\end{eqnarray}
where we defined $\Lambda=(l-1) l (l+1) (l+2)$, $\delta_r:= \delta(r-r_p)$, $\delta_{\sigma}:= \delta(\sigma -m\Omega_p)$ and the angular functions are defined as:
\begin{align}
X_{l m}(\theta,\phi)&= 2\frac{\partial}{\partial \phi}
\left(\frac{\partial}{\partial \theta}-\cot \theta \right)Y_{l m}(\theta,\phi) \,, \\
W_{l m}(\theta,\phi)&= \left(\frac{\partial ^2}{\partial \theta ^2}
-\cot \theta \frac{\partial}{\partial \theta}
-\frac{1}{\sin ^2 \theta}
\frac{\partial ^2}{\partial \phi ^2}
\right)Y_{l m}(\theta,\phi) \,.
\end{align}
We also used a tilde to emphasize that the functions shown above are frequency-domain quantities, i.e. $\tilde{\cal A}^{(0)}_{l m}(\sigma,r)$ is the Fourier transform of ${\cal A}^{(0)}_{l m}(t,r)$ and similarly for the other functions.

\subsubsection{Ingoing Eddington-Finkelstein coordinates}
Let us now perform the same computation as above but considering a generic spherically symmetric metric in ingoing Eddington-Finkelstein coordinates written as
\begin{equation}\label{EF_metric_general}
ds^{2} = -A(r)dv^{2} + 2H(r)dvdr + r^{2}d\theta^{2}+r^{2}\sin^{2}\theta d\phi^{2}\,.
\end{equation}
Notice that we recover the Schwarzschid metric in ingoing Eddington-Finkelstein coordinates, Eq.~\eqref{Sch_metric_EF}, when $A(r)=1-2M/r$, $H(r)=1$. On the other hand, for the case of the metric describing a BH surrounded by a spherical scalar cloud, Eq.~\eqref{EF_perts}~\footnote{As argued in Sec.~\ref{sec:Oq1epsilon2}, here we neglect the dependence on the advanced time for the functions $\delta M$ and $\delta\lambda$.}, one has $A(r)=(1-2[M+\epsilon^2 \delta M (r)]/r)e^{2\epsilon^2\delta\lambda(r)}$ and $H(r)=e^{\epsilon^2\delta\lambda(r)}$.

Circular, equatorial orbits in this geometry are computed in App.~\ref{app:app_geo}. The expressions for the particle's Lorentz factor, energy $E$, angular momentum $L$ and orbital frequency $\Omega_p$ are the same as the ones given above, i.e., they only depend on the function $A(r)$. Applying the same method as above, we find
\begin{eqnarray}
    \tilde{\cal A}^{{\rm EF}(0)}_{l m} &=& \frac{E A(r)}{r^2}\delta_r\delta_{\sigma}Y_{lm}^*(\pi/2,0)\label{comp_Tmunu_EF_1}\,,\\
    \tilde{\cal A}^{{\rm EF}(1)}_{l m} &=& i\frac{\sqrt{2} E H(r)}{r^2}\delta_r\delta_{\sigma}Y_{lm}^*(\pi/2,0)\,,\\
    \tilde{\cal A}^{{\rm EF}}_{l m} &=& \frac{E H(r)^2}{A(r)r^2}\delta_r\delta_{\sigma}Y_{lm}^*(\pi/2,0)\,,
\end{eqnarray}
\begin{eqnarray}
    \tilde{\cal B}^{{\rm EF}(0)}_{l m } &=& \frac{\sqrt{2}m\Omega_p E}{\sqrt{l(l+1)}r} \delta_r\delta_{\sigma}Y_{lm}^*(\pi/2,0)\,,\\
     \tilde{\cal B}^{{\rm EF}}_{l m }  &=&  -\frac{i\sqrt{2}m\Omega_p E H(r)}{\sqrt{l(l+1)}A(r)r} \delta_r\delta_{\sigma}Y_{lm}^*(\pi/2,0)\,,\\
    \tilde{\cal Q}^{{\rm EF}(0)}_{l m } &=& \frac{\sqrt{2}\Omega_p E}{\sqrt{l(l+1)}r} \delta_r\delta_{\sigma}\partial_\theta Y_{lm}^*(\pi/2,0)\,,\\
     \tilde{\cal Q}^{{\rm EF}}_{l m } &=& \frac{i\sqrt{2}\Omega_p E H(r)}{\sqrt{l(l+1)}A(r)r} \delta_r\delta_{\sigma}\partial_\theta Y_{lm}^*(\pi/2,0)\,,\\
    \tilde{\cal D}^{{\rm EF}}_{l m } &=& -\frac{i E\Omega_p^2}{\sqrt{2\Lambda} A(r)} \delta_r\delta_{\sigma}X_{lm}^*(\pi/2,0)\,,\\
    \tilde{\cal G}^{{\rm EF}}_{l m } &=& \frac{E \Omega_p^2}{\sqrt{2} A(r)}\delta_r\delta_{\sigma}Y_{lm}^*(\pi/2,0)\,,\\
    \tilde{\cal F}^{{\rm EF}}_{l m } &=& -\frac{E \Omega_p^2}{\sqrt{2\Lambda} A(r)}\delta_r\delta_{\sigma}W_{lm}^*(\pi/2,0)\,,\label{comp_Tmunu_EF_2}
\end{eqnarray}
where we added the superscript ``EF'' just to remind ourselves that those expressions are valid when considering the metric~\eqref{EF_metric_general}. 

We notice that, if we set $H(r)=\sqrt{A(r)/B(r)}$, the metrics~\eqref{Sch_metric_general} and~\eqref{EF_metric_general} are related by a coordinate transformation $x_{\rm Sch}^{\mu}=\{t(v,r),r,\theta,\phi\} \to x_{\rm EF}^{\mu}=\{v,r,\theta,\phi\}$, where $t(v,r)=v-\bar{r}_*$ and $\bar{r}_*$ is defined through $d\bar{r}_*/dr=H(r)/A(r)$. Therefore, under this coordinate transformation and setting $H(r)=\sqrt{A(r)/B(r)}$, one should be able to obtain the stress-energy tensor in ingoing Eddington-Finkelstein coordinates from the one in Schwarzschild coordinates using $T^{\rm EF}_{\mu\nu}(v,r,\theta,\phi)=\frac{\partial x_{\rm Sch}^{\rho}}{\partial x_{\rm EF}^{\mu}}\frac{\partial x_{\rm Sch}^{\sigma}}{\partial x_{\rm EF}^{\nu}}T^{\rm Sch}_{\rho\theta}(t,r,\theta, \phi)$. As a check of our computation, we verified that Eqs.~\eqref{comp_Tmunu_EF_1}~--~\eqref{comp_Tmunu_EF_2} can be obtained from Eqs.~\eqref{comp_Tmunu_Sch_1}~--~\eqref{comp_Tmunu_Sch_2} after applying this coordinate transformation.

\section{Metric perturbations for a point particle in circular orbit}\label{app:grav_perts}
In this Appendix we review the main tools we used to compute metric perturbations. Perturbations in a Schwarzschild BH background have been widely studied, therefore we only briefly review some important formulas and refer the reader to Refs.~\cite{Regge:1957td,Zerilli:1970se,Zerilli:1970wzz,Sago:2002fe,Martel:2005ir,Cardoso:2022whc} for more details. All the formulas shown below assume circular, equatorial orbits.
\subsection{Regge-Wheeler gauge}
Metric perturbations can also be decomposed using tensor spherical harmonics~\cite{Regge:1957td,Zerilli:1970se,Zerilli:1970wzz}. Here we work in the Regge-Wheeler gauge~\cite{Regge:1957td,Zerilli:1970se,Zerilli:1970wzz} for which the polar and axial perturbations appearing in Eq.~\eqref{decom} can be written as:
\begin{equation}\label{polar}
h_{\mu\nu}^{{\rm polar}, lm} = 
\begin{pmatrix}
H_{0}Y_{lm}& H_{1} Y_{lm} & 0 & 0\\
*&H_{2} Y_{lm} & 0 & 0\\
*&*&r^{2} K Y_{lm}& 0\\
*&*&*&r^{2} \sin^{2} \theta K Y_{lm} 
\end{pmatrix}\,,
\end{equation}
and
\begin{equation}\label{axial}
    h_{\mu\nu}^{{\rm axial}, lm} = \begin{pmatrix}
0 & 0 & -h_{0} Y_{,\phi}^{lm}/\sin\theta  &  h_{0}\sin\theta Y_{,\theta}^{lm} \\
* & 0 & -h_{1} Y_{,\phi}^{lm}/\sin\theta  &  h_{1}\sin\theta Y_{,\theta}^{lm}\\
* & * & 0  &  0\\
* & * & * & 0
\end{pmatrix}\,,
\end{equation}
where $Y_{lm} := Y_{lm}(\theta, \phi)$  are the usual scalar spherical
harmonics and $(H_0, H_1, H_2, K, h_0,h_1)$ are radial functions that depend on the frequency and on the angular momentum numbers $l$ and $m$, e.g., $H_{0} := H_{0}^{lm}(\sigma;r)$. The $*$ entries indicate symmetric components, such that $h_{\mu\nu} = h_{\nu\mu}$.

\subsection{Polar perturbations of a Schwarzschild black hole}
For $l\geq 2$, polar perturbations can be described in terms of a single gauge-invariant scalar function, the Zerilli-Moncrief function $\Psi_{\rm pol}^{lm}$, which can be computed from the metric perturbations and that satisfies the Zerilli equation~\cite{Martel:2005ir}. Considering a Schwarzschild BH background written in Schwarzschild coordinates, the Fourier transform of the Zerilli-Moncrief function, $\psi_{\rm pol}^{lm}(\sigma;r)$ [see Eq.~\eqref{fourier}], is related to the metric functions appearing in Eq.~\eqref{polar} by~\cite{Martel:2005ir}:
\begin{align}
&\psi^{lm}_{\rm pol}(r) =\frac{2r}{l(l+1)} \nonumber \\
&\times \left\{K^{lm}(r) + \frac{r f(r)}{\lambda r+ 3M}\left[f(r) H^{lm}_{2}(r) - r \left(K^{lm}\right)'(r)\right]\right\}\,, \label{zerilli_function}
\end{align}
where we recall that $\lambda:=(l+2)(l-1)/2$ and $f(r) = 1 - 2M/r$. 

Using the perturbed Einstein field equations with a stress-energy tensor given by~\eqref{harmonicTmunu} one can obtain a single differential equation for $\psi^{lm}_{\rm pol}(r)$, Eq.~\eqref{mastereven}, where the source term reads:
%
\begin{align}
&S_{\rm pol}^{lm} =
\frac{4 \pi}{\lambda  (\lambda +1) (3 M+\lambda  r)^2}\Bigg\{2 (2 M-r) (3 M+\lambda  r)\nonumber\\
&\times\left[\sqrt{2} \sqrt{\lambda  (\lambda +1)} \tilde{\cal F}_{l m}(r) (3 M+\lambda  r)-\lambda  r^2 \left(\tilde{\cal A}^{(0)}_{l m}\right)'(r)\right]\nonumber\\
& -2 \lambda  r \tilde{\cal A}^{(0)}_{l m}(r) \left[24 M^2+(7 \lambda -9) M r+(\lambda -1) \lambda 
   r^2\right]\Bigg\}\,.
\label{eq:sourcepol}
\end{align}
%
Notice that, from Eqs.~\eqref{comp_Tmunu_Sch_1} and~\eqref{comp_Tmunu_Sch_2}, one finds that $S_{\rm pol}^{l,m}(\sigma;r)^*=(-1)^m S_{\rm pol}^{l,-m}(-\sigma;r)$, which implies that $\psi^{l,m}_{\rm pol}(\sigma; r)^*=(-1)^m \psi^{l,-m}_{\rm pol}(-\sigma; r)$ as well.

After finding a solution for $\psi_{\rm pol}(r)$ and its first derivative $\psi_{\rm pol}'(r)$, one can also reconstruct the polar metric perturbations (see e.g. Appendix of Ref.~\cite{Sago:2002fe}). From the perturbed Einstein equations we find:
\begin{widetext}
    \begin{eqnarray}
    H_0(r)&=& \left[\frac{(r-2 M) \left(9 M^3+9 \lambda  M^2 r+3 \lambda ^2 M r^2+\lambda ^2 (\lambda +1) r^3\right)}{r^3 (3 M+\lambda r)^2}-r \sigma ^2\right]\psi_{\rm pol}(r)\nonumber\\
    &+&\frac{(2 M-r) \left[3 M^2+\lambda  r (3 M-r)\right]}{r^2 (3 M+\lambda  r)}\psi_{\rm pol}'(r)
    +\frac{8 \pi  r^2 \left(3 M^2+\lambda  r (3 M-r)\right)}{(\lambda +1) (3 M+\lambda  r)^2}\tilde{\cal A}^{(0)}\,,\\
    H_1(r)&=& \frac{i \sigma  \left(3 M^2+\lambda  r (3 M-r)\right)}{(r-2 M) (3 M+\lambda  r)}\psi_{\rm pol}(r)-i\sigma r \psi_{\rm pol}'(r)
    +\frac{8 i \pi  r^5 \sigma }{(\lambda +1) (r-2 M) (3 M+\lambda  r)}\tilde{\cal A}^{(0)}\,,\\
    H_2(r)&=&\frac{\frac{(r-2 M) \left(9 M^3+9 \lambda  M^2 r+3 \lambda ^2 M r^2+\lambda ^2 (\lambda +1) r^3\right)}{(3 M+\lambda  r)^2}-r^4 \sigma ^2}{r (r-2 M)^2} \psi_{\rm pol}(r)
    +\left[1+M \left(\frac{1}{2 M-r}-\frac{3}{3 M+\lambda  r}\right)\right]\psi_{\rm pol}'(r)
    \nonumber\\
    &+&\frac{8 \pi  r^4 \left(3 M^2+\lambda  r (3 M-r)\right)}{(\lambda +1) (r-2 M)^2 (3 M+\lambda  r)^2}\tilde{\cal A}^{(0)}-\frac{8 \sqrt{2} \pi  r^3}{\sqrt{\lambda  (\lambda +1)} (r-2 M)}\tilde{\cal F}\,,\\
    K(r)&=&  \frac{6 M^2+3 \lambda  M r+\lambda  (\lambda +1) r^2}{r^2 (3 M+\lambda  r)}\psi_{\rm pol}(r)+(1-\frac{2M}{r})\psi_{\rm pol}'(r)
    -\frac{8 \pi  r^3}{(\lambda +1) (3 M+\lambda  r)}\tilde{\cal A}^{(0)}\,.
    \end{eqnarray}
\end{widetext}
We checked that, using the conservation equations for the particle's stress-energy tensor, our results are equivalent to the ones found in Refs.~\cite{Martel:2005ir,Sago:2002fe}. From these expressions one finds that $H_0^{l,m}(\sigma; r)^*=(-1)^m H_0^{l,-m}(-\sigma; r)$ with equivalent expressions for all the other radial polar functions.
\subsection{Axial perturbations of a Schwarzschild black hole}
Axial perturbations with $l\geq 2$ can be described in terms of a single gauge-invariant scalar function $\Psi_{\rm ax}^{lm}$. Several gauge-invariant axial scalar functions can be found in the literature, but here we choose to work with the Cunningham-Price-Moncrief function, as defined in Ref.~\cite{Martel:2005ir}. Namely, using the Regge-Wheeler gauge and working in Schwarzschild coordinates, the Fourier transform of $\Psi_{\rm ax}^{lm}$ is related to the metric functions appearing in Eq.~\eqref{axial} by~\cite{Martel:2005ir}:
\begin{align}
&\psi^{lm}_{\rm ax}(r) =-\frac{2r}{(l-1)(l+2)} \nonumber \\
&\times \left[i\sigma h_1^{lm}(r) +\left(h_0^{lm}\right)'(r)-\frac{2h_0^{lm}(r)}{r}\right]\,. \label{CPM_function}
\end{align}
The function $\psi_{\rm ax}(r)$ satisfies Eq.~\eqref{masterodd} with a source term given by
\begin{align}
S_{\rm ax}^{lm} = &
-\frac{16 \sqrt{2} \pi  r f(r) \left[r \left(\tilde{\cal Q}^{(0)}_{l m }\right)'(r)+\tilde{\cal Q}^{(0)}_{l m }(r)\right]}{\sqrt{l (l+1)} \left(l^2+l-2\right)}\,.
\label{eq:sourceax}
\end{align}
Similarly to the polar sector, one can also reconstruct the axial metric functions using $\psi_{\rm ax}(r)$, $\psi_{\rm ax}'(r)$ and $\tilde{\cal Q}^{(0)}_{l m}(r)$. From the perturbed Einstein equations we find:
\begin{align}
    h_0(r)&= -\frac{f(r)}{2}\frac{d\left[r \psi_{\rm ax}(r)\right]}{dr}-\frac{8 \sqrt{2} \pi  r^3\tilde{\cal Q}^{(0)}_{l m }(r)}{\sqrt{l (l+1)} \left(l^2+l-2\right)}\,,\\
    h_1(r)&= \frac{i r \sigma   \psi_{\rm ax}(r)}{2 f(r)}\,.
\end{align}
These expressions reproduce the ones found in Ref.~\cite{Lousto:2005xu} (see their App. B), after noticing that there is an overall minus sign difference between Eq.~(B.21) of Ref.~\cite{Lousto:2005xu} and the formulas shown here. This is due to the fact that our normalization for $\psi^{lm}_{\rm ax}$ and our definition of $\tilde{\cal Q}^{(0)}_{l m }(r)$ differ from Ref.~\cite{Lousto:2005xu} by an overall minus sign. As in the polar sector, from these relations it follows that $h_0^{l,m}(\sigma; r)^*=(-1)^m h_0^{l,-m}(-\sigma; r)$ and $h_1^{l,m}(\sigma; r)^*=(-1)^m h_1^{l,-m}(-\sigma; r)$.

\subsection{Axial perturbations of a black hole surrounded by a spherical scalar cloud}\label{app:axial_epsilon2}
Let us now consider axial perturbations of the metric~\eqref{EF_perts}. As argued in Sec.~\ref{sec:Oq1epsilon2}, we neglect the $v$ dependence of $\delta M$ and $\delta\lambda$. Since the spacetime is spherically symmetric we also use the decomposition~\eqref{decom} and~\eqref{axial}, but with $t \to v$ in Eq.~\eqref{decom}. Using this decomposition and Fourier transforming the point particle's stress-energy tensor, we find that the $v\theta$, $r\theta$ and $\theta\phi$ components of the perturbed Einstein's equations give the following inhomogeneous equations, respectively:
\begin{widetext}
\begin{align}
&\left(1-\frac{2M+2\epsilon^2\delta M(r)}{r}\right)h_0''(r)+\left[i \sigma  \left(\epsilon ^2 \delta \lambda(r)-1\right)-\frac{\epsilon ^2 (r-2 M) \delta \lambda'(r)}{r}\right]h_0'(r)+\frac{i \sigma  \left(r-2 M-2 \epsilon ^2 \delta M(r)\right)}{r}h_1'(r)\nonumber\\
&+\frac{4 M-r \left(l^2+l-2 i r \sigma \right)}{r^3}h_0(r)
 +2\epsilon^2\Big\{2 \delta M(r)+r \left\{-3 M \delta \lambda '(r)+r \left[(2 M-r) \delta \lambda ''(r)-i \sigma  \delta \lambda (r)+\delta M''(r)\right]\right\}\nonumber\\
&-2r^2\left[\tilde R'(r) \left((r-2 M) \tilde R'(r)^*+i r \omega  \tilde R(r)^*\right)+r \tilde R(r) \left(\mu ^2 \tilde R(r)^*-i \omega  \tilde R'(r)^*\right)\right]\Big\}\frac{h_0(r)}{r^3}\nonumber\\
&+ \left[\sigma  \left(\sigma +\frac{2 i (r-2 M)}{r^2}\right)\right]h_1(r)+\epsilon^2 \frac{i \sigma  \left\{-4 \delta M(r)+r \left[(2 M-r) \delta \lambda '(r)+i r \sigma  \delta \lambda (r)\right]\right\}}{r^2}h_1(r)=\frac{8 \sqrt{2} \pi  r}{\sqrt{l (l+1)}}\tilde{\cal Q}^{{\rm EF}(0)}_{l m }(r)\,,\label{EFEax1}\\
&\left(\epsilon ^2 \delta \lambda (r)-1\right)h_0''(r)+\epsilon ^2 \delta \lambda '(r) h_0'(r)+i \sigma  \left(\epsilon ^2 \delta \lambda (r)-1\right)h_1'(r)+\frac{2-2 \epsilon ^2 \left(r \delta \lambda '(r)+\delta \lambda (r)\right)}{r^2}h_0(r)\nonumber\\
&-\frac{l^2+l+2 i r \sigma -2}{r^2}h_1(r)+\epsilon^2 \Big\{\left[i r^2 \sigma -2 (M+r)\right] \delta \lambda '(r)+2 r \left[(2 M-r) \delta \lambda ''(r)+\delta M''(r)\right]+2 i r \sigma  \delta \lambda (r)\nonumber\\
&-4 r\left[\tilde R'(r) \left((r-2 M) \tilde R'(r)^*+i r \omega  \tilde R(r)^*\right)+r \tilde R(r) \left(\mu ^2 \tilde R(r)^*-i \omega  \tilde R'(r)^*\right)\right]\Big\}\frac{h_1(r)}{r^2}
=\frac{8 i \sqrt{2} \pi  r}{\sqrt{l (l+1)}}\tilde{\cal Q}^{{\rm EF}}_{l m }(r)\,,\label{EFEax2}\\
&\left(1-\epsilon ^2 \delta \lambda (r)\right)h_0'(r)+\left(1-\frac{2M+2\epsilon^2\delta M(r)}{r}\right)h_1'(r)+\left(\frac{2 M}{r^2}-i \sigma\right)h_1(r)\nonumber\\
&+\epsilon^2\frac{2 \delta M(r)+r \left[(r-2 M) \delta \lambda '(r)+i r \sigma  \delta \lambda (r)-2 \delta M'(r)\right]}{r^2}h_1(r)=\frac{8 i \sqrt{2} \pi  r^2}{\sqrt{(l-1) l (l+1) (l+2)}}\tilde{\cal D}^{{\rm EF}}_{l m }(r)\label{EFEax3}\,.
\end{align}
\end{widetext}
These equations should be solved jointly with Eqs.~\eqref{eq_radial_KG_EF},~\eqref{dMr_final} and~\eqref{dlambdar_final}, which can be used to eliminate derivatives of $\delta M(r)$, $\delta \lambda(r)$, $\tilde R'(r)$ and $\tilde R'(r)^*$. This system of equations can be turned into a single inhomogeneous differential equation by defining a generalized master function for axial perturbations. Inspired by Eq.~\eqref{CPM_function} we define a new radial function given by
\begin{align}
&\bar\psi^{lm}_{\rm ax}(r) =-\frac{2r e^{-i \sigma \bar{r}_{*}} e^{-\epsilon^{2} \delta \lambda(r)}}{(l-1)(l+2)} \nonumber \\
&\times \left[i\sigma h_1^{lm}(r) +\left(h_0^{lm}\right)'(r)-\frac{2h_0^{lm}(r)}{r}\right]\,,
\label{psiodd_epsilon2}
\end{align}
where $\bar{r}_{*}$ is the generalized tortoise coordinate defined below Eq.~\eqref{masterodd_epsilon2}. Eliminating $h_1^{lm}$ from Eq.~\eqref{EFEax1} using~\eqref{psiodd_epsilon2}, gives us an equation that relates $h_0^{lm}$ with $\bar\psi^{lm}_{\rm ax}(r)$, $(\bar\psi^{lm}_{\rm ax})'(r)$ and $\tilde{\cal Q}^{{\rm EF}(0)}_{l m }(r)$. Using the resulting equation for $h_0^{lm}$ and~\eqref{psiodd_epsilon2} we can then eliminate $h_1^{lm}$ and $h_0^{lm}$ from Eq.~\eqref{EFEax2}, which becomes a single differential equation for $\bar\psi^{lm}_{\rm ax}$ that can be written in the form of Eq.~\eqref{masterodd_epsilon2} with a source term given by
\begin{align}
\bar{S}_{\rm ax}^{lm}(r) = &\frac{16 \sqrt{2} \pi r e^{-i \sigma \bar{r}_{*}}F_*(r)}{\sqrt{l (l+1)} (l^2+l-2)}\nonumber\\
&\times \frac{\left[r \sigma  \tilde{\cal Q}^{{\rm EF}}_{l m }(r)-r\left(\tilde{\cal Q}^{{\rm EF}(0)}_{l m }\right)'(r)-\tilde{\cal Q}^{{\rm EF}(0)}_{l m }(r)\right]}{\sqrt{l (l+1)} (l^2+l-2)}\,, \label{axial_source_epsilon2}
\end{align}
where we recall that Eq.~\eqref{masterodd_epsilon2} is formally only valid up to order $\epsilon^2$. 
As a consistency check, one can show that~\eqref{EFEax3} is automatically satisfied after eliminating $h_1^{lm}$ and $h_0^{lm}$ from~\eqref{EFEax3} using the procedure just outlined, and then making use of Eq.~\eqref{masterodd_epsilon2} and the conservation of the stress-energy tensor that allows to write $\tilde{\cal D}^{{\rm EF}}_{l m }$ in terms of $\tilde{\cal Q}^{{\rm EF}}_{l m }$ and $\tilde{\cal Q}^{{\rm EF}(0)}_{l m }$. We also note that the homogeneous part of Eq.~\eqref{masterodd_epsilon2} agrees with Ref.~\cite{Bamber:2021knr}, where axial perturbations of the metric~\eqref{EF_perts} were also discussed, but in the context of the computation of quasinormal modes.
\section{Coefficients of the scalar perturbation equation}\label{app:scalar_source}
In this Appendix we show the explicit form of the radial functions appearing in Eq.~\eqref{l1_source}. 

\subsection{Using the Regge-Wheeler and Zerilli gauge}
In Regge-Wheeler gauge, the radial functions appearing in Eq.~\eqref{l1_source} take the form
\begin{widetext}
\begin{align}
P_{lm} & =  H_0\frac{(f-1) f R'-r \omega  R (\sigma +2 \omega )}{2 f}+ i H_1 f \left[r (\sigma +2 \omega) R'+2 \omega  R\right]- K\frac{R \left[f \ell_i (\ell_i+1)+r^2 \sigma  \omega \right]}{r}- K' R' f^2 r
\nonumber\\
& -H_2\frac{f \left[R \left\{r^2 \omega  (\sigma +2 \omega )-2 f \left(\ell_i^2+\ell_i+\mu^2 r^2\right)\right\}+(f-1) f r R'\right]}{2 r} + \frac{1}{2}H_0' R'f r + i \omega H_1' R f r + \frac{1}{2} H_2' R' f^3 r\,,\label{Plm}\\
\hat P_{lm} & = \frac{R \left(H_0-f^2 H_2\right)}{2 r} \,,\\
A_{lm} & = -\frac{f \left[h_1 \left\{(f-1) R-2 f r R'\right\}-f r R h_1'\right]-i r
   h_0 R (\sigma +2 \omega )}{r^2}\,.\label{Alm}
\end{align}
\end{widetext}
In the expressions above $f:=f(r)=1-2M/r$, and it is implicitly assumed that the functions $\{H_0,H_1,H_2,K\}$ are computed with angular numbers $l,m$ and $R$ with $n_i,\ell_i$. As explained in Sec.~\ref{sec:GRperts}, for the particular case in which $l=1$, one can use the Zerilli gauge to simplify the source term by setting $K^{l=1}(r)=0$ in~\eqref{Plm} and $h^{l=1}_1(r)=0$ in~\eqref{Alm}. If $l=0$ instead, the Zerilli gauge corresponds to setting $H^{l=0}_1(r)=0$ and $K^{l=0}(r)=0$ in~\eqref{Plm}. Finally, we should also note that when deriving the expressions for the source term of Eq.~\eqref{KG_ep1q1}, we made use of Eq.~\eqref{eq_radial_KG} and the second-order differential equation that spherical harmonics satisfy [see Eq.~\eqref{ddY} in Appendix~\ref{app:mass_vs_charge}], in order to substitute second-order derivatives of $R(r)$ and of the spherical harmonics.
\subsection{Using the singular gauge for polar, dipolar metric perturbations}
As mentioned in Sec.~\ref{sec:GRperts}, another useful gauge for polar metric perturbations with $l=1$ is the singular gauge in which the metric perturbations take the form of Eqs.~\eqref{polar_singular}~--~\eqref{eq:eta1s}. In this gauge, the $l=1$ polar radial functions appearing in Eq.~\eqref{l1_source} are given by:
\begin{widetext}
\begin{align}
P^s_{1m} & = i H^s_1 f \left[r (\sigma +2 \omega) R'+2 \omega  R\right]-H^s_2\frac{f \left[R \left\{r^2 \omega  (\sigma +2 \omega )-2 f \left(\ell_i^2+\ell_i+\mu^2 r^2\right)\right\}+(f-1) f r R'\right]}{2 r}\nonumber\\
&-\frac{2 f^2 R' \eta^s_1}{r}
+ i \omega (H^s_1)' R f r + \frac{1}{2} (H^s_2)' R' f^3 r\,,\\
\hat P^s_{1m} & =-R\frac{f \left[f r H_2^s-2 f r (\eta_1^s)'+2 (f-1) \eta_1^s\right]}{2 r^2}+\frac{2 f^2 R' \eta^s_1}{r} \,,
\end{align}
\end{widetext}
where we used the superscript ``$s$'' to emphasize that these quantities are in the singular gauge. Results for the scalar power when using this gauge are shown in App.~\ref{app:fluxsingular}.

\section{Integral of the product of three spherical harmonics}\label{app:sphharmonics}
In this Appendix we show how to explicitly compute the integrals~\eqref{IntP}~--~\eqref{IntA} in terms of Wigner 3-j symbols. The integral~\eqref{IntP} is a particular case of the generic integral between three spin-weighted spherical harmonics~\cite{Spiers:2023mor}:
\begin{equation}\label{Cdef}
C^{ \ell_j m_j s_j}_{l m s \ell_i m_i s_i}:=\int {}_{s_j}Y^*_{\ell_j m_j}\,{}_{s}Y_{l m}\,{}_{s_i}Y_{\ell_i m_i}d\Omega\,,
\end{equation}
where ${}_{s}Y_{\ell m}$ are spin-weighted spherical harmonics with spin weight $s$. This integral can be explicitly evaluated using (see Sec. IV.A in Ref.~\cite{Spiers:2023mor})
\begin{align}\label{Cdef_3j}
C^{ \ell_j m_j s_j}_{l m s \ell_i m_i s_i} &= (-1)^{m_j+s_j}\sqrt{\frac{(2\ell_j+1)(2l+1)(2\ell_i+1)}{4\pi}}\nonumber\\
&\quad	\times \begin{pmatrix} \ell_j &  l &  \ell_i \\ s_j & -s & -s_i\end{pmatrix}
\begin{pmatrix} \ell_j &  l &  \ell_i \\ -m_j & m & m_i\end{pmatrix}\!,\!
\end{align}
where the arrays are Wigner 3-j symbols. From the properties of the Wigner 3-j symbols we find that the integral above vanishes unless the following conditions are satisfied: $-m_j+m+m_i=0$, $s_j-s-s_i=0$, $|\ell_j-\ell_i|\leq l\leq \ell_j+\ell_i$ and $\ell_j+\ell_i + l$ is an integer (see Chapter 34 in Ref.~\cite{DLMF}). From the properties of the 3-j symbols it also follows that (see Ref.~\cite{Spiers:2023mor} for a full set of symmetries)
\begin{equation}\label{C_symmetry}
C^{\ell_j m_j s_j}_{l m s \ell_i m_i s_i} = (-1)^{\ell_j+l+\ell_i}C^{\ell_j m_j -s_j}_{l m -s \ell_i m_i -s_i}\,.
\end{equation}

In particular, the integral~\eqref{IntP} can be evaluated by computing $C^{ \ell_j m_j 0}_{l m 0 \ell_i m_i 0}$, and it follows directly from the properties of $C^{\ell_j m_j s_j}_{l m s \ell_i m_i s_i}$ just mentioned above, that~\eqref{IntP} vanishes unless it satisfies the selection rules discussed in Sec.~\ref{sec:KGperts}. In particular the property~\eqref{C_symmetry} implies that $C^{ \ell_j m_j 0}_{l m 0 \ell_i m_i 0}$ vanishes unless $\ell_j+\ell_i + l$ is an even number. The integrals~\eqref{IntHatP} and~\eqref{IntA} can also be similarly computed by using the following relations between vector spherical harmonics and spin-weighted spherical harmonics (see Sec. IV.B in Ref.~\cite{Spiers:2023mor}):
\begin{align}
\mathbf{Y}^{\ell m}_{a} &=\frac{\lambda_{\ell,1}}{2}
\left({}_{-1}Y_{\ell m}\tilde m_a-{}_1Y_{\ell m}\tilde m^*_a\right)\,,\\
\mathbf{X}^{\ell m}_{a} &= -\frac{\lambda_{\ell,1}}{2}i\left({}_{-1}Y_{\ell m}\tilde m_a+{}_1Y_{\ell m}\tilde m^*_a\right)\,,
\end{align}
where, following~\cite{Spiers:2023mor}, we defined the complex null covector $\tilde m_a=(1,i \sin\theta)$ and $\lambda_{\ell,1}=\sqrt{(\ell+1)!/(\ell-1)!}$. Inserting in the integral~\eqref{IntHatP} we find 
\begin{widetext}
\begin{eqnarray}
\int \, Y^*_{\ell_j m_j}\mathbf{Y}_a^{lm}\mathbf{Y}_b^{\ell_i m_i}\gamma^{ab} d\Omega&=&-\frac{\lambda_{l,1}\lambda_{\ell_i,1}}{2}\left(\int \, Y^*_{\ell_j m_j}\,{}_{-1}Y_{l m}\,{}_{1}Y_{\ell_i m_i} d\Omega + \int \, Y^*_{\ell_j m_j}\,{}_{1}Y_{l m}\,{}_{-1}Y_{\ell_i m_i} d\Omega\right)\nonumber\\
&=&-\frac{\lambda_{l,1}\lambda_{\ell_i,1}}{2}\left(C^{ \ell_j m_j 0}_{l m -1 \ell_i m_i 1}+C^{ \ell_j m_j 0}_{l m 1 \ell_i m_i -1}\right)\nonumber\\
&=&-\frac{\lambda_{l,1}\lambda_{\ell_i,1}}{2}\left(C^{ \ell_j m_j 0}_{l m -1 \ell_i m_i 1} +(-1)^{\ell_j + l+ \ell_i} C^{ \ell_j m_j 0}_{l m -1 \ell_i m_i 1}\right)\,,
\end{eqnarray}
\end{widetext}
where we used Eq.~\eqref{C_symmetry} in the last step. Similarly for the integral~\eqref{IntA} we find
\begin{widetext}
\begin{eqnarray}
\int \, Y^*_{\ell_j m_j}\mathbf{X}_a^{lm}\mathbf{Y}_b^{\ell_i m_i}\gamma^{ab} d\Omega&=&i\frac{\lambda_{l,1}\lambda_{\ell_i,1}}{2}\left(\int \, Y^*_{\ell_j m_j}\,{}_{-1}Y_{l m}\,{}_{1}Y_{\ell_i m_i} d\Omega - \int \, Y^*_{\ell_j m_j}\,{}_{1}Y_{l m}\,{}_{-1}Y_{\ell_i m_i} d\Omega\right)\nonumber\\
&=&i\frac{\lambda_{l,1}\lambda_{\ell_i,1}}{2}\left(C^{ \ell_j m_j 0}_{l m -1 \ell_i m_i 1} - C^{ \ell_j m_j 0}_{l m 1 \ell_i m_i -1}\right)\nonumber\\
&=&i\frac{\lambda_{l,1}\lambda_{\ell_i,1}}{2}\left(C^{ \ell_j m_j 0}_{l m -1 \ell_i m_i 1} - (-1)^{\ell_j + l+ \ell_i} C^{ \ell_j m_j 0}_{l m -1 \ell_i m_i 1}\right)\,.
\end{eqnarray}
\end{widetext}
From the expressions above it follows that~\eqref{IntHatP} vanishes unless $\ell_j+l+\ell_i$ is an even number, whereas~\eqref{IntA} vanishes unless $\ell_j+l+\ell_i$ is an odd number. Together with the remaining properties of $C^{\ell_j m_j s_j}_{l m s \ell_i m_i s_i}$ mentioned above, this implies that~\eqref{IntHatP} and ~\eqref{IntA} vanish unless they satisfy the selection rules discussed in Sec.~\ref{sec:KGperts}.
\section{Eigenfrequency shifts induced by the metric perturbations}\label{app:freq_shift}
In Sec.~\ref{sec:KGperts} we argued that Eq.~\eqref{KG_genericl} for $\ell_j=\ell_i$ and $m_j=m_i$, can be used to compute corrections to the cloud's eigenfrequencies, which are induced by the metric perturbations. Let us see this more explicitly in this Appendix. 

As done in quantum mechanics, the usual approach to compute the corrections to the eigenfrequencies is to expand them as $\omega=\omega^{(0)}+q\omega^{(q)}+\ldots$, where again  $\omega^{(0)}=\omega_{n_i \ell_i m_i}$ are the eigenfrequencies in the background Schwarzschild BH, and $\omega^{(q)}$ are the corrections we are interested in. If we do this from the very beginning, the perturbation scheme in Sec.~\ref{sec:pert_scheme} needs to be slightly modified in order to take into account that the scalar field depends on the eigenfrequency: $\Phi_{\omega}=\epsilon\Phi_{\omega}^{(1)}+\epsilon q\Phi_{\omega}^{(q,1)} +\ldots$, where we added the subscripts $\omega$ in order to emphasize the dependence on the frequency. Upon inserting in the Klein-Gordon [see Eq.~\eqref{KGeq}] and expanding up to order $\mathcal{O}(\epsilon^1,q^1)$, as done in Sec.~\ref{sec:pert_scheme} we find~\cite{Hussain:2022ins}
\begin{align}
    &\epsilon\left(\Box_{\omega^{(0)}}^{(0)}- \mu^2\right)\Phi_{\omega^{(0)}}^{(1)}+\epsilon q\left(\Box_{\omega^{(0)}}^{(0)}- \mu^2\right)\Phi_{\omega^{(0)}}^{(q,1)} \nonumber\\
    &+ \epsilon q \omega^{(q)}\frac{\partial \Box_{\omega}^{(0)}\Phi_{\omega}^{(1)}}{\partial\omega}|_{q=0} = \epsilon q S_{\omega^{(0)}}^{\Phi}[h^{(0)},\Phi_{\omega^{(0)}}^{(1)}]\,.
\end{align}
Here the term of order $\mathcal{O}(\epsilon^1,q^0)$ gives $\left(\Box_{\omega^{(0)}}^{(0)}- \mu^2\right)\Phi_{\omega^{(0)}}^{(1)}=0$ which is simply Eq.~\eqref{KG_ST_10} (solved in Sec.~\ref{sec:QBSs}). Instead, the terms of order $\mathcal{O}(\epsilon^1,q^1)$ reduce to Eq.~\eqref{KG_ep1q1} but with the addition of the term that depends on $\omega^{(q)}$. For $\Phi^{(1)}$ in a given eigenstate $\{n_i,\ell_i,m_i\}$ of Eq.~\eqref{KG_ST_10}, the operator $\partial \Box_{\omega}^{(0)}\Phi_{\omega}^{(1)}/\partial\omega|_{q=0}$ is defined by
\begin{equation}~\label{eq:shift_omegaq}
    \frac{\partial \Box_{\omega}^{(0)}\Phi_{\omega}^{(1)}}{\partial\omega}|_{q=0}= 2\omega^{(0)} f(r)^{-1} R_{n_i\ell_i}(r) Y_{\ell_i m_i}(\theta,\phi) e^{-i\omega t}\,,
\end{equation}
with $f(r)=1-2M/r$. Following the same procedure as in Sec.~\ref{sec:KGperts}, one can see that the terms of order $\mathcal{O}(\epsilon^1,q^1)$ therefore reduce to Eq.~\eqref{KG_nonsepar} but with the additional term $\omega^{(q)}\partial \Box_{\omega}^{(0)}\Phi_{\omega}^{(1)}/\partial\omega|_{q=0}$. Upon projecting in the spherical harmonics, as done below Eq.~\eqref{KG_nonsepar}, one concludes that if $\ell_j\neq \ell_i$ or $m_j\neq m_i$ the final equation we need to solve is Eq.~\eqref{KG_genericl}, as we assumed in the main text. On the other hand, if $\ell_j=\ell_i$ and $m_j=m_i$ one ultimately gets
\begin{align}\label{eq:omegap}
&\left[\frac{d^2}{dr^2_*}+\left(\omega^{(0)}\right)^2 -V_i\right]\tilde{Z}^{\ell_i m_i}_{+}
+2\omega^{(0)}\omega^{(q)} r R_{n_i\ell_i}\nonumber\\
&=\tilde{S}^{\ell_i,\ell_i}_{m_i,m_i}(\omega^{(0)},r)|_{\sigma=0}\,,
\end{align}
where we used the fact that $Z^{\ell_i m_i}_{+}= \tilde Z^{\ell_i m_i}_{+}\delta\left(\sigma\right)$, and similarly for the metric perturbations (see Sec.~\ref{sec:KGperts}), and therefore the resulting equation only as support at $\sigma=0$.
In analogy with quantum mechanics, we can isolate $\omega^{(q)}$ from this equation by defining an appropriately chosen product $\langle A| B\rangle$ such that: (i) it vanishes if $A$ and $B$ correspond to two different eigenstates $|n_i\ell_i m_i\rangle$; (ii) it is finite if $A=B=|n_i\ell_i m_i\rangle$ are the same eigenstate and (iii) the order $\mathcal{O}(q^0)$ Klein-Gordon equation is self-adjoint with respect to this product (discussions on how to define such product in similar situations can be found in e.g. Refs.~\cite{Leung:1997was,Mark:2014aja,Hussain:2022ins,Green:2022htq}), which for our purposes ultimately means
\begin{equation}
\left\langle r R_{n_i\ell_i} |\mathcal{\hat O}\tilde{Z}^{\ell_i m_i}_{+}\right\rangle=
\left\langle \mathcal{\hat O}(r R_{n_i\ell_i})|\tilde{Z}^{\ell_i m_i}_{+}\right\rangle =0\,,
\end{equation}
where we defined the operator $\mathcal{\hat O}:=\left[\frac{d^2}{dr^2_*}+\left(\omega^{(0)}\right)^2 -V_i\right]$ and, in the last step, we used Eq.~\eqref{eq_radial_KG}. Then, in analogy with quantum mechanics, we can act on the left of Eq.~\eqref{eq:omegap} with $\langle r R_{n_i\ell_i}|$ and use the self-adjoint property discussed above to eliminate the first term~\cite{Mark:2014aja,Hussain:2022ins}, which then allows to find 
\begin{equation}\label{eq:omegaq}
\omega^{(q)}=\frac{\langle r R_{n_i\ell_i}|\tilde{S}^{\ell_i,\ell_i}_{m_i,m_i}(\omega^{(0)},r)\rangle}{2\omega^{(0)}\langle r R_{n_i\ell_i}|r R_{n_i\ell_i} \rangle}\,.
\end{equation}
If we go to higher orders in perturbation theory, a similar procedure can also be used to compute frequency shifts of order $\epsilon^{2}$ that arise due to the metric perturbation $g_{\mu\nu}^{(2)}$ [cf.~\eqref{eq:metric_expansion}]. Here we will not study this issue further, leaving a more detailed study of these shifts for future work. 

An equivalent approach, is to follow the perturbation scheme in Sec.~\ref{sec:pert_scheme} as is, and only expand $\omega$ {\it a posteriori}, i.e., use the expansion $\omega=\omega^{(0)}+q\omega^{(q)}$ in Eq.~\eqref{KG_genericl}, where again $\omega^{(0)}=\omega_{n_i \ell_i m_i}$. Doing so, we find~\footnote{Here we are already using the fact that $Z^{\ell_j m_j}_{+}= \tilde Z^{\ell_j m_j}_{+}\delta\left(\sigma-(m_j-m_i)\Omega_p\right)$, and similarly for the metric perturbations inside the source term of Eq.~\eqref{KG_genericl}.}:
\begin{align}
&\left[\frac{d^2}{dr^2_*}+\left(\omega^{(0)}_+\right)^2 -V_j\right]\tilde{Z}^{\ell_j m_j}_{+}
+2q\omega^{(0)}_+\omega^{(q)} \tilde{Z}^{\ell_j m_j}_{+}\nonumber\\
&=\tilde{S}^{\ell_j,\ell_i}_{m_j,m_i}(\omega,r)|_{\sigma=(m_j-m_i)\Omega_p}\,,
\end{align}
where we defined $\omega^{(0)}_+=\omega^{(0)}+(m_j-m_i)\Omega_p$.
For $\ell_j\neq \ell_i$ and $m_j\neq m_i$, we can obtain a finite solution using~\eqref{full_sol_KG}, where terms related to $\omega^{(q)}$ only end up contributing at higher order in $q$ and therefore we can neglect them, as we argued in the main text. If instead $\ell_j=\ell_i$, $m_j=m_i$, one finds an equation identical to~\eqref{eq:omegap} but with $r R_{n_i\ell_i}$ replaced by $q\tilde{Z}^{\ell_i m_i}_{+}$ in the left-hand side. We can use the resulting equation to find $\omega^{(q)}$ by redefining $\tilde{Z}^{\ell_i m_i}_{+}$ as $\tilde{Z}^{\ell_i m_i}_{+} \to r R_{n_i\ell_i}/q +\tilde{Z}^{\ell_i m_i}_{+}$~\footnote{Notice that the term  $r R_{n_i\ell_i}/q$ simply contributes to a renormalization of the amplitude in the $\mathcal{O}(q^0)$ solution, therefore we are always allowed to do this redefinition.}. After using Eq.~\eqref{eq_radial_KG}, and discarding higher-order terms in $q$ we recover again Eq.~\eqref{eq:omegap}. We can then again find $\omega^{(q)}$ using Eq.~\eqref{eq:omegaq}.
\section{Relation between mass, angular momentum and Noether charge of a scalar cloud}\label{app:mass_vs_charge}
%
In this Appendix we provide a derivation of the relation between the total mass, angular momentum and Noether charge of the scalar cloud that we used in the main text, namely $M_b=\omega Q$ and $J_b=m_i Q$. Since the cloud's energy density is singular at the horizon in Schwarzschild coordinates (see Ref.~\cite{Bamber:2021knr} and Sec.~\ref{sec:Oq1epsilon2}), we work with Eddington-Finkelstein coordinates $(x^0,x^1,x^2,x^3)=(v,r,\theta,\phi)$, although we checked that our proof also works when using Schwarzschild coordinates. We decompose the scalar field as in Eq.~\eqref{background_scalar_EF} and take the background metric to be given by Eq.~\eqref{Sch_metric_EF}. As done in the main text, we neglect the slow decay of the cloud due to absorption at the horizon, i.e., we take $\omega \approx \Re(\omega)$. The Noether charge of the cloud, Eq.~\eqref{KG_charge}, then reads
\begin{equation}\label{Q_EF}
    Q = i \epsilon^2\int_V dr d\Omega\,r^2|Y(\theta,\phi)|^2 \mathcal{W}(r)= i \epsilon^2\int_{2M}^{\infty} dr\,r^2 \mathcal{W}(r)\,,
\end{equation}
where we used the orthonormality properties of the spherical harmonics and defined
%
$\mathcal{W}(r)=\tilde R(r)\frac{d\tilde R^*(r)}{dr}-\tilde R^*(r)\frac{d\tilde R(r)}{dr}$.
%
On the other hand, the cloud's angular momentum is given by~\cite{Schunck:1996he}
\begin{equation}
    J_b = \int_{\Sigma}d^3x \sqrt{-g}\,T^{0}_{3} = m_i Q\,,
\end{equation}
where the last step follows from directly evaluating $T^{0}_3$ using Eq.~\eqref{KG_ST}. The cloud's total mass is instead given by
\begin{widetext}
    \begin{eqnarray}\label{Mb_EF}
    M_b =&& -\int_{V}d^3x \sqrt{-g}\,T^{0}_0 =\nonumber\\ 
    &&\epsilon^2 \int_V dr d\Omega\,\left\{|Y(\theta,\phi)|^2 \left[|\tilde R(r)|^2 \left(r^2\mu^2+\frac{m_i^2}{\sin^2\theta}\right)+f(r)r^2\left|\frac{d\tilde R(r)}{dr}\right|^2\right]+|\tilde R(r)|^2 \left|\frac{\partial Y(\theta,\phi)}{\partial \theta}\right|^2\right\}\,.
    \end{eqnarray}
\end{widetext}
Although this expression seems very different from Eq.~\eqref{Q_EF} we will now show that it can be simplified by performing integration by parts. Let us first integrate the angular part of the last term in the integrand of Eq.~\eqref{Mb_EF} by parts: 
\begin{widetext}
 \begin{eqnarray}\label{int_theta}
    \int d\Omega\, \left|\partial_\theta Y(\theta,\phi)\right|^2 = &-& \int_0^{2\pi} d\phi\int_{0}^{\pi} d\theta\,Y^*(\theta,\phi)\partial_\theta[\sin\theta\,\partial_\theta Y(\theta,\phi)]\nonumber\\
    = &-& \int_0^{2\pi} d\phi\int_{0}^{\pi} d\theta\,\sin\theta|Y(\theta,\phi)|^2\left(\frac{m_i^2}{\sin^2\theta}-\ell_i(\ell_i+1)\right)\,,
\end{eqnarray}
\end{widetext}
where $\partial_\theta := \partial/\partial\theta$, we used the fact that boundary terms vanish since $\sin\theta=0$ at $\theta=0,\pi$, and we used the differential equation that spherical harmonics satisfy
\begin{equation}
    \frac{1}{\sin\theta}\partial_\theta[\sin\theta\,\partial_\theta Y(\theta,\phi)]=\left(\frac{m_i^2}{\sin^2\theta}-\ell_i(\ell_i+1)\right) Y(\theta,\phi)\,.\label{ddY}
\end{equation}
On the other hand, integrating the radial part of the second term by parts gives
\begin{widetext}
 \begin{eqnarray}\label{int_r}
   \int_{2M}^{\infty} dr\,f(r)r^2\left|\frac{d\tilde R(r)}{dr}\right|^2 = && \frac{1}{2}\int_{2M}^{\infty} dr\,f(r)r^2\left|\frac{d\tilde R(r)}{dr}\right|^2 + \frac{1}{2}\int_{2M}^{\infty} dr\,f(r)r^2\left|\frac{d\tilde R(r)}{dr}\right|^2 \nonumber\\
     = && -\frac{1}{2}\int_{2M}^{\infty} dr\,\tilde R^*(r)\frac{d[f(r)r^2\tilde R'(r)]}{dr}-\frac{1}{2}\int_{2M}^{\infty} dr\,\tilde R(r)\frac{d[f(r)r^2 (\tilde R^*)'(r)]}{dr} \nonumber\\
     = && -\int_{2M}^{\infty} dr\, \left\{|\tilde R(r)|^2 \left[\ell_i(\ell_i+1)+r^2\mu^2\right]-ir^2 \omega\mathcal{W}(r)\right\}\,,
\end{eqnarray}
\end{widetext}
where we used the fact that the boundary terms vanish since $f(2M)=0$ and $\lim_{r\to\infty}\tilde R(r) =0$, and we used Eq.~\eqref{eq_radial_KG_EF}. 
Substituting~\eqref{int_theta} and~\eqref{int_r} in Eq.~\eqref{Mb_EF} we then find 
\begin{equation}
    M_b = i \epsilon^2\omega \int_{2M}^{\infty} dr\, r^2 \mathcal{W}(r) =\omega Q\,,
\end{equation}
as promised. Our derivation assumed a non-spinning BH background and for simplicity we neglected the slow decay of the cloud. However, we conjecture that a similar derivation can be done in a Kerr BH background, where we recall that true bound states satisfying $\omega=m_i\Omega_H$ exist. In fact, it was shown in~\cite{Herdeiro:2017phl} using BH thermodynamics that ``Kerr BHs with bosonic hair'' satisfy $M_b=\omega J_b/m_i=\Omega_H J_b$ in the test field limit, in agreement with our expectation.
%
\section{Using the singular gauge for polar, dipolar metric perturbations to compute the scalar power}\label{app:fluxsingular}
\begin{figure*}[thb!]
    \centering
    \includegraphics[width=0.48\textwidth]{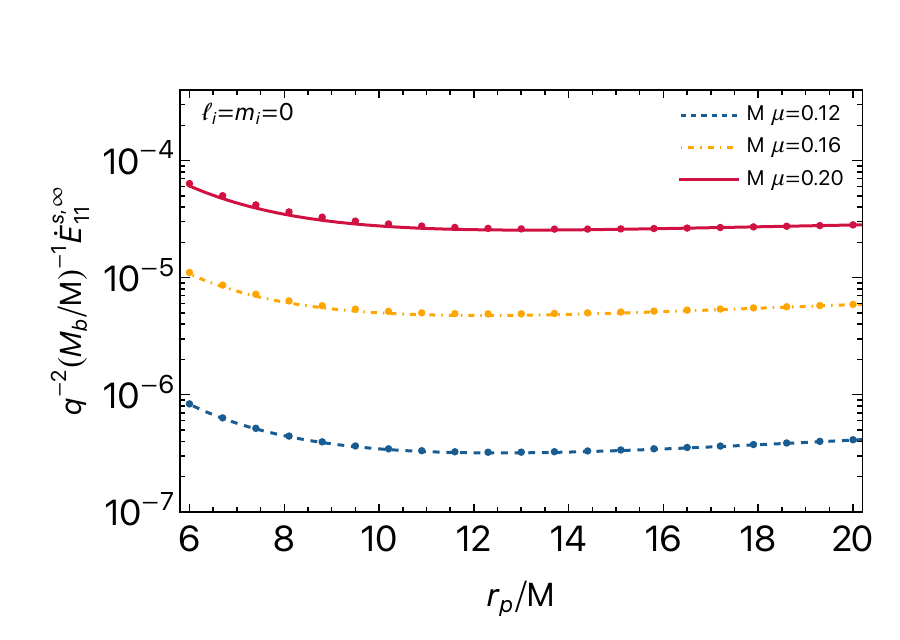}
     \includegraphics[width=0.48\textwidth]{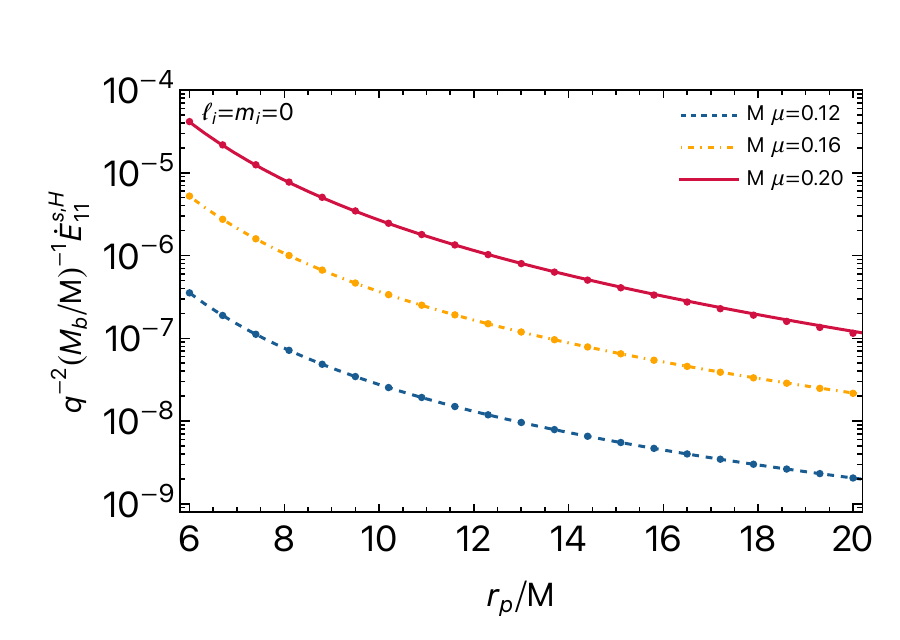}
    \includegraphics[width=0.48\textwidth]{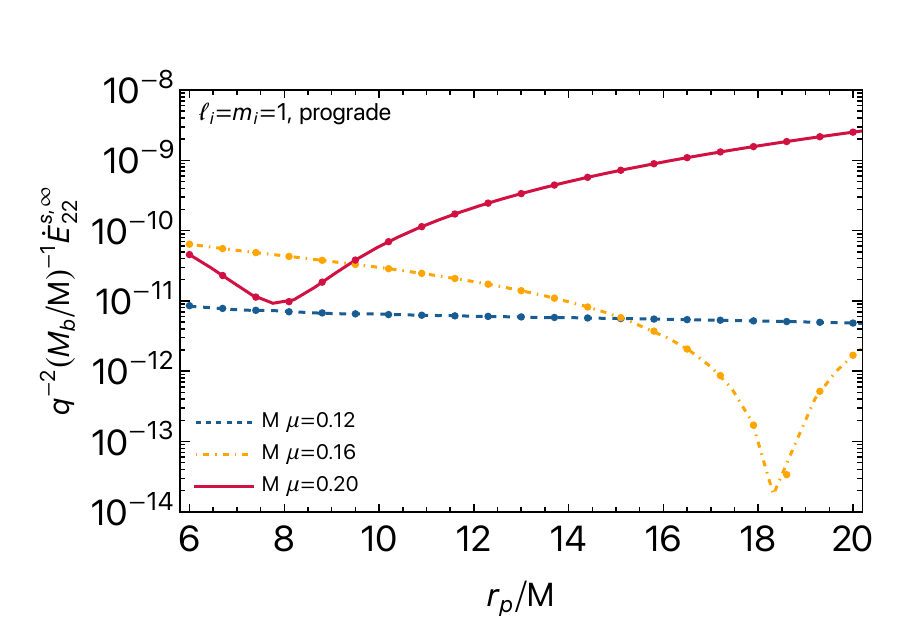}
     \includegraphics[width=0.48\textwidth]{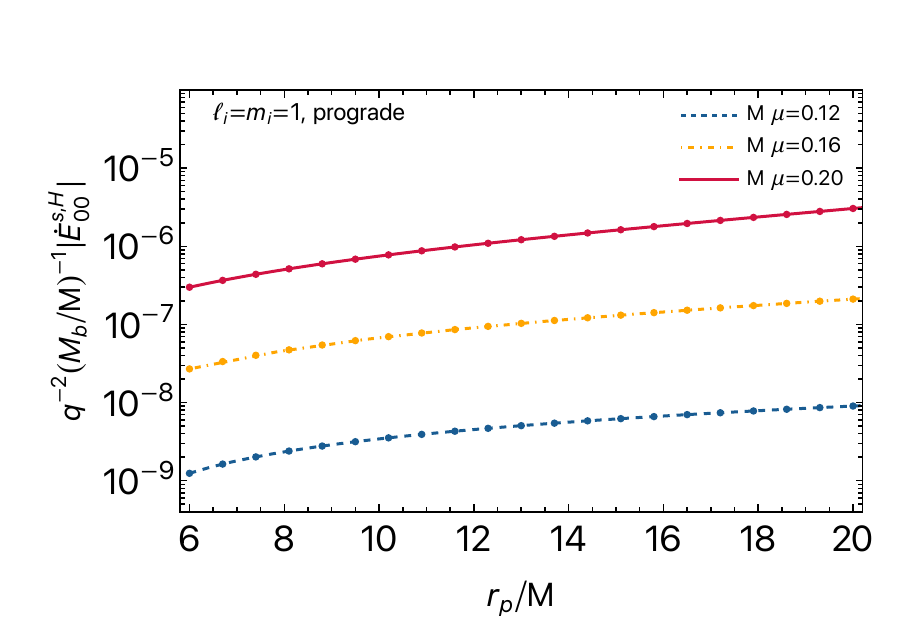}
     \includegraphics[width=0.48\textwidth]{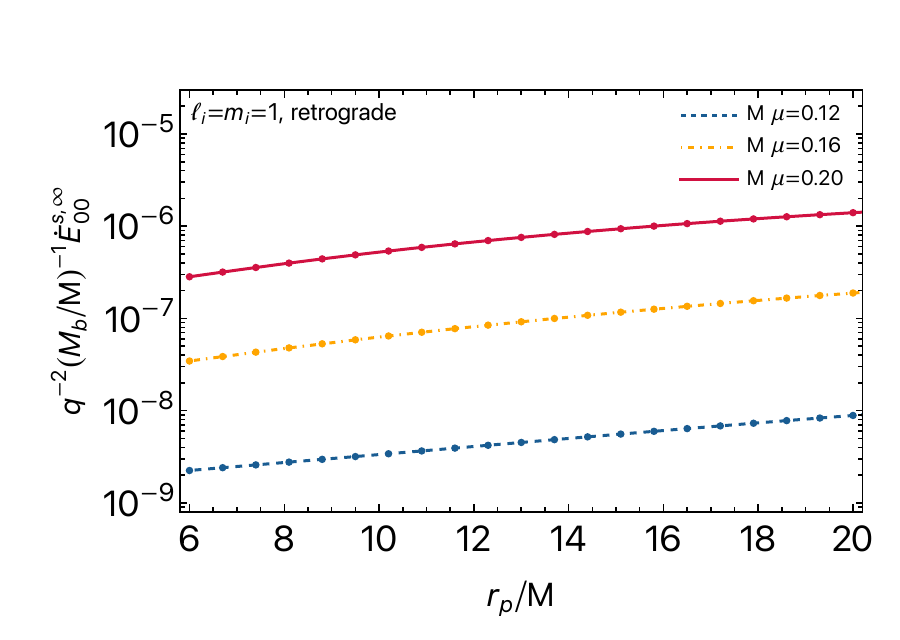}
     \includegraphics[width=0.48\textwidth]{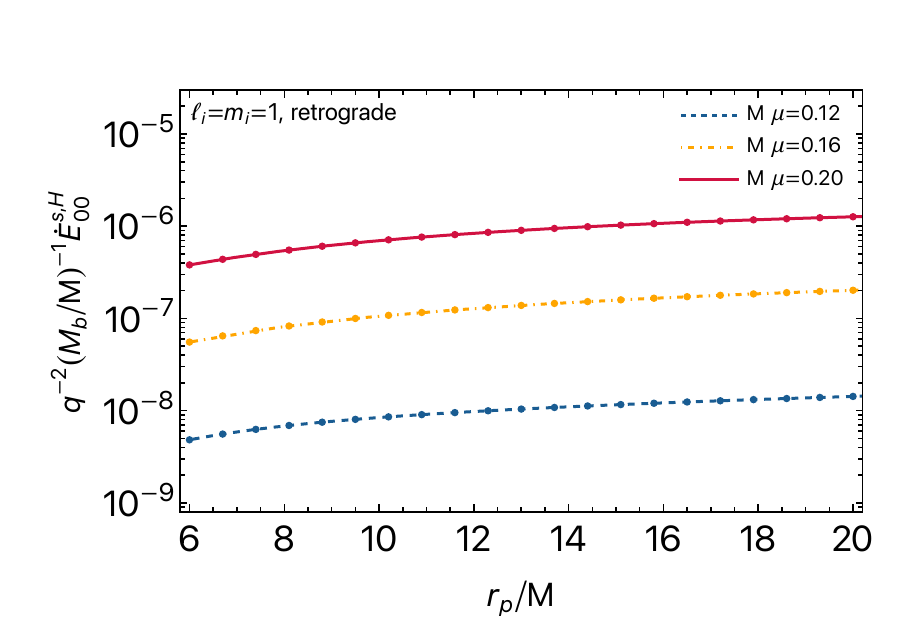}
    \caption{Comparison between the scalar power $\dot E_{\ell_j m_j}^{s}$ computed when using the Zerilli gauge for the $l=1$ polar metric perturbations (solid lines) against using the singular gauge (data points). Within numerical accuracy, the two gauge choices give the same results. Only the most relevant multipoles that are sourced by $l=1$ polar metric perturbations are shown. Namely, for a $\ell_i=m_i=0$ background scalar cloud, we show the $\ell_j=m_j=1$ multipole of the scalar power at infinity (top left panel) and at the horizon (top right panel), for a $\ell_i=m_i=1$ background scalar cloud and prograde orbits, we show the $\ell_j=m_j=2$ multipole of the scalar power at infinity (center left panel) and the $\ell_j=m_j=0$ multipole of the scalar power at the horizon (center right panel), whereas for a $\ell_i=m_i=1$ background scalar cloud and retrograde orbits, we show the $\ell_j=m_j=0$ multipole of the scalar power at infinity (bottom left panel) and at the horizon (bottom right panel). The results are shown as a function of the orbital radius $r_p/M$ and for different values of $M\mu$.}
    \label{fig:singVSzer}
\end{figure*}
As explained in Sec.~\ref{sec:GRperts}, throughout the main text we always used the Zerilli gauge for the $l=1$ polar metric perturbations. As a check of the robustness of our results, for the most relevant multipoles of the scalar power that depend on $l=1$ polar perturbations, we also checked our computations using the singular gauge that we introduced in Sec.~\ref{sec:GRperts}. Comparisons between results obtained with the Zerilli and the singular gauge are shown in Fig.~\ref{fig:singVSzer}. For the data points shown in the plots the relative difference between the two gauge choices is typically at the percent level or smaller, indicating that the results are in good agreement between the two gauge choices, as expected.
\section{Circular orbits around a black hole surrounded by a spherical scalar cloud}\label{app:app_geo}
In Sec.~\ref{sec:Oq1epsilon2} we showed how to compute the leading-order corrections to the Schwarzschild BH metric induced by a spherically symmetric scalar cloud. As we argued there, it is convenient to use ingoing Eddington-Finkelstein coordinates for this computation. Here we are interested in computing timelike circular geodesics in a metric of the form~\eqref{EF_perts}. Although the computation of circular geodesics around generic spherically symmetric spacetimes can be found in the literature (see e.g. Ref.~\cite{Cardoso:2008bp}), this is typically done using a standard Schwarzschild coordinate system $(t,r,\theta,\phi)$. Let us then generalize this computation to the case where the metric takes the generic form:
\begin{equation}
    ds^2 = g_{vv} dv^2 + 2 g_{vr} dv dr + r^{2} d\Omega^2\,.
\end{equation}
Following Refs.~\cite{Chandrasekhar:579245,Cardoso:2008bp} we can describe geodesic motion on this spacetime using the Lagrangian:
 \begin{equation}\label{Lag_PP}
     \mathcal{L} = \frac{1}{2}g_{\mu\nu}\dot x_p^{\mu} \dot x_p^{\nu}\,,
 \end{equation}
where $x_p^{\mu}=[v(\tau),r(\tau),\theta(\tau),\phi(\tau)]$ is the particle's worldline and an overdot denotes a derivative with respect to the proper time such that for a timelike geodesic we have $2\mathcal{L}=-1$. Focusing on equatorial geodesics where $\theta(\tau) = \pi/2$ and $\dot \theta(\tau) = 0$, the Lagrangian reads
 \begin{equation}\label{Lag_PP_eq}
     2\mathcal{L} = g_{vv} \dot v^{2} + 2 g_{vr} \dot v \dot r + r^{2} \dot \phi^{2}\,.
 \end{equation}
From this Lagrangian we can derive generalized momenta given by
\begin{equation}
    p_{v} := \frac{\partial \mathcal{L}}{\partial \dot v}  = g_{vv} \dot v + g_{vr} \dot r\,,\quad
    p_{\phi} := \frac{\partial \mathcal{L}}{\partial \dot \phi}  = r^2 \dot \phi\,,
\end{equation}
besides, $p_{r} = g_{vr} \dot v$ and $p_{\theta}=0$, since we are fixing $\dot \theta = 0$. Given that the Lagrangian~\eqref{Lag_PP_eq} does not depend on $v$ and $\phi$, it follows that $p_v$ and $p_{\phi}$ are constants of motion that describe the energy and angular momentum (per unit rest mass) of the point particle: $p_{v} = -E/m_p \equiv -\tilde E$, $p_{\phi} = L/m_p\equiv \tilde L$. 

Plugging these results back in the Lagrangian~\eqref{Lag_PP_eq} and equating $2\mathcal{L}=-1$, we find, after some algebra,
\begin{equation}
     \dot r^2 = V_r \equiv \frac{\tilde E^{2}}{g_{vr}^{2}} + \frac{\tilde  L^{2} g_{vv}}{r^{2} g_{vr}^2} + \frac{g_{vv}}{g_{vr}^{2}}\,.
\end{equation}
Circular orbits, which satisfy $\dot r=\ddot r=0$, exist when $V_r=dV_r/dr=0$. This requirement yields: 
 \begin{equation}
   \tilde E^{2} = \frac{2 g_{vv}(r_p)^2}{r_p g_{vv}'(r_p)-2
   g_{vv}(r_p)}\,,\, \tilde L^2 = \frac{r_p^3 g_{vv}'(r_p)}{2 g_{vv}(r_p)-r_p
   g_{vv}'(r_p)}\,,
\end{equation}
where we fixed $r=r_p$ and the prime stands for a derivative with respect to $r_p$. On the other hand, the orbital angular frequency associated with these circular orbits is given by
\begin{equation}
    \Omega_p := \frac{d\phi}{dv}= \frac{\dot \phi}{\dot v} =\sqrt{-\frac{g_{vv}'(r_p)}{2r_p}}\,.
\end{equation}
In the spacetime metric given by Eq.~\eqref{EF_perts}, these expressions reduce to Eqs.~\eqref{E_epsilon2}~--~\eqref{Omega_epsilon2}, where the order $\mathcal{O}(\epsilon^2)$ corrections are given by
\begin{widetext}
\begin{eqnarray}
\delta E(r_p) &=& \frac{(r_p-6 M)\delta M(r_p)+(r_p-2 M) \left[(6 M-2 r_p) \delta \lambda(r_p)+r_p(2 M-r_p) \delta\lambda'(r_p)+r_p\delta M'(r_p)\right]}{2 (3 M-r_p) \sqrt{r_p (r_p-3 M)}}\,,\label{deltaE}\\
\delta L(r_p) &=& \frac{r_p^2(2 M-r_p) \left[(2 M-r_p) \delta\lambda'(r_p)+\delta M'(r_p)\right]+r_p^2\delta M(r_p)}{2(r_p-3 M) \sqrt{M (r_p-3 M)}}\,,\label{deltaL}\\
\delta \Omega(r_p) &=& \frac{r_p (r_p-2 M) \delta\lambda'(r_p)+2 M \delta\lambda(r_p)-r_p\delta M'(r_p)+\delta M(r_p)}{2 \sqrt{M r_p^3}}\,.\label{deltaOmega}
\end{eqnarray}
\end{widetext}

Circular orbits are stable if $V_r''(r_p)<0$~\cite{Cardoso:2008bp}. For a spacetime perturbatively close to Schwarzschild, stable circular orbits should exist for $r_p>r_{\rm ISCO}$. Therefore the ISCO radius $r_{\rm ISCO}$ in this spacetime can be found by looking at the inflection point $V_r''(r_{\rm ISCO})=0$. Assuming an {\it ansatz} of the form $r_{\rm ISCO} = 6M +\epsilon^2\delta r_{\rm ISCO}$, and solving the equation perturbatively in $\epsilon^2$, we find that $\delta r_{\rm ISCO}$ is given by
\begin{align}\label{eq:deltarISCO}
   \delta r_{\rm ISCO} & = 6 \left[\delta M(6M)-96 M^3 \delta\lambda''(6 M)- 32 M^2\delta\lambda'(6 M) \right. \nonumber\\
   &\left. + 24 M^2 \delta M''(6M)-4 M \delta M'(6M)\right]\,.
\end{align}

\bibliography{bibliography.bib}

\end{document}